\documentclass[10pt,preprint]{aastex}

\usepackage{psfig}

\newcommand{\citepeg}[1]{\citep[{e.g.,}][]{#1}}

\slugcomment{Submitted to the Astrophysical Journal}

\begin{document}

\title{Towards a More Standardized Candle Using GRB Energetics and Spectra}

\author{Andrew S. Friedman\altaffilmark{1} and Joshua S. Bloom
\altaffilmark{1}$^,$\altaffilmark{2}$^,$\altaffilmark{3}}

\altaffiltext{1}{Harvard Smithsonian Center for Astrophysics, 
	60 Garden Street, Cambridge, MA 02138}
\altaffiltext{2}{Department of Astronomy, 601 Campbell Hall, 
	University of California at Berkeley, Berkeley, CA 94720}
\altaffiltext{3}{Harvard Society of Fellows, 78 Mount Auburn Street, 
	Cambridge, MA 02138}

\begin{abstract}

The use of $\gamma$-ray bursts (GRBs) energetics for cosmography has
long been advanced as a means to probe out to high redshifts, to the
epoch of deceleration. However, though relatively immune to systematic
biases from dust extinction, the prompt energy release in GRBs, even
when corrected for jetting geometry, is far from being a standard
candle. In this work, we explore the cosmographic potential of a GRB
standard candle based on the newly-discovered relation by Ghirlanda et
al.\ between the apparent geometry-corrected energies ($E_{\gamma}$)
and the peak in the rest frame prompt burst spectrum ($E_p$).  We
present an explicit, self-consistent formalism for correcting GRB
energies with a thorough accounting for observational
uncertainties. In contrast to other work, we demonstrate that the
current sample of 19 GRBs is not yet cosmographically competitive to
results from Type Ia supernovae, large-scale structure, and the
microwave background.  Although the $E_p$--$E_{\gamma}$ relation is a
highly significant correlation across a range of cosmologies [$0
\le\Omega_M,\ \Omega_{\Lambda} \le 2$], the goodness of fit of the
data to a power law ($E_p \propto E_{\gamma}^\eta$), depends strongly
on input assumptions. The most important of these assumptions concern
the unknown density (and density profile) of the circumburst medium,
the efficiency of converting explosion energy to $\gamma$--rays, data
selection choices for individual bursts (some of which were not
included in similar work), and assumptions in the error analysis.
Independent of assumptions, with very few low-$z$ bursts, the current
sample is most sensitive to $\Omega_M$ but essentially {\it
insensitive} to $\Omega_\Lambda$ (let alone the dark energy equation
of state $w$).  The proper use of the relation clearly brings GRBs an
impressive step closer toward a standardizable candle, but until the
physical origin of the $E_p$--$E_{\gamma}$ relation is understood,
calibrated with a training set of low redshift (e.g., cosmology
independent) bursts, and the major potential systematic uncertainties
and selection effects are addressed, we urge caution concerning claims
of the utility of GRBs for cosmography.

\end{abstract}
\keywords{cosmological parameters --- cosmology:\ observations --- gamma-rays:\ bursts}

\section{Introduction}

As ultra-luminous explosions from the death of massive stars,
gamma-ray bursts (GRBs) can, in principle, occur and be detected from
redshifts at the epoch of reionization, serving as unique
probes of the gas and metal-enrichment history in the early universe
\citepeg{loeb01,mesz03,inou04}. At such redshifts beyond $z \sim 7$,
the dynamics of the universal expansion is not yet dominated by the
cosmological constant $\Lambda$, so the construction of an
early-universe Hubble diagram using GRBs would complement cosmography
results found in the $\Lambda$-dominated regime at lower
redshifts. Since $\gamma$-rays penetrate dust, a standard candle
derived from GRB energetics could avoid potential systematic errors
inherent in supernovae (SNe) due to uncertainties in dust
extinction. Cosmological $k$-corrections for GRBs \citep{bloo01b} are,
in principle, more tractable than traditional optical $K$-corrections
for SNe; where GRB spectra are devoid of emission/absorption features
at $\gamma$-ray wavelengths, SNe spectra --- due to the variety of
filter bands used and uncertainty in the intrinsic spectral shape ---
are generally considered to contribute a redshift-dependent systematic
error to SNe Ia magnitudes \citep{wangy01}. Still, both samples
necessarily contend with unknown evolution of the standard candle;
but, owing to very different physics in the emission mechanisms, any
such evolution would unlikely be the same for GRBs and SNe Ia.

Early attempts to meaningfully constrain cosmological parameters using
GRB energetics were stymied \citepeg{derm92,rutl95,cohe97} by what is
now known \citep{bloo01b,schmidt01a} as a wide distribution --- more
than 3 orders-of-magnitude --- in the intrinsic isotropic-equivalent
energies ($E_{\rm iso}$) and luminosities of GRBs. The realization
that GRBs are a jetted phenomena \citep{harr99,stan99} led to the
discovery that the geometry-corrected prompt energy release
($E_{\gamma}$) in GRBs appears nearly constant
\citep[$\sim$10$^{51}$\,erg $\equiv 1$ foe;][]{frai01,pira01}. This,
along with the possibility of inferring GRB redshifts from the
$\gamma$-ray properties alone \citepeg{reic01,norr02}, renewed
enthusiasm for the cosmographic utility of GRBs
\citep{scha03a,taka03}. \citet{frai01} had noted this apparent
constancy for what was then the current sample of 17 GRBs with known
redshifts. \citet{bloo03}, with an expanded sample of 29 GRBs with
known $z$, later argued that even geometry-corrected energetics were
not sufficient for cosmography on both conceptual and empirical
grounds. First, while the physical motivation for a standard energy
release is plausible, the geometry correction of $E_{\rm iso}$ is
highly model dependent, requiring an inference of the nature of the
circumburst environment and assumptions about the structure of the
jet. This problem persists even with new energy corrections detailed
herein.  Second, the cosmographic utility of $E_{\gamma}$ was limited
by the presence of several low energy and high energy outliers,
comprising upwards of 20\% of the sample, spanning three orders of
magnitude (GRB 980425 aside). \citet{bloo03} argued that without an
energy-independent discriminant (such as properties of the afterglow)
such outliers could not be excluded (or re-calibrated) {\it a priori}
when constructing a GRB Hubble Diagram. A regularization of $E_{\gamma}$
would require a universal relation between $E_{\gamma}$ and other
observables.

The recent discovery of a connection between $E_{\gamma}$ and the peak
energy ($E_p$) in the rest-frame prompt burst spectrum \citep{ghir04}
is apparently such a universal relation spanning the hardest,
brightest bursts to the softest, faintest X-ray Flashes (XRFs;
\citealt{heis01}). In this paper, we demonstrate that this
$E_{\gamma}$--$E_p$ (``Ghirlanda'') relation can serve as an approximate
empirical correction to GRB energies, advancing GRB energetics towards
a more standardized candle. In presenting the formalism for correcting
GRB energetics, we draw a strong analogy between our corrections and
the empirical light-curve shape corrections (based upon the peak
brightness--decline rate correlation) used to standardize the peak
magnitudes of Type Ia SNe
\citep{phil93,hamu95,hamu96,ries95,ries96,perl97,tonr03}. In
\S~\ref{sec:cor} we confirm the $E_p$--$E_{\gamma}$ relation and show
that although the goodness of fit to the simple power-law relation is
highly sensitive to input assumptions, the correlation itself is still
highly significant over a variety of plausible cosmologies. In
\S~\ref{sec:standard} we introduce a new formalism, with an explicit
accounting for observational uncertainties, for correcting GRB
energies. In \S~\ref{sec:compare} we discuss similar work from
\citet{dai04} and \citet{ghir04b}, noting critical differences in our
respective methodologies and datasets. We then attempt to lay the
groundwork for identifying relevant systematic errors and selection
effects in \S~\ref{sec:sys} and end with a discussion of the future
prospects of an even more standardized GRB energy. Unless otherwise
noted, we assume a standard cosmology of ($\Omega_M$,
$\Omega_\Lambda$, $h = H_0 / 100 $km s$^{-1}$ Mpc$^{-1}$) $=$
($0.3,0.7,0.7$).

\section{GRB Energetics and the $E_p$--$E_{\gamma}$ Relation Revisited}
\label{sec:cor}

We compute the geometry-corrected prompt energy release in gamma-rays
($E_{\gamma}$) following \citet{bloo03} and the associated uncertainties
with a slightly improved formalism in \S~\ref{sec:egamma_err}. All
energies are computed using the ``top hat'' model prescription for the
jet: the energy per steradian in the jet is assumed to be uniform
inside some half-angle $\theta_{\rm jet}$ and zero outside
\citep{rhoa97b,sari99}. Following \citet{frai01}, the total beaming
corrected gamma-ray energy can be written as
\begin{equation}
E_{\gamma} = E_{\rm iso} f_b = 
\frac{4\pi S_{\gamma}\, k\,  Dl_{\rm th}^2}{1+z} 
[1 -\cos (\theta_{\rm jet})],
\label{eq:egamma}
\end{equation} 
where $f_b = 1 -\cos(\theta_{\rm jet})$ is the beaming fraction, $z$
is the observed redshift, $Dl_{\rm th}$ is the theoretical luminosity
distance for a given cosmology, $S_{\gamma}$ is the $\gamma$-ray
fluence in the observed bandpass, and $k$ is the ``cosmological
$k$-correction'' \citep{bloo01b}, a correction factor of order unity
which blueshifts the observed redshifted GRB spectrum back into some
``bolometric'' cosmological rest-frame bandpass which we take as
[20,2000] keV \citep{bloo03}. See \S~\ref{sec:kcor} for a
justification of this choice of bandpass.  Following \citet{sari99},
in the case of a homogeneous circumburst medium (``ISM''),
\begin{eqnarray}
\label{eq:theta_jet}
\theta_{\rm jet} &=& 0.101 {\ \rm rad} \left(\frac{t_{\rm jet}}{1\ {\rm day}}\right)^{3/8}\left(\frac{\xi}{0.2}\right)^{1/8}\left(\frac n{10 \ {\rm cm}^{-3}}\right)^{1/8} \nonumber \\
& & 
\times \left(\frac{1+z}{2}\right)^{-3/8}\left(\frac{E_{\rm iso}}{10^{53} \ {\rm erg}}\right)^{-1/8},
\end{eqnarray}
\noindent where $z$ is the redshift, $t_{\rm jet}$ is the afterglow
jet break time, $n$ is the density of the ambient medium (ISM),
$E_{\rm iso}$ is defined via eq.~\ref{eq:egamma}, and $\xi$
is the efficiency for converting the explosion energy to $\gamma$-rays.
For simplicity in the later formalism, we write eq.~\ref{eq:theta_jet}
as $\theta_{\rm jet} = B\ t_{\rm jet}^{3/8} \xi^{1/8}
{n}^{1/8} (1+z)^{-3/8} E_{\rm iso}^{-1/8}$ defining the constant $B =
0.101 (1\,{\rm day})^{-3/8} (0.2)^{-1/8} (10$ cm$^{-3})^{-1/8}
(2)^{3/8} (10^{53}$ erg$)^{1/8} = 5.08 \times 10^5$ erg$^{1/8}$
cm$^{3/8}$ day$^{-3/8}$ which absorbs the relevant units.  See
\S~\ref{sec:wind} for a discussion of how the analysis changes for a
circumburst medium that is not homogeneous, for example, a wind
profile from a massive star \citep{chev99,chev00}.

\subsection{Error Analysis}
\label{sec:egamma_err}

We estimate the uncertainty in $E_{\gamma}$ under the assumption of no
covariance between the measurement of the observables $S_{\gamma}$,
$k$, $t_{\rm jet}$, and $ n$, and the inference of $\theta_{\rm jet}$.
We assume the error in the determination of the redshift $z$ is
negligible.  We also assume priors on the Hubble constant ($h=0.7$)
and $\gamma$--ray efficiency ($\xi=0.2$), each with no
error.  Under these assumptions, the fractional uncertainty in
$E_{\gamma}$ is given by
\begin{eqnarray}
\left(\frac{\sigma_{E_{\gamma}}}{E_{\gamma}}\right)^2 & = &  \left(1 - \sqrt{C_{\theta_{\rm jet}}}\right)^2\left[\left(\frac{ \sigma_{S_{\gamma}} }{S_{\gamma}}\right)^2 \ + \ \left(\frac{\sigma_k}{k}\right)^2\right]\ \nonumber \\
& & \ + \ C_{\theta_{\rm jet}}\left[9\left(\frac{ \sigma_{t_{\rm jet}} }{t_{\rm jet}}\right)^2 \ + \left(\frac{\sigma_{n}}{n}\right)^2\right]\ \nonumber \\
\label{eq:E_gamma_err}
\end{eqnarray} 
\noindent where $C_{\theta_{\rm jet}}$ is defined in eq.\ 5 of
\citet{bloo03}.  The above expression is slightly modified from eq. 4
of \citet{bloo03} which also assumed no covariance, but in contrast,
employed the approximation of ignoring the implicit $E_{\rm iso}$
dependence inside of $f_b$.  This changes the multiplicative factor
for the terms on the first line of eq.~\ref{eq:E_gamma_err} from the
old term $(1+C_{\theta_{\rm jet}})$ to the new term $\left(1 -
\sqrt{C_{\theta_{\rm jet}}}\right)^2$, indicating that eq. 4 of
\citet{bloo03} was, at worst, conservatively {\it overestimating} the
error by about 25\% for a typical burst. 

While the above expression makes fewer assumptions than previous work,
the assumption of no covariance (also discussed in \citealt{bloo03})
still requires justification, which we defer to
\S~\ref{sec:covar}. However, using the triangle inequality, we can
place a firm upper limit on $\sigma_{E_{\gamma}}$ even assuming
maximal covariance.
\begin{eqnarray}
\left(\frac{\sigma_{E_{\gamma}}}{E_{\gamma}}\right) & \leq &  \left(1 - \sqrt{C_{\theta_{\rm jet}}}\right)\left[\left(\frac{ \sigma_{S_{\gamma}} }{S_{\gamma}}\right) \ + \ \left(\frac{\sigma_{k}}{k}\right)\right]  \ \nonumber \\
& & \ + \ \sqrt{C_{\theta_{\rm jet}}}\left[ 3\left(\frac{ \sigma_{t_{\rm jet}} }{t_{\rm jet}}\right) \ + \ \left(\frac{\sigma_{n}}{n}\right)\right]
\label{eq:egamma_err_max}
\end{eqnarray}
\noindent Evaluating this expression for a typical burst tells us that
even maximal covariance (we argue it is nowhere near maximal in
\S~\ref{sec:covar}) would mean we are underestimating the errors by at
most a factor of $\lesssim 2$.  As such, we believe the assumption of
no covariance is a reasonable starting point, although, in the extreme
case, a factor of two increase in the error bars would significantly
affect the results.

\subsection{Dataset Compilation}
\label{sec:data_selection2}

Computing $E_{\gamma}$, $\sigma_{E_{\gamma}}$, and constructing the
Ghirlanda relation requires a compilation of all available data. The
observables of interest include $z$, $S_{\gamma}$, $t_{\rm jet}$, $n$,
and $\xi$, as defined in \S~\ref{sec:cor}.  Also needed are
the observed peak energy $E^{\rm obs}_p$ [$E_p = E^{\rm obs}_p(1+z)$],
as well as the low energy and high energy spectral slopes $\alpha$ and
$\beta$ of the Band function, respectively \citep{band93}. Ideally,
high-energy measurements would be derived from a single satellite and
inferences of afterglow parameters would be construed from consistent
modeling with homogeneously-acquired data. In practice, however, we
must compile a heterogeneous dataset with varying degrees of accuracy
on parameters derived from different models and different
instrumentation.

Still, in the interest of obtaining from the literature the highest
quality and most homogeneous dataset possible, we abide by several
guidelines. First, we preferentially choose $E^{\rm obs}_p$
measurements with reported error bars that have accompanying reports
of $\alpha$ and $\beta$ with errors. Second, we use input fluence
measurements with reported errors with priority over fluence
measurements in wider bandpasses. Third, we choose the best sampled
afterglow light curve with the smallest errors on the best fit value
of $t_{\rm jet}$, preferring those estimations that use the earliest
available afterglow data before the break. Measurements reported in
published papers are assumed to supersede those given in GCN or IAU
Circulars. Notes on the data selection for individual bursts are given
in Appendix~\ref{sec:data_selection}.

Often, measurements on some non-critical input parameters to the
energetics are not available (we of course exclude bursts from our
analysis where no redshift, fluence, or jet-break time is known). For
these, we choose a single value for every burst with an associated
``measurement error''. In the absence of reported values of $\alpha$
or $\beta$ (there are no cases of both missing in our sample), we set
$\alpha=-1$ and $\beta=-2.3$ as described in
Appendix~\ref{sec:data_selection}. Following \citet{frai01}, we also
assume $\xi=0.2$ (20\%) for all bursts (see
\S~\ref{sec:eta_gamma} for a critique of this assumption).  Following
\citet{bloo03}, we assume $n = 10 \pm 5$ cm$^{-3}$ (the 50\% error
assumption is new to this work) in the absence of constraints from
broadband afterglow modeling (see \S~\ref{sec:n_dist} for a discussion
of this choice).  We note, however, that the analysis is very
sensitive to the assumptions for the circumburst density (and to a
lesser extent, the $\gamma$--ray efficiency), as we show in sections
\S~\ref{sec:prev}, \S~\ref{sec:outliers}. In the absence of reported
errors, we assume errors of 10\% for $S_{\gamma}$ and 20\% for $E^{\rm
obs}_p$.  These errors are reflective of those for bursts with
reported errors (see Table~\ref{table1}). Errors on $t_{\rm jet}$ are
available for all bursts in the set we use (although see
Appendix~\ref{sec:data_selection}).  All errors on the cosmological
$k$-correction are computed via the formalism in \citet{bloo01b}.
These implicitly depend on the low energy slope $\alpha$, the high
energy slope $\beta$, and the break energy $E^{\rm obs}_o = E^{\rm
obs}_p(2 +\alpha)$, of the Band Function \citep{band93}, and we assume
20\% errors on these parameters when they are not reported -- as these
are also typical of reported errors. When asymmetric fluence or peak
energy errors are reported in the literature (e.g., {\it HETE II}
bursts; \citealt{saka04b}), we assume $\sigma_{S_{\gamma}} =
\sqrt{\sigma_{S_{\gamma}}^{+}\sigma_{S_{\gamma}}^{-}}$, and the
$\sigma_{E^{\rm obs}_p} = \sqrt{\sigma_{E^{\rm
obs}_p}^{+}\sigma_{E^{\rm obs}_p}^{-}}$ (i.e., the geometric mean).
This assumption has little effect on the overall analysis. Finally, we
assume $h=0.7$ with no error to calculate the energetics.

The most current input data and reported errors are listed in
Table~\ref{table1}.  Again, see Appendix~\ref{sec:data_selection}
concerning data selection for individual bursts.  Since the
\citet{bloo03} energetics compilation, spectroscopic redshifts have
been determined for 10 additional bursts: XRF 020903, GRB 030226, GRB
030323, GRB 030328, GRB 030329, XRF 030429, GRB 031203, XRF 040701,
GRB 040924, and GRB 041006 for a total of 39 bursts with $z$, along
with at least 4 upper limits: XRF 020427, GRB 030324, GRB 030528, and
XRF 030723. Of these 14 bursts, 10 have measurements or constraints on
$t_{\rm jet}$, along with 7 bursts where constraints have been added
or updated from the \citet{bloo03} sample.  We use this updated list
of GRB observables\footnotemark\footnotetext{See also {\it
http://www.cosmicbooms.net}, which contains data links to the
compilation found in Table~\ref{table1} of this paper. It is our
intention to keep data at this site up--to--date as new bursts are
observed.} as inputs to the energetics calculations which follow.  The
$E_{\gamma}$ values (with errors) of these new bursts, and updates to
the previous compilation are given in Table~\ref{table2} for the
standard cosmology.

\subsection{Refitting the $E_p$-$E_{\gamma}$ relation}
\label{sec:energetics}

Limited to only those 23 bursts with redshifts and observed jet break
times without upper or lower limits (hereafter, ``Set E''), the median
value of $\log(E_{\gamma} {\rm [erg]})$ is 50.90 ($\sim$ 1 foe) with an RMS
scatter of 0.55 dex.  Under our assumptions, the average fractional
error on $E_{\gamma}$ for these bursts is $\sim 26\%$. Including 10
more bursts with upper or lower limits taken at face value does not
significantly affect the median, yielding 50.91 ($\sim$ 1 foe), with
an RMS of 0.55 dex.  It should be noted that the RMS scatter actually
is an overestimate of the true 1-$\sigma$ error on the median since
the distribution is only approximately gaussian with a broad tail
extending to low energies: as recognized by a number of authors, the
low redshift burst GRB 030329 ($z = 0.1685$), as well as GRB 990712,
GRB 021211, and XRF 030429 all appear to be under-energetic by around
1 order of magnitude.  Moreover the low redshift GRB 031203
($z=0.1055$) along with XRFs 030723 ($z < 2.1$) and 020903 ($z=0.251$)
also appear under-luminous by at least $2$--$3$ orders-of-magnitude
--- even assuming an isotropic explosion --- as the geometry
correction is not known for these bursts.

\citet{ghir04} recognized that these under-luminous bursts appeared
systematically softer in the prompt burst spectrum than bursts of
apparent higher $E_{\gamma}$.  Expanding upon the much discussed
correlation \citep{amat02} between the isotropic-equivalent energy
$E_{\rm iso}$ and the restframe peak energy in the GRB spectrum
($E_p$), the authors discovered a remarkably strong correlation
between $E_{\gamma}$ and $E_p$, which can be represented as a power-law:
\begin{equation}\label{ghir}
E_p = \kappa\left(\frac{E_{\gamma}}{E^{*}}\right)^{\eta}
\label{eq:epegamma}
\end{equation}
The scaling $E^{*}$ is a constant which we choose in order to minimize
the covariance between $\eta$ and $\kappa$ when fitting for this
two-parameter relation, simplifying future error analyses. This choice
of $E^{*}$ does not affect the values of the best fit slope $\eta$ (or
the goodness of fit) and the parameter $\kappa$ simply scales as $(E^{*})^\eta$.

Using an updated set including 19 bursts with redshifts, $E_p = E^{\rm
obs}_p(1+z)$, and reported $t_{\rm jet}$ measurements without
upper/lower limits (hereafter ``Set A''), we confirm the strong
correlation in the standard cosmology for the set of assumptions
detailed in \S~\ref{sec:cor}, finding the best fit
values\footnotemark\footnotetext{Unless otherwise noted, all
uncertainties on derived parameters reported hereafter are 1-$\sigma$
derived from $\chi^2$ analysis.  They do not reflect any covariance
with other parameters nor are the uncertainties scaled by
$\sqrt{\chi^2/{\rm dof}}$, as is customary under the assumption that
the data {\it should} be well-fit by the model.} of $\eta = $ 0.669 $\pm$
0.034, $\kappa = $ 252 $\pm$ 11 keV ($E^{*} = 4.14 \times 10^{50}$
erg), with a Spearman $\rho$ correlation coefficient of 0.86 (null
hypothesis probability of $2.9 \times 10^{-6}$).

\begin{figure*}[htp]
\centerline{\psfig{file=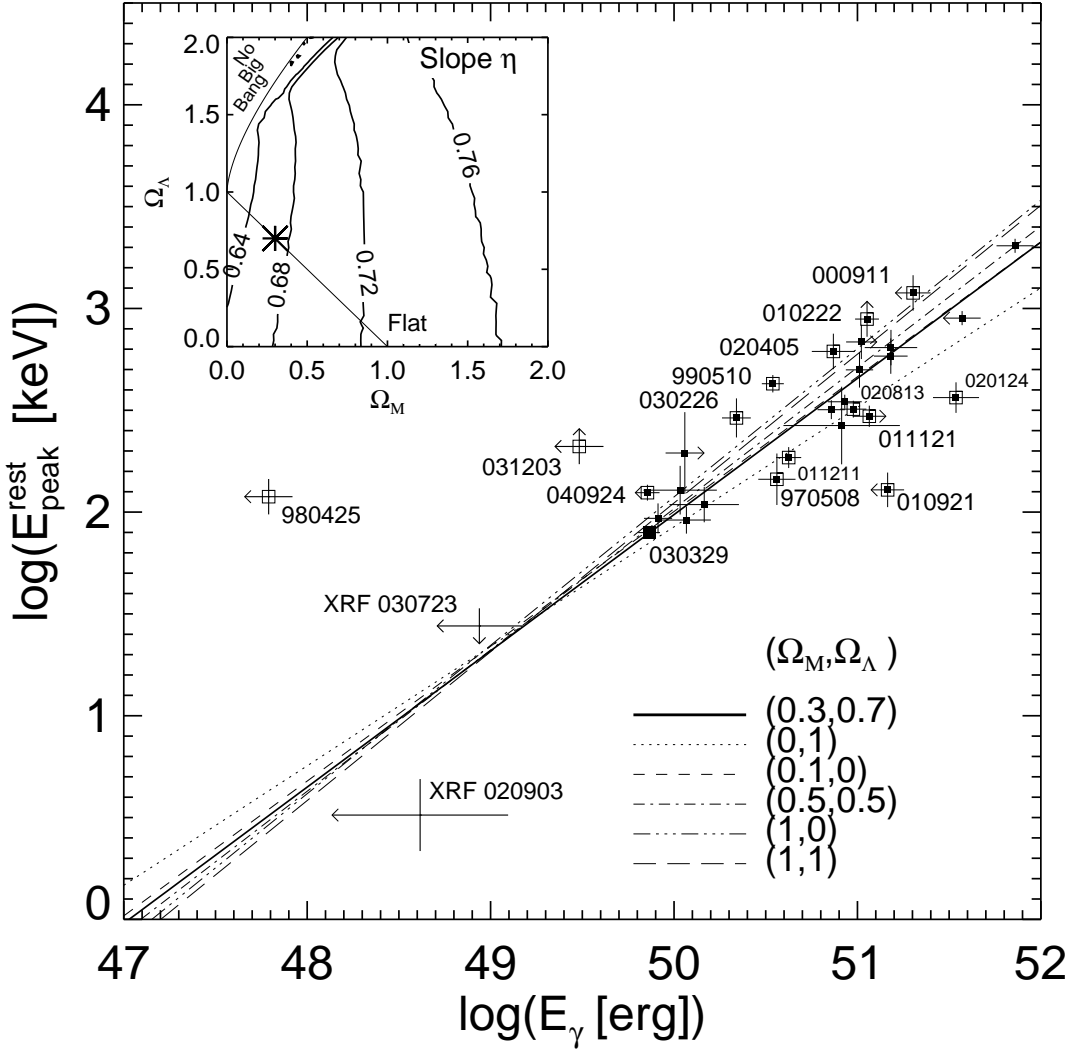,width=6in}}
\caption[]{\footnotesize The (weak) cosmological dependence of the
$E_p$--$E_{\gamma}$ relation.  The best fit power-law relation for a
representative set of cosmologies are shown as a series of lines.
Only the derived $E_{\gamma}$ values in standard cosmology of
($\Omega_M,\Omega_{\Lambda},h$)=($0.3,0.7,0.7$) are shown for clarity
with upper/lower limits indicated with arrows.  Set A is comprised of
those bursts with no upper/lower limits (small filled squares), with a
goodness of fit $\chi^2/$dof$= 3.71$ (17 dof).  Data are calculated
for a bolometric rest frame bandpass of $[20,2000]$ keV, assuming a
$\gamma$-ray production efficiency of $\xi = 0.2$, and an external
homogeneous ISM density of $n = 10 \pm 5$ cm$^{-3}$ when there are no
reliable constraints from broadband afterglow modeling.  Notable
outliers under these assumptions are indicated with a large square
surrounding the data points (small filled squares). GRBs 980425 and
031203 are major outliers regardless of their geometry correction or
external density. The only outliers for which density is constrained
are GRBs 970508 and 990510.  As such, all other nominal outliers can
be made consistent with the relation simply by changing the density
(or increasing the error on the density), as discussed in
\S~\ref{sec:outliers}. With the current data, and our assumptions,
XRFs 020903 and 030723 are consistent with the relation. Note GRB
030329, a large outlier in $E_{\gamma}$ (large filled black square),
falls directly on the relation.  The best fit value of the slope
($\eta$) is shown inset as a contour plot over the cosmological
parameters ($\Omega_M$, $\Omega_{\Lambda}$). Over a wide range of
cosmologies [$0 \le \Omega_M$, $\Omega_{\Lambda}, \le 2$], $\eta$
falls in a narrow range from $\sim0.6$--$0.8$, with typical errors
$\sim0.03$--$0.5$ ($5$--$6\%$).  Note that the data for a standard
cosmology with best fit $\eta = $ 0.67 $\pm$ 0.03, (indicated by the
asterisk in the inset contour plot, and the thick black best fit line
in the outer plot), essentially brackets the fits across all
cosmologies in the range [$0 \le \Omega_M$, $ \Omega_{\Lambda} \le
1$], $\eta \in [0.64,0.70]$, excepting only the extreme cosmologies
with $\Omega_M \sim 0$ and $\Omega_{\Lambda} \sim 1$ or $\Omega_M \sim
1$.  Thus, for a wide range of reasonable cosmologies, (not to mention
$k$-correction bandpasses and external densities) the slope of the
relation is $\sim 2/3$.}
\label{fig1}
\end{figure*}

The relation for a standard cosmology is shown in Figure~\ref{fig1},
with inset panel indicating the cosmology dependence, which is
discussed in detail in \S~\ref{sec:cosmo}.  Although the correlation
is clearly significant, we find a reduced $\chi^2_{\nu} \equiv
\chi^2/$dof = 3.71 (for 17 degrees of freedom; dof), suggesting that a
single power-law does not adequately accommodate the data, given the
assumptions and dataset compilation: we will address the dependence of
$\chi^2/$dof on various assumptions in detail in
\S~\ref{sec:prev}, \S~\ref{sec:outliers}.

Despite the poor fit, \citet{ghir04} correctly noted that this
power-law fit is better than the fit to the $E_p$--$E_{\rm iso}$
``Amati'' relation $E_p = A\left(E_{\rm iso}/E^{\prime}\right)^{m}$
($E^{\prime} = 10^{52}$ erg).  Indeed, for the subset of 29 bursts
with measurements of redshift $z$, and $E_{p}$ without upper/lower
limits (excluding GRB 980425), we find best fit $m = 0.496 \pm 0.037$,
$A=90 \pm 8$ keV and a Spearman $\rho$ correlation coefficient of
$\rho=0.88$, with a null hypothesis probability of no correlation of
$4.9\times10^{-10}$.  As originally recognized by \citet{amat02}, the
correlation is clearly significant.  However the goodness of fit found
here: $\chi^2_{\nu} \sim 9.48$ (27 dof), is clearly poorer than for
the new Ghirlanda relation, and cannot easily be improved by changing
input assumptions.  Recent work \citep{naka04b}, indicates that a
significant fraction of GRBs without known redshifts cannot fall on
the Amati relation, which --- due to selection effects --- may be better
understood as a demarcation of an upper limit (where burst energies
can be no greater than their isotropic equivalents).  This implies
that any intrinsic spectra-Energy connection is more closely related
to the Ghirlanda relation than the Amati relation; this is not
surprising given the more physically motivated, beaming-corrected
energy, rather than the poor approximation of energy inferred for a
spherical explosion.  However, see \citet{band05} for a similar
analysis of the Ghirlanda relation which raises the possibility that
the relation itself could arise due to selection effects --- mostly
concerning the measurement of $t_{\rm jet}$, $E^{\rm obs}_p$, and $z$.

\subsection{Comparison with Other Work}
\label{sec:prev}

The $E_p$--$E_{\gamma}$ relation has been fit in several other works
\citep{ghir04,ghir04b,dai04} using different data sets and a range of
input assumptions different from those assumed herein.  We focus on
comparing our results with the sample of \citet{ghir04} and
\citet{ghir04b}: 15 bursts, hereafter ``Set G'', which uses a more
complete sample than that of \citet{dai04}: 12 bursts, hereafter ``Set
D''.  In contrast, our Set A contains 19 bursts (see our
Table~\ref{table1}).  The bursts belonging to each of these samples
are noted in the leftmost column of our Table~\ref{table2}.  For
clarification, the Set name A, G, or D simply refers to the {\it
names} of the bursts in the sample, not to the assumptions used by
various groups or the individual references chosen for the data for a
given burst, differences which are detailed in
Appendix~\ref{sec:ghir_comp}.  In referring below to the
\citeauthor{ghir04} data, we refer to the overlapping subset of {\it
our data} (Table~\ref{table1}) as G, and refer to the \citet{ghir04}
data itself (their Tables 1-4), as G$^*$, which uses {\it their data
selection and assumptions}, and likewise for D (our
Table~\ref{table1}) and D$^{*}$ (Table 1 of \citealt{dai04}).

The parameterization of the $E_p$--$E_{\gamma}$ relation and the
reported errors on the slope that we find; $\eta = 0.669 \pm 0.034$,
$\chi^2_{\nu}$ =3.71 (17 dof), are consistent with those given in
\citet{ghir04} and \citet{ghir04b}; $\eta = 0.706 \pm 0.047$,
$\eta^{-1} = 1.416 \pm 0.09$, respectively. Both fits are performed in
the standard ($\Omega_M$, $\Omega_{\Lambda}$, $h$) = (0.3, 0.7, 0.7),
cosmology, but differ somewhat owing to the slightly larger sample
used here to construct the fit (19 vs.\ 15 bursts), data selection
differences for the bursts common to both samples (again, see
Appendix~\ref{sec:data_selection}), differing assumptions for the
density and its fractional error, as well as the different energy
bandpass used for $E_{\gamma}$.  We compute the energy in the
restframe $[20,2000]$ keV band as opposed to $[1,10^{4}]$ keV in
\citet{ghir04} and \citet{ghir04b}, although we find similar results
for our data Set A by adopting the $[1,10^{4}]$ keV bandpass: $\eta =
0.647 \pm 0.034$, $\chi^2_{\nu}=4.15$ (17 dof), Spearman $\rho= 0.83$
(null prob. $1.02 \times 10^{-5}$), and in fact, for an even wider
range of bandpasses (see \S~\ref{sec:kcor} for a detailed discussion
of bandpass choice).  Despite these differences, the value of the
slope $\eta$ and the high significance of the correlation coefficient
are remarkably {\it insensitive} to these assumptions and the sample
selection --- although $\eta$ itself {\it does} depend on the
cosmology (see \S~\ref{sec:cosmo}).

Although the slope range $0.6 < \eta < 0.8$ (consistent with $\eta
\sim 2/3$), and high correlation significance appear robust in our
standard cosmology for a variety of input assumptions, the value of
the goodness of the fit, however, is not.  The value of $\chi^2_{\nu}$
is not reported in either \citet{ghir04} or \citet{ghir04b}, hindering
a direct comparison. However, that group has since reported
$\chi^2_{\nu}=1.27$ (13 dof) for the fit to the $E_p$--$E_{\gamma}$
relation \citep{ghir_rome04}.  After discussing the differences
between our input assumptions (G.\ Ghirlanda \& D.\ Lazzati -- private
communications), we re-fit the data directly from Tables 1--4 of
\citet{ghir04}, using their assumptions and confirm
$\chi^2_{\nu}=1.27$.  As such, we attempt here to reconcile their
marginally good fit with our unacceptable fit by comparing our data
and assumptions.

\begin{figure*}[htp]
\centerline{\psfig{file=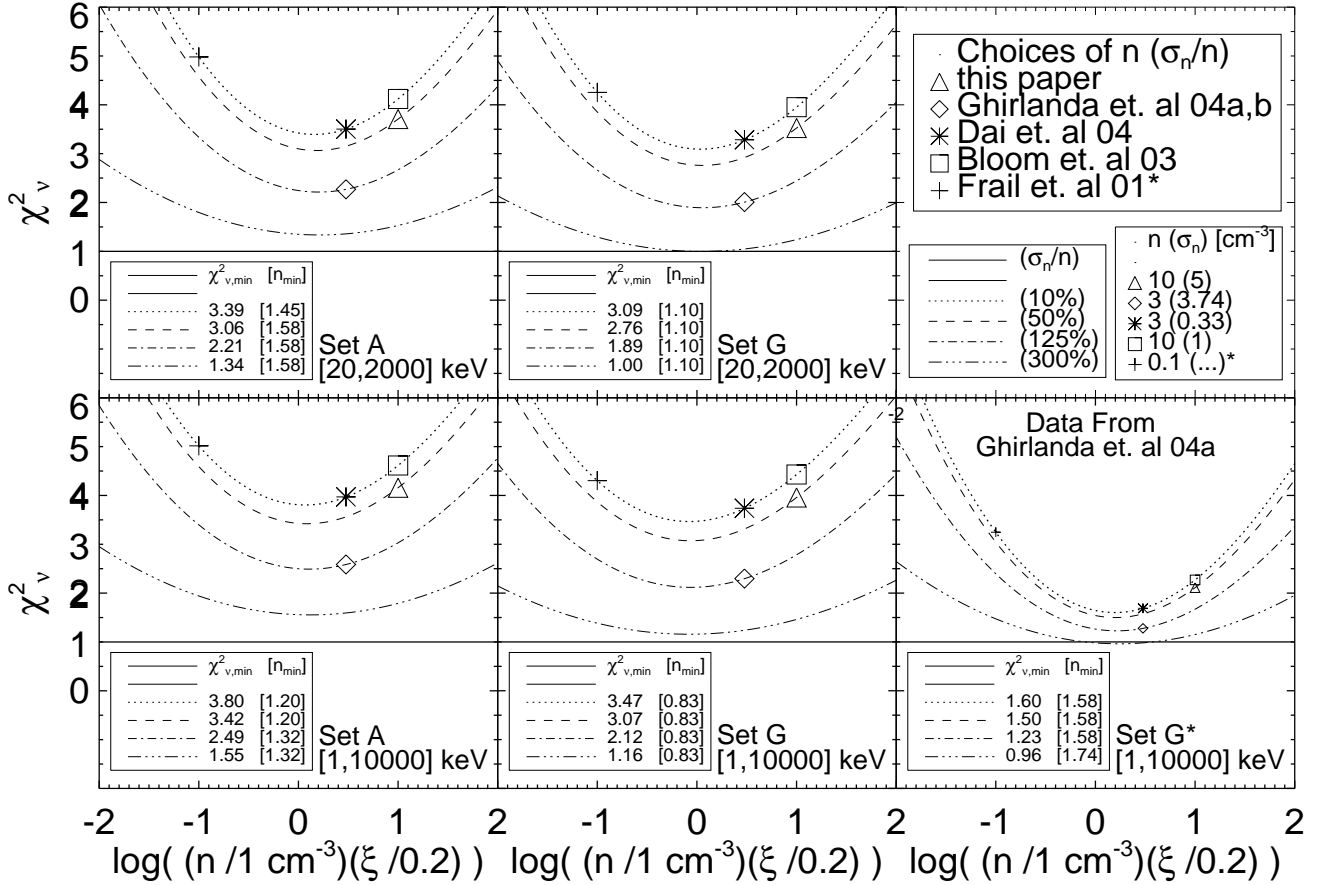,width=7in}}
\caption[]{\footnotesize The extreme sensitivity of the goodness of
fit of the Ghirlanda relation to input parameters (density and
$k$-correction bandpass) and data selection criteria. Here, we compare
the effects of different assumptions on the 15 bursts common to our
sample Set A, and the \citet{ghir04} and \citet{ghir04b} sample Set --
denoted by: G$\rightarrow$our data, and G$^*$$\rightarrow$their
data. Plotted are the reduced $\chi^{2}_{\nu} = \chi^2/{\rm dof}$ vs.\
${\rm log}[(n/1$ cm$^{-3})(\xi/0.2)]$, with all plots assuming
$\xi=0.2$.  Burst by burst comparison indicates only a few significant
differences in the references, noted in detail in
\S~\ref{sec:outliers} and Appendix~\ref{sec:ghir_comp}.  However,
using the \citet{ghir04} data (Set G$^*$) results in an significantly
improved goodness of fit in comparison to the same 15 bursts using our
data (Set A), indicating the strong sensitivity of the fit to data
selection choices.  In an individual plot, we vary only the density
assumed for all bursts without reliable density constraints. Various
assumptions for $n$ and $\sigma_n/n$ from previous work are indicated
with plot symbols referenced in the upper right
panel. \citet{frai01}$^*$ did not assume an error on the density, and
we mark their density assumption of $n = 0.1$ cm$^{-3}$ on the 10\%
error curve for display purposes only. For all datasets A, G, G$^*$, a
density choice of $n_{\rm min} \sim$ 1--2 cm$^{-3}$ minimizes the
goodness of fit, essentially independent of the fractional error,
$k$-correction bandpass. Clearly, an increased fractional error on the
density improves the fit for any choice of density, as seen for each
curve which corresponds to an increasing fractional error on the
density. The top (bottom) plots show the results for a restframe
$k$-correction of $[20,2000]$ keV ($[1,10^{4}]$ keV). Different
$k$-correction bandpass change $\chi^{2}_{\nu}$ slightly, but the fits
actually {\it worsen} going from our $[20,2000]$ keV to their
$[1,10^{4}]$ keV bandpass for our data.  The addition of the 4 new
bursts after the original Ghirlanda 15 slightly improves
$\chi^2_{\nu}$. Ultimately, there exist a certain set of input
assumptions which lead to a good fit for the $E_p$--$E_{\gamma}$
relation (also see \S~\ref{sec:n_dist}). However, these assumptions
are not favored {\it a priori} over many other equally-plausible
assumptions which yield poor fits.}
\label{fig2}
\end{figure*}

Ultimately, both values for $\chi^2_{\nu}$ follow from data
compilation and input assumptions.  However, the large discrepancy
indicates that $\chi^2_{\nu}$ is {\it highly sensitive to assumptions
and individual parameter measurements}.  Figure~\ref{fig2}
illustrates the sensitivity of the goodness of fit to the assumptions
which differ between our analyses including density and its fractional
error, $\gamma$--ray efficiency, $k$-correction bandpass, sample size,
and data selection differences for the 15 bursts common to both
samples. The dominant factors are the assumptions on density and its
fractional error and the choice of references for individual bursts
common to both samples.  Although the $\gamma$--ray efficiency (set to
$\xi=0.2$ in Fig.~\ref{fig2}) plays the same role as
density in eq's.~\ref{eq:egamma},~\ref{eq:theta_jet}, the former is
less important as it is, by definition, constrained to values between
$[0,1]$ (likely $\sim1\%$--$90\%$ in practice), whereas the density
can range over several orders-of-magnitude (see \S~\ref{sec:n_dist},
\S~\ref{sec:eta_gamma}). Although we must assume values for the
fluence error, $\alpha$, $\beta$, and their errors for some bursts, we
single out the assumptions for $n$ and $\xi$ in particular
because {\bf i)} they apply to most/all of the bursts in the sample,
and {\bf ii)} they have a much stronger effect on computing
$E_{\gamma}$, $\sigma_{E_{\gamma}}$.  Changing the $k$-correction
bandpass alters $\chi^2_{\nu}$ slightly, but we find that the fits
actually {\it worsen} going from our $[20,2000]$ keV to their
$[1,10^{4}]$ keV bandpass (see Fig.~\ref{fig2}).  The addition of 4
new bursts to our sample slightly improves $\chi^2_{\nu}$, as do data
updates to older bursts from the most current literature
(e.g., \citealt{saka04b}).

More specifically, following \citet{bloo03}, we assume $n = 10$
cm$^{-3}$ (with 50\% error), in the absence of constraints from
broadband afterglow modeling, which applies to most bursts (13/19) in
our sample. In contrast, \citet{ghir04b} assume $n = 3$ cm$^{-3}$,
with errors where they ``allow $n$ to cover the full [1-10] cm$^{-3}$
range''.  Re-fitting their data directly from Tables 1-4 of
\citet{ghir04}, we determine that the error assumption as described in
\citet{ghir04b} translates to $n=3^{+7}_{-2}$ cm$^{-3}$, where the
geometric mean of the asymmetric errors is then used to approximately
symmetrize the errors, giving $n = 3 \pm 3.74$ cm$^{-3}$ (i.e.,
$\sigma_{n} \approx \sqrt{7\times2} = \sqrt{14} = 3.74$ cm$^{-3}$).
Only with this assumption for the fractional error on the density can
we recover $\chi^2_{\nu}=1.27$ from their data. This is a fractional
error of about 125\%, in contrast with our assumption of 50\% error.

In fact, we find that for $\sigma_n/n$ between 10\% and 300\% (which
covers the range of fractional errors on density which have been
assumed in previous work --- \citealt{bloo03,dai04,ghir04,ghir04b}),
the choice of density that minimizes $\chi^2_{\nu}$ for our data Set A
and our data Set G, is around 1-2 cm$^{-3}$, for either the
$[20,2000]$ keV or $[1,10^{4}]$ keV bandpass (see Figure~\ref{fig2}).
Although the choice of $n = 3$ cm$^{-3}$ \citep{ghir04,ghir04b}
does not optimize the fit for our data sets A and G or their data set
G$^*$, (for either bandpass) it improves the fit dramatically as
compared to our choice of $n = 10$ cm$^{-3}$.  Clearly an increase in
$\sigma_n/n$ also improves the goodness of fit, as shown in
Figure~\ref{fig2}.

We also have different references for $S_{\gamma}$, $E_p$, $t_{\rm
jet}$, and $n$, for several bursts common to both samples, although
the majority of the input data are identical. See our
Table~\ref{table1} compared to Tables 1-4 of \citet{ghir04} as well as
Appendix~\ref{sec:ghir_comp} for detailed burst by burst comparison.
As an example of the most notable differences, consider the jet break
time for 020124: We use $t_{\rm jet} = 15 \pm 5$ days (e.g., $t_{\rm
jet}=10$--$20$ days; \citealt{berg02}), vs. their reference of $t_{\rm
jet} = 3 \pm 0.4$ days, (also citing \citealt{berg02} jointly with
\citealt{goro_gcn_1224,bloo03} - see their Table 2), although we can
not verify this number from those sources or anywhere else in the
literature.  For our data Set A, this single burst has a strong effect
on the fit improving it from $\chi^2_{\nu} =3.71$ to $2.80$, simply by
changing this jet break reference from our reference to their
reference.  As seen in the lower right panel of Figure~\ref{fig2},
$\chi^2_{\nu}$ is very sensitive to these data selection differences
for the bursts common to both samples, worsening dramatically for our
slightly different references, the most sensitive of which we believe
are either more current (e.g., \citealt{saka04b}), or more reliable
(e.g., \citealt{berg02}) than those cited in \citet{ghir04} for the
bursts in question.  In fact, the data in \citet{ghir04} for their Set
G$^*$ yields marginally acceptable fits for a much larger range of
assumed densities and fractional errors (again, see
Figure~\ref{fig2}).

\citet{dai04} also re-examine the Ghirlanda relation, and do not
include GRBs 990510 and 030226 (in addition to 970508, 021004, 021211,
which were known at the time, as well as 030429 and 041006, which were
discovered later), keeping only 12 bursts (hereafter ``Set D'').
Using those 12 bursts, a slightly different ($\Omega_M$,
$\Omega_{\Lambda}$, $h$) = (0.27, 0.73, 0.71) cosmology, a
$k$-correction bandpass of $[1,10^{4}]$ keV (as in
\citealt{ghir04,ghir04b}), and $n = 3 \pm 0.33$ cm$^{-3}$
(i.e., D$^{*}$; their Table 1), \citet{dai04} report $\chi^{2}_{\nu} =
0.53$ ($\eta^{-1} = 1.5 \pm 0.08$), a very good fit to a power law.
Using Set D, with our assumptions, we find, $\eta = $ 0.659 $\pm$
0.034, and a reduced $\chi^2_{\nu} = 2.70$ (10 dof), which is much
worse than the \citet{dai04} fit.  Since this comparison is for the
same 12 bursts, again, the large discrepancy comes primarily from
different density assumptions.  As mentioned, \citet{dai04} assume $n
= 3 \pm 0.33$ cm$^{-3}$, a choice which improves the fit relative to
our choice of $n = 10$ cm$^{-3}$, even though they assume a fractional
error (11\%) that is smaller than our assumption (50\%), which, all
other things being equal, would tend to worsen their fit. The
different $k$-correction bandpasses, and the slightly different
cosmology they use compared to our standard cosmology --- ($\Omega_M$,
$\Omega_{\Lambda}$, $h$) = (0.27, 0.73, 0.71) vs. our (0.3, 0.7, 0.7)
choice --- has little effect on the goodness of fit.

Under our assumptions, the fit to our Set D ($\chi^2_{\nu} = 2.70$) is
much better than the fit for our Set A ($\chi^2_{\nu} = 3.71$),
although both are poor.  This discrepancy arises due to data
selection, as the \citet{dai04} sample {\it does not include two of
the major outliers to the Ghirlanda relation}, 990510 and 030226, as
seen in Figure~\ref{fig1}.  \citet{dai04} specifically argue that
these bursts should be left out on grounds which are somewhat
controversial.  The strong affect of removing only two bursts in such
a small sample is not surprising, as we have already seen that the
data are sensitive to reference choices for individual bursts (e.g.,
the jet break for 020124).  Ultimately, the difference between sets D
and A comes from data selection while the larger difference between
fits for sets D$^{*}$ and D comes from differing assumptions.  The
combination of both leads to the largest difference between fits for A
($\chi^2_{\nu} = 3.71$) and D$^{*}$ ($\chi^{2}_{\nu} = 0.53$),
although, as with the comparison to the data Set G$^*$
\citep{ghir04,ghir04b}, the best fit slopes themselves remain largely
unchanged.

The sample selection critique (i.e., excluding outlier bursts) does
not apply to the Set G$^*$ \citep{ghir04,ghir04b}, or to our Set G, as
the fit could have been improved by including some of the bursts in
our Set A and/or removing some burst from their set
G$^*$. Nevertheless, the realization that individual data selection
choices can change the fit from a good one to a poor one, gives us
great pause in believing a standard candle derived from the relation,
which requires that the relation is well fit by a power law.  To
quantify this, we identify and discuss the role of outliers further in
the following section.

\subsection{Identifying $E_p$--$E_{\gamma}$ Outliers}
\label{sec:outliers}

If the $E_p$--$E_{\gamma}$ correlation holds, then
eq.~\ref{eq:epegamma} can be rewritten to yield a dimensionless
number, the GRB standard candle $A_\gamma$, which should be
a constant of order unity from burst to burst, constructed as 
\begin{equation}
A_{\gamma} = \left(\frac{ E_{\gamma} }{ E^{*} }\right)\left(\frac{ \kappa }{ E_p }\right)^{1/\eta}
\label{eq:agamma}
\end{equation}
\noindent with error (neglecting the uncertainty in redshift, and
assuming no covariance ) given by
\begin{eqnarray}
\label{eq:a_gamma_err}
\left(\frac{ \sigma_{A_{\gamma}} }{A_{\gamma}}\right)^2 &=& \left(\frac{ \sigma_{E_{\gamma}} }{E_{\gamma}}\right)^2 \ + \ \left(\frac{1}{\eta}\right)^2 \left\{ \left(\frac{ \sigma_{E_{p}} }{E_{p}}\right)^2 
+ \left(\frac{ \sigma_{\kappa}}{\kappa}\right)^2\right\}  \nonumber \\
& & \ + \ \left(\frac{1}{\eta}\right)^2\left\{\left(\frac{ \sigma_{\eta} }{\eta}\right)^2\left[ \ln\left(\frac{E_p}{\kappa}\right)\right]^2\right\}
\end{eqnarray}
\noindent Since the combination $\kappa^{1/\eta}/E^{*}$ (or
equivalently $\kappa/(E^{*})^{\eta}$) is a constant for the fit, we
are free to choose $E^{*}$ to minimize the covariance between $\eta$
and $\kappa$ without affecting $A_{\gamma}$, as $\kappa$ changes to
compensate. As such, we can safely neglect the related covariance
terms in eq.~\ref{eq:a_gamma_err}. Certainly $E_p$ and $E_{\gamma}$
themselves are correlated -- this is the central point of interest in
this work -- however, this correlation is likely an {\it intrinsic}
correlation (possibly due to local physics), not observational
covariance, which must be dealt with in the error analysis (although
see \citealt{band05}).  As before, we assume no covariance and delay
further justification until \S~\ref{sec:covar}, although again, even
assuming maximal covariance changes the errors by at most a factor of
$\lesssim 2$, and simply indicates that certain bursts which were
minor outliers may actually be consistent with the relation.

Computing $A_{\gamma}$ provides a quick diagnostic to determine
which bursts deviate from the Ghirlanda relation.  Bursts that fall
significantly off the relation (outliers) will have an $A_{\gamma}$
value that significantly deviates from unity (within the errors).  A
list of $A_{\gamma}$ values for all bursts in our sample (Set A) in a
standard cosmology, using our assumptions for density, etc..., can be
found in Table~\ref{table2}, where 7 bursts from Set A (970508,
990510, 011211, 020124, 020405, 020813, and 030226) have computed
$A_{\gamma}$ values at least 1-$\sigma$ from $A_{\gamma}=1$ (assuming
no covariance). Of these, 020124 and 020405 are between 2-$\sigma$ and
3-$\sigma$ away, whereas 990510 and 030226 are at more than 3-$\sigma$
away from $A_{\gamma}=1$, respectively. Also see Fig.~\ref{fig1},
where these nominal outliers are indicated on the plot.  See
Appendix~\ref{sec:ghir_comp} for a detailed burst by burst comparison
of the outliers between Sets A, G, and D. Again, note that Set D
excludes 990510 and 030226, the two {\it largest outliers} to the
relation in our set.

Additionally, there are several bursts with upper/lower limits on
$t_{\rm jet}$, $E^{\rm obs}_p$, or $z$, not included in Set A, which
can be identified as outliers by considering limiting cases.  Of
course, one must assume the values of $\eta$ and $\kappa$ {\it derived for
Set A} in order to place other bursts on the relation.  As noted by
\citet{ghir04}, the very low-redshift GRB 980425 falls well-off the
relation. \citet{berger04} recently noted that GRB 031203 also falls
off the relation, with an $E_p > 210$ keV \citep{sazo04}. Although we
cannot compute $\sigma_{A_{\gamma}}$ for these bursts, since neither
have a jet break constraint, even assuming isotropy (i.e., $f_b = 1$),
these bursts appear as major outliers in the Ghirlanda relation,
completely independent of any assumptions concerning circumburst
density (see Fig.~\ref{fig1}).  Other bursts not in Set A (010222,
010921, 011121, 000911, 040924) are also minor (1-$\sigma$) to major
(2--3 $\sigma$) outliers, depending on the assumptions involving
$t_{\rm jet}$, $E_p$, and $z$. Several bursts with uncertain redshift
(980326, 980519, 030528) also are outliers under reasonable
assumptions. See Appendix~\ref{sec:ghir_comp} for details.

Despite the apparent ubiquity of outliers to the relation, in light of
the results highlighted in Figure~\ref{fig2}, a major caveat must be
stressed.  For most of these bursts, the ambient density is unknown,
and any discussion about bursts being outliers is only meaningful {\it
modulo assumptions made concerning the density and the $\gamma$--ray
efficiency}. In fact, only for the bursts 980425 and 031203 (and to a
lesser extent, 990510)\footnote{The density for 990510 has been
constrained ($n=0.29^{+0.11}_{-0.15}$ cm$^{-3}$; \citealt{pana02}), so
it is an outlier regardless of our density assumption for other
bursts, although there is some freedom as the model uncertainty in
deriving the constraint is likely to far exceed the reported
statistical uncertainty shown here (see \S~\ref{sec:n_dist}).  030226,
with unknown density, is an even greater outlier (compared to our
$n=10 \pm 5$ cm$^{-3}$ assumption) if one applies the \citet{dai04}
assumption of $n=3 \pm 0.33$ cm$^{-3}$, which further reduces the
energy (see Fig.~\ref{fig1}).} can we be relatively certain that they
are still outliers independent of the circumburst environment or
$\gamma$--ray efficiency.  More quantitatively, $E_{\gamma} \propto
[(n/10$ cm$^{-3})(\xi/0.2)]^{1/4}$ in the small $\theta_{\rm
jet}$ limit [$1-{\rm cos}(\theta_{\rm jet}) \approx \theta_{\rm
jet}^2/2$].  As an example, simply changing the assumed density from,
say, 10 cm$^{-3}$ to 1 cm$^{-3}$ (while keeping $\xi$=0.2)
leads to a decrease in inferred energy by a factor of
$\sim(1/10)^{1/4} = 0.56$ ($\sim$ 50\%), or vice versa.  Ultimately,
while the product ($n\xi$) can not be tuned {\it
arbitrarily} --- given existing constraints on $n$
(\S~\ref{sec:n_dist}) and $\xi$ (\S~\ref{sec:eta_gamma}) ---
it does provide enough freedom to make most outlier bursts consistent
with the relation (980425 and 031203 aside).  As such, we now conclude
that without reliable density (and efficiency) estimates, GRB
cosmology using the $E_p$--$E_{\gamma}$ relation becomes prohibitively
uncertain.  On the other hand, the current data {\it do not rule out}
an eventual good fit to the relation, as there still exist reasonable
density assumptions which yield good fits (see
\S~\ref{sec:n_dist}). However, since these assumptions are not favored
{\it a priori} over equally reasonable assumptions which yield poor
fits, only an improved sample can determine the true goodness of fit
to the relation.

\subsection{Cosmology Dependence of the Relation}
\label{sec:cosmo_dep}

Although there are several significant outliers under our assumptions,
and the goodness of fit of the $E_p$--$E_{\gamma}$ relation is sensitive
to these assumptions, the relation does appear to be a significant
correlation, for our standard cosmology, with a slope between roughly
0.6 and 0.8 independent of any density assumptions, with most choices
of $n$ and $\sigma_n$ giving a slope $\sim$2/3.  As also recently
suggested by \citet{dai04} and \citet{ghir04b}, the correlation could
provide a means to correct the energetics and use GRBs for
cosmography. However, without any knowledge of the slope of the power
law {\it a priori}, in the cosmographic context, it is imperative to
demonstrate that the power-law fit to the correlation is statistically
acceptable over the range of plausible cosmologies --- this is
non-trivial, given the complex dependence of $E_{\gamma}$ upon the
luminosity distance (eq's.~\ref{eq:egamma},~\ref{eq:theta_jet}).

Figure \ref{fig1} shows the correlation for Set A for a variety of
cosmologies, placing emphasis on the outliers. The inset of
Figure~\ref{fig1} shows the best fit values of $\eta$ as a contour
plot in the ($\Omega_M$--$\Omega_{\Lambda}$) plane, with data
calculated for our standard assumptions.  Over a wide range of
cosmologies [$0 \le \Omega_M$, $\Omega_{\Lambda} \le 2$], $\eta$ falls
in a narrow range from $\sim0.6$--$0.8$, with typical errors
$\sim0.03$--$0.5$ ($5$--$6\%$) that are essentially invariant to the
cosmology.  Recalling eq.~\ref{eq:epegamma}, by choosing the
normalization parameter $E^{*}$ that minimizes the covariance between
the slope $\eta$ and the intercept $\kappa$, we find that log($E^{*}
{\rm [erg]}$) remains in a small range $[50.3-50.8]$ across the entire
grid [$0 \le \Omega_M$, $\Omega_{\Lambda} \le 2$], and that with this
choice for $E^{*}$, $\kappa$ remains essentially a constant in the
range $[247-256]$ keV.  Along with associated 1-$\sigma$ error, the
best fit value of $\eta = 0.669 \pm 0.034$ in a standard cosmology
brackets the best fit values in all but the most extreme cosmologies
in the range [$0 \le \Omega_M$, $\Omega_{\Lambda} \le 1$]. We thus
confirm the claim by \citet{dai04} that the slope of the
$E_{p}-E_{\gamma}$ relation is relatively insensitive to $\Omega_M$,
as it changes by no more than 25\% across the entire grid, and by
closer to 5 or 10\% in what is arguably the most plausible region of
the ($\Omega_M$--$\Omega_{\Lambda}$) plane.  However, even these small
changes in the slope, along with the uncertainty involved in
determining it from the data, must be taken into account
self-consistently to avoid circularity in the cosmography analysis.

Previously, we reported a poor fit ($\chi^2_{\nu}=3.71$) to the
Ghirlanda relation for our Set A in the standard cosmology.
Re-fitting the relation for Set A over many cosmologies shows that a
power-law also provides an unacceptable fit ($5 > \chi^2_{\nu}> 3$)
over the range [$0 \le \Omega_M$, $\Omega_{\Lambda} \le 2$].  The fit
also remains poor over this same region of the ($\Omega_M$,
$\Omega_{\Lambda}$) plane for subsets G and D.  Thus, based on our
assumptions, the relation can not be well fit simply by changing the
cosmology.  However, as discussed for our standard cosmology, good
fits exist for different density assumptions and, ultimately, this
remains the case for every cosmology in our grid.

\section{Formalizing the Standardized GRB Energy}
\label{sec:standard}

Despite the apparent {\it intrinsic scatter} in the Ghirlanda
relation, and the uncertainties in the assumptions used to fit it, the
correlation is highly significant, and can be used to standardize GRB
energetics with a simple empirical correction.  By constructing the
GRB standard candle $A_{\gamma}$, which should be identically unity if
the Ghirlanda relation exactly holds for all bursts, we can derive an
expression for the GRB luminosity distance ($Dl_{\gamma}$) and the GRB
distance modulus ($DM_{\gamma}$).  Although it is perfectly possible
to solve for $Dl_{\gamma}$ numerically without employing the small
angle approximation for the beaming fraction (as in \citealt{ghir04b},
and outlined briefly below), such a choice leaves the formalism less
explicit and not much is gained as the small angle approximation
yields values of $Dl_{\gamma}$ that are accurate to within $\lesssim
1\%$ of the numerical result even for the widest jets in the sample
($\sim20- 30^{\circ}$), making the approximation much less important
than the sensitivity due to input assumptions or the propagated
observational errors. Although we do calculate these quantities
numerically {\it in practice} for the subsequent analysis and for the
values reported in Table~\ref{table2}, we still feel it is instructive
to additionally present the formalism with the small angle
approximation.  As such, to derive $Dl_{\gamma}$ analytically, we can
approximate $\theta_{\rm jet}$ as a small angle (i.e., $f_b \approx
\theta_{\rm jet}^2/2$), yielding
\begin{equation}
E_{\gamma} \approx  \left(\frac{B^2}{2}\right)\left(\frac{4 \pi S_{\gamma} k t_{\rm jet} (n \xi)^{1/3}}{(1+z)^2}\right)^{3/4} Dl_{th}^{3/2}h_{70}^{-3/2}
\label{eq:egamma_small}
\end{equation}
\noindent where all variables are defined in \S~\ref{sec:cor} and we
assume $h_{70} = h/0.7 = 1$.  This expression is accurate to within
$\lesssim 1\%$ of the exact expression for $E_{\gamma}$
(eq.~\ref{eq:egamma}) for all bursts in the sample.

Under the standard candle assumption $A_{\gamma} \equiv 1$ (or
equivalently, $E_{\gamma} \equiv
E^{*}\left(E_p/\kappa\right)^{1/\eta}$), the GRB luminosity distance
$Dl_{\gamma}$ is found by solving for $Dl_{\rm th}$ in
eq.~\ref{eq:egamma_small}. Thus, if $A_{\gamma} \equiv 1$ is true for
each burst, then $Dl_{\gamma} \equiv Dl_{\rm th}$.  Making these
substitutions and solving for $Dl_{\gamma}$, we find
\begin{equation}
Dl_{\gamma} \approx \left(\frac{2 E^{*}}{B^2}\right)^{2/3}\left(\frac{E_p}{\kappa}\right)^{2/3\eta}\left(\frac{(1+z)^2 }{4 \pi S_{\gamma} k t_{\rm jet}(n \xi)^{1/3}}\right)^{1/2} h_{70}
\label{eq:Dl_gamma}
\end{equation}
This is similar to the derived quantity in \citet{dai04} (who take
$\eta \equiv 2/3$). 

As shown, $\sigma_{E_{\gamma}}$ can be derived analytically without
the small angle approximation.  As with $E_{\gamma}$,
$\sigma_{E_{\gamma}}$ is also well approximated by direct error
propagation of eq.~\ref{eq:egamma_small}, which assumes the small
angle limit.\footnote{From eq. 5 of \citet{bloo03}, the
$C_{\theta_{\rm jet}}$ term in our eq.~\ref{eq:E_gamma_err} is given
by $C_{\theta_{\rm jet}} = [\theta_{\rm jet}{\rm sin}(\theta_{\rm
jet})/\{8(1-{\rm cos}(\theta_{\rm jet}))\}]^2$ .  In the small angle
limit, $C_{\theta_{\rm jet}} \approx 1/16$.  By taking this limit in
eq.~\ref{eq:E_gamma_err} or computing the result of direct error
propagation of eq.~\ref{eq:egamma_small} we find
$\left(\sigma_{E_{\gamma}}/E_{\gamma}\right)^2 \approx
{(9/16)}\left[\left( \sigma_{S_{\gamma}} /S_{\gamma}\right)^2 \ + \
\left(\sigma_{k}/k\right)^2 \ + \ \left( \sigma_{t_{\rm jet}} /t_{\rm
jet}\right)^2 \ + \ {1/9}\left(\sigma_{n}/n\right)^2\right]$. One can
show that this expression is equivalent to within $\lesssim 1\%$ of
eq.~\ref{eq:E_gamma_err}, which does not use the small angle
approximation.}. While $Dl_{\gamma}$ {\it can not} be derived
analytically without the small angle or some other approximation
(e.g., \citealt{bloo03}), as mentioned, the small angle expression for
$Dl_{\gamma}$ (eq.~\ref{eq:Dl_gamma}) is accurate to within $\lesssim
1\%$ of the numerical result. As with $\sigma_{E_{\gamma}}$, the error
$\sigma_{Dl_{\gamma}}$ can be derived analytically without the small
angle approximation (see \S~\ref{sec:error}).  However, unlike the
case for $\sigma_{E_{\gamma}}$, the expression derived by propagating
the errors in eq.~\ref{eq:Dl_gamma} (as done similarly in
\citealt{dai04}) actually {\it overestimates} the error in
$Dl_{\gamma}$ by a factor of $\sim4/3$ for each burst compared to
direct error propagation of the RHS and LHS of the following equation,
derived by combining eq's.~\ref{eq:egamma} and~\ref{eq:theta_jet}, and
setting $Dl_{\rm th} \equiv Dl_{\gamma}$.
\begin{equation}
Dl_{\gamma}\sqrt{fb(Dl_{\gamma})} = \left(\frac{E^{\rm obs}_{p}(1+z)}{\kappa}\right)^{1/2\eta}\left(\frac{E^{*}(1+z)}{4 \pi S_{\gamma}k}\right)^{1/2} h_{70}
\label{eq:Dl_gamma_numerical}
\end{equation}
\noindent where $fb(Dl_{\gamma}) = 1-{\rm cos}(\theta_{\rm
jet}(Dl_{\gamma}))$.  The more tractable terms not involving
$Dl_{\gamma}$ are grouped on the RHS.  This equation makes explicit
how to solve for $Dl_{\gamma}$ numerically.  Simply evaluate the
RHS and vary $Dl_{\gamma}$ in the LHS until $|1-({\rm LHS/RHS})| <
\epsilon$, where $\epsilon$ can be tuned to achieve the desired
accuracy.

Returning to the small angle approximation, using an alternative
approach, we recast eq.~\ref{eq:Dl_gamma} in cgs units with an analogy
to astronomical magnitudes, and derive the ``apparent GRB distance
modulus'', $DM_\gamma = 5\, \log (Dl_{\gamma}/10{\rm \,pc})$, finding
\begin{equation}
DM_{\gamma} \approx -2.5 \log \left(\frac{4 \pi S_{\gamma} k t_{\rm jet}(n\xi)^{1/3}}{(1+z)^2}\right) + C_{\gamma} + {\rm zp}
\label{eq:DM_gamma}
\end{equation}
\noindent with the ``GRB energy correction'' term in [mag]:
\begin{equation}
C_{\gamma} = \frac{10}{3\eta}\log\left(\frac{E^{\rm
obs}_{p}(1+z)}{\kappa}\right)
\label{eq:C_gamma}
\end{equation}
\noindent and the zero point ${\rm zp} = (10 / 3) \log \left(2 E^{*} /
B^2 \right) - 5\log (3.085 \times 10^{19}{\rm cm}) + 5\log (h_{70})$.
The zero point ${\rm zp}$ contains unit conversion terms so that the
first term + ${\rm zp}$ is in [mag], as well as the scaled Hubble
constant $h_{70}$, and the normalization $E^{*}$ of the Ghirlanda
relation (in [erg]), chosen to minimize the covariance between $\eta$
and $\kappa$.  In principle, ${\rm zp}$ could also be defined in terms
of $\kappa$ rather than $E^{*}$ since the quantity
$\kappa/(E^{*})^{\eta}$ is related to the true ``y-intercept'' in the
2-parameter fit (with $\eta$) to the Ghirlanda relation.  Note that
since the parameters $\kappa/(E^{*})^{\eta}$ and $\eta$ are {\it fit
from the data} for each cosmology, they do not need to be marginalized
over, nor does $E^{*}$ via its role in ${\rm zp}$.  In contrast to SNe
Ia work, the Hubble constant does not need to be marginalized over,
precisely because of the cosmology dependence of the GRB standard
candle.  In other words, while assuming a prior on $h$
(e.g., $h_{70}=1$) is necessary to calculate $E_{\gamma}$,
$A_{\gamma}$, and $DM_{\gamma}$ for a given cosmology, it is
unnecessary for cosmography as its effect cancels in the upcoming
eq.~\ref{eq:chi2}, shown in \S~\ref{sec:cosmo}.

As discussed previously, the above analysis differs from the analysis
in \citealt{bloo03} (their eq's.\ 2, 3), where the assumption was to
neglect the implicit $Dl_{\rm th}$ dependence inside $f_b$, which
itself is a reasonable approximation (see their footnote 7), but is
not as accurate as the small angle jet approximation.  The other main
assumption in \citet{bloo03} was a different standard candle
assumption, namely $\epsilon_{\gamma} = E_{\gamma}/\bar{E_{\gamma}}
\equiv 1$, where $\bar{E_{\gamma}}$ is the median energy for all the
bursts in the sample, which, for self-consistency, {\it must be
recalculated for every cosmology in the same way that the Ghirlanda
relation must be re-fit for every cosmology} to determine the best fit
$\eta$ and $\kappa$.  Fitting for $\bar{E_{\gamma}}$ (or $\eta$ and
$\kappa \rightarrow E^{*}$) thus represents the freedom in determining
the cosmological zero point for each cosmology from a sample of
high-$z$ bursts in the absence of a low-$z$ ``training set'' to
calibrate the relation in a cosmology independent way (see
\S~\ref{sec:dis}).

Based on the differences between the assumptions in \citet{bloo03}
and those herein, we can write an expression involving the
``uncorrected'' GRB distance modulus ($DM_{\gamma,{\rm unc}}$), which is
related to the ``corrected'' GRB distance modulus
(eq.~\ref{eq:DM_gamma}) in the small angle limit by
\begin{equation}
DM_{\gamma} - DM_{\gamma,{\rm unc}} \approx C_{\gamma}  - 
\frac{10}{3}\log\left(\frac{\bar{E_{\gamma}}}{E^{*}}\right) = 
\frac{10}{3}\log\left(\frac{\epsilon_{\gamma}}{A_{\gamma}}\right).
\label{eq:DM_diff}
\end{equation}
\noindent The value $C_{\gamma}$ corrects for the now untenable
assumption of a standard energy.  Note that $DM_{\gamma} -
DM_{\gamma,{\rm unc}} \approx C_{\gamma}$ iff $\bar{E_{\gamma}}
\approx E^{*}$.  While the correction term $C_{\gamma}$ differs from
burst to burst, the $(10/3)\log\left(\bar{E_{\gamma}}/E^{*}\right)$
term is simply a constant for all bursts in a given cosmology.
Although $C_{\gamma}$ is defined in the context of the small angle
approximation, it is still appropriate to think of the exact
correction (where the difference $DM_{\gamma} - DM_{\gamma,{\rm unc}}$
is calculated numerically) as a magnitude correction, for which
$C_{\gamma} \approx DM_{\gamma} - DM_{\gamma,{\rm unc}} +
(10/3)\log\left(\bar{E_{\gamma}}/E^{*}\right)$ is a reasonable
approximation.  This can be seen by comparing the relevant columns in
Table~\ref{table2}.  For reference,
$(10/3)\log\left(\bar{E_{\gamma}}/E^{*}\right) = 0.95$ mag in the
standard cosmology.  For self-consistency, the comparison in
eq.~\ref{eq:DM_diff} {\it should} be derived for the {\it same} set of
bursts used to define both the standard candles $\epsilon_{\gamma}$
and $A_{\gamma}$.  Although there are 23 bursts (Set E) which can be
used to compute $E_{\gamma}$ ($\bar{E_{\gamma}}$), and only 19 bursts
(Set A) with all the data necessary for the $E_p$--$E_{\gamma}$
relation ($\eta$, $\kappa$), we still use all the bursts to compute
$\bar{E_{\gamma}}$. In practice, this is a small point, since
log($\bar{E_{\gamma}}$ [erg]) = 50.85 and 50.90 for sets A and E
respectively.

Ultimately, we derive the formalism in terms of the distance modulus
$DM_{\gamma}$ (rather than only the luminosity distance
$Dl_{\gamma}$), and cast $C_{\gamma}$ in magnitudes to highlight a
direct analogy to various empirical (magnitude) corrections for Type
Ia supernovae (i.e., $\Delta m_{15}$, \citealt{phil93,hamu95,hamu96};
the Multicolor Light Curve Shape (MLCS) method,
\citealt{ries95,ries96}; (MLCS2k2: S. Jha et. al in preparation); the
Stretch method, \citealt{perl97}, and the BATM method
\citealt{tonr03}). From the Ghirlanda relation, a large,
more positive, $C_\gamma$ is obtained for bursts with larger inferred
$E_{\gamma}$, and $C_\gamma < 0$ for bursts that are under-energetic
from the median. As seen in Table~\ref{table2}, the spread in
$C_{\gamma}$ is rather large, $\sim$8 mag, reflecting the intrinsic
scatter of more than 3 orders-of-magnitude in $E_{\gamma}$.  In
contrast to typical, 1-parameter, peak luminosity corrections for SNe
Ia involving factors of $\sim2$--$3$, the GRB energy correction
involves factors of $ \gtrsim 10^{3}$.  This alone requires more
rigorous support to justify using GRBs for precision cosmology.

\begin{figure*}[htp]
\centerline{\psfig{file=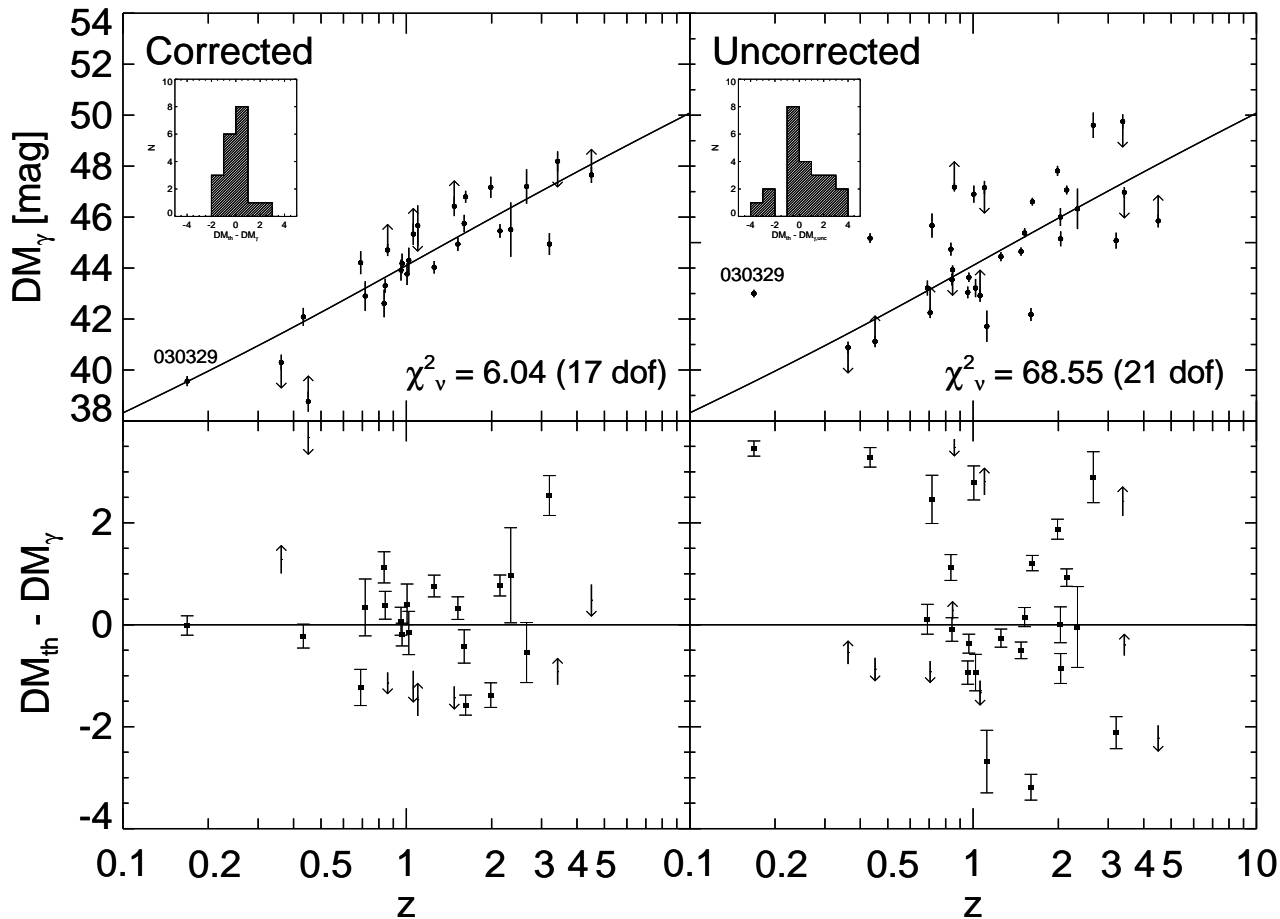,width=6.8in}}
\caption[]{\footnotesize ({\bf Top Row}) The GRB Hubble Diagram with
(left) and without (right) the GRB energy correction term $C_\gamma$.
The solid curve is the theoretical distance modulus ($DM_{\rm th}$) in
the standard cosmology of ($\Omega_M$, $\Omega_{\Lambda}$, $h$) =
(0.3,0.7,0.7).  The data are $DM_{\gamma}$ (left) and $DM_{\gamma,{\rm
unc}}$ (right) with associated errors, respectively, both derived
assuming the standard cosmology, and a density of $n = 10 \pm 5$
cm$^{-3}$ for bursts without constraints. Although $C_{\gamma}$ is
defined using the small angle approximation, the data are calculated
numerically. Bursts without upper or lower limit arrows (Set A) were
used to fit the Ghirlanda relation to obtain the cosmological zero
point to $DM_{\gamma}$ (left), while those with $z$, $t_{\rm jet}$ and
no upper/lower limits (Set E) are used analogously for
$DM_{\gamma,{\rm unc}}$ (right).  The potential utility of $C_\gamma$
in a cosmographic context is best seen by the effect on GRB 030329
(lowest $z$ data shown): without the correction, that burst is
significantly discrepant from the best fit by $\sim$4 magnitudes (a
factor of $\sim$15 in energy), yet is consistent to within $\sim$0.2
mag with the $C_\gamma$ correction.  ({\bf Bottom Row}).  The Hubble
diagram residuals, $DM_{\gamma} - DM_{\rm th} \approx
(10/3\eta)\log(A_{\gamma})$, and $DM_{\gamma,{\rm unc}} - DM_{\rm th}
\approx (10/3\eta)\log(\epsilon_{\gamma})$, plotted for the corrected
(left) and uncorrected (right) standard candles, respectively.
Histograms of the residuals are also shown inset in the top row.  The
scatter about a constant value of 0 is a measure of the goodness of
the standard candle.  Clearly the correction improves the scatter
about a ``standard'' GRB magnitude.  Only 3 bursts (left) appear to be
more than 2-$\sigma$ away from 0.  While the highest redshift burst
without upper/lower limits (020124, $z = 3.198$) appears as major
outlier, this can be remedied simply by changing the density from
$n = 10$ to $n = 1$ cm$^{-3}$.  As such, there is no apparent
evolution with redshift, even out to $z=4.5$, but, ultimately,
redshift evolution of a the standard candle $A_{\gamma}$ can not be
probed accurately without better density constraints.  Note that any
cosmological parameter determination requires a separate plot like the
upper left panel {\it for each cosmology}.  The global minimum
$\chi^2_{\nu}$ over all cosmologies then gives the favored
cosmological parameters.  However, for this cosmology, the fit is poor
($\chi^2_{\nu} \sim 6$).  In fact, for Set A, $\chi^2_{\nu} \gtrsim 5$
for all cosmologies in our grid (see Fig.~\ref{fig4}), undermining any
cosmographic utility of the Ghirlanda relation at present, at least
under the assumptions we have made.}
\label{fig3}
\end{figure*}

Figure \ref{fig3} shows the effect of the correction term on the
effective absolute GRB magnitude, as a function of redshift. The
improvement in the scatter about the Hubble diagram is
apparent. Equivalently, the corrected distribution of residuals,
$DM_{\rm th} - DM_{\gamma} \approx
(10/3)\log\left(A_{\gamma}/1\right)$, is clearly much narrower than
the distribution of ``uncorrected'' residuals, $DM_{\rm th} -
DM_{\gamma,{\rm unc}} \approx
(10/3)\log\left(\epsilon_{\gamma}/1\right)$, (see Figure \ref{fig3}
inset plots) reflecting the relative superiority of the standard
candle assumption $A_{\gamma} \equiv 1$ vs. $\epsilon_{\gamma} \equiv
1$.  As seen in Table~\ref{table2}, the $A_{\gamma}$ distribution for
Set A has a spread of only a factor of 2-3 as compared to several
orders-of-magnitude for $\epsilon_{\gamma}$.  Since the {\it same}
assumptions for density apply to both the $\epsilon_{\gamma}$ and
$A_{\gamma}$ standard candles, it is clear that the latter is far
superior, independent of the relevant input assumptions.

\subsection{Error Estimates}
\label{sec:error}

As with $\sigma_{E_{\gamma}}$, we estimate the error in the inferred
GRB luminosity distance $Dl_{\gamma}$ under the assumption that there
is no covariance between the {\it measurement} of the observables
$S_{\gamma}$, $k$, $E^{\rm obs}_{p}$, $t_{\rm jet}$, $n$ and the
inference of $\theta_{\rm jet}$. Under these assumptions, and the
approximation of Gaussian errors, the fractional uncertainty in
$Dl_{\gamma}$, which can be derived analytically without the small
angle approximation, is given by
\begin{eqnarray}
\left(\frac{\sigma_{Dl_{\gamma}}}{Dl_{\gamma}}\right)^2 & = & \frac{1}{4}\left(\frac{ \sigma_{E_{\gamma}} }{E_{\gamma}}\right)^2 \ + \ \left(\frac{1}{2\eta}\right)^2\left\{\left(\frac{ \sigma_{E_p} }{E_p}\right)^2 \ + \left(\frac{ \sigma_{\kappa} }{\kappa}\right)^2\right\}  \ \nonumber \\
& & + \ \left(\frac{1}{2\eta}\right)^2\left(\frac{ \sigma_{\eta} }{\eta}\right)^2\left[ {\rm ln}\left(\frac{E_p}{\kappa}\right)\right]^2  \\
&=& \frac{1}{4}\left(\frac{ \sigma_{A_{\gamma}} }{A_{\gamma}}\right)^2 \nonumber
\label{eq:Dl_gamma_err}
\end{eqnarray} 
\noindent Equation~\ref{eq:Dl_gamma_err} shows an implicit
relationship between the intrinsic scatter in the Ghirlanda relation
and the measurement errors in $E_p$. Note we have also treated the
errors on $\eta$ and $\kappa$ as statistical, rather than
systematic. See \S~\ref{sec:covar} for a discussion of possible
systematic errors from neglecting nonzero covariance, although,
as discussed, even assuming maximal covariance --- using the triangle
inequality --- implies that eq.~\ref{eq:Dl_gamma_err} is underestimating
the errors by at most a factor of $\lesssim 2$. The error on the
apparent GRB distance modulus is then obtained from
$\sigma_{DM_\gamma} = (5/\ln
10)\left(\sigma_{Dl_{\gamma}}/Dl_{\gamma}\right) \approx
2.17\left(\sigma_{Dl_{\gamma}}/Dl_{\gamma}\right)$.

Similarly, the errors on $C_{\gamma}$ (which uses the small angle
approximation) are given by
\begin{equation}
\label{eq:c_gamma_err}
\left(\sigma_{C_{\gamma}}\right)^2 = \left(C_{\gamma}\right)^2 \left(\frac{\sigma_{\eta}}{\eta}\right)^2 + \left(\frac{10}{3\eta {\rm ln}10}\right)^2\left[\left(\frac{\sigma_{E_p}}{E_p}\right)^2 + \left(\frac{\sigma_{\kappa}}{\kappa}\right)^2\right].
\end{equation}
Figure \ref{fig3} shows the GRB Hubble diagram for a standard
cosmology with the $C_\gamma$ term and without. It is clear that the
inclusion of the $C_\gamma$ term {\bf i)} accommodates bursts that are
highly-discrepant in $E_{\gamma}$ (e.g., 030329) and {\bf ii)}
significantly reduces the scatter about the luminosity distance,
redshift relation. Under our assumptions, typical fractional errors
are $\left(\sigma_{S_{\gamma}} / S_{\gamma} \right)\sim 9\%$,
$\left(\sigma_k/\rm k\right)\sim 6\%$, $\left(\sigma_{t_{\rm
jet}}/t_{\rm jet}\right)\sim 21\%$, $\left(\sigma_{n}/n\right)\sim
63\%$, $\left(\sigma_{E_{\gamma}} / E_{\gamma} \right)\sim 26\%$,
$\left(\sigma_{E_{p}} / E_{p} \right)\sim 17\%$, $\left(
\sigma_{\eta}/ \eta\right)\sim 5\%$ and $\left( \sigma_{\kappa} /
\kappa\right)\sim4\%$.  In order of decreasing importance, typical
error terms in eq.~\ref{eq:Dl_gamma_err} are
$\left(1/2\right)\left(\sigma_{E_{\gamma}}/E_{\gamma}\right)\sim
0.132$, $\left(1/2\eta\right)\left( \sigma_{E_{p}}/E_{p}\right)\sim
0.059$, $\left(1/2\eta\right)\left( \sigma_{\kappa} /\kappa\right)\sim
0.014$, and
$\left(1/2\eta\right)\left(\sigma_{\eta}/\eta\right)\left[{\rm
ln}(E_{p}/\kappa)\right]\sim 0.002$, where the first of these terms
implicitly includes sub terms from eq.~\ref{eq:E_gamma_err}. The
quadrature sum of these numbers gives a typical fractional error on
$Dl_{\gamma}$ of $\left(\sigma_{Dl_{\gamma}}/Dl_{\gamma}\right)\sim
19\%$ or an error in the apparent GRB distance modulus of
$\sigma_{DM_\gamma}\sim 0.42$ magnitudes, slightly more than a factor
of 2 larger than the typical error in determining the distance modulus
of Type Ia SNe ($\sim$ 0.2 mag, see Table 5 of \citet{ries04b}, which
uses the MLCS2k2 algorithm - S. Jha et. al in preparation).

Under our assumptions, the dominant terms come from the errors on the
jet-break time and external density (which contribute to the error on
$E_{\gamma}$), and the error on the observed spectral peak energy;
assuming no error on $z$, the fractional error on $E_p$ is the same as
the error on $E^{\rm obs}_p = E_p/(1+z)$.  Although non-negligible,
the intrinsic scatter in the fit to the Ghirlanda relation (via $\eta$
and $\kappa$), the fluence, and $k$-correction have the least
important error terms.  The relative unimportance of the statistical
error $\sigma_{k}/k$ in determining the distance highlights an
advantage of GRBs over SNe Ia, where the latter suffers from both
statistical and additional {\it systematic} errors in determining the
$K$--correction.  However, as discussed, we can increase the errors
arbitrarily by increasing the fractional error on density, which is
implicit inside the error term for $E_{\gamma}$ in
eq.~\ref{eq:Dl_gamma_err}.  We also assume no error on the efficiency
$\xi$, an assumption critiqued in \S~\ref{sec:eta_gamma}.

\subsection{Are GRBs Useful as Cosmological Distance Indicators in Principle?}
\label{sec:indicators}

Given the preceding formalism, one can construct a GRB standard
candle and use it to test cosmological models. However, a crucial
point not yet addressed is whether GRBs are actually competitive as
cosmological distance indicators {\it in principle}.

Of the main advantages --- high-redshift detection, immunity to dust,
more tractable $k$-corrections, orthogonal evolution to SNe Ia --- the
first is arguably the most important.  While $z_{\rm max} \sim 1.7$ is
essentially the upper limit for currently feasible SNe Ia redshift
detection with {\it HST} (e.g., SN 1997ff; \citealt{ries01}), and
future SNe Ia detection with {\it SNAP} \citep{lind04}, 10 GRBs out of
the sample of 39 with known $z$ already have measured redshifts $
\gtrsim 2$ (see Table~\ref{table1}).  While this is clearly promising
for future high-$z$ detections with {\it Swift}, it is not obvious
that the $z>1.7$ region is an interesting part of the Hubble diagram
since it is in the matter dominated epoch, and at first blush, does
not strongly constrain the dark energy. However, \citet{lind03} argue
that a full survey in the range $0 < z < 2$ is necessary for revealing
the nature of dark energy because, while low-$z$ measurements are
crucial for determining $\Omega_{\Lambda}$ and $w$, certain inherent
systematic errors and degeneracies due to the dark energy and its
possible time variation are only resolvable at high redshift.  This is
particularly evident in Figures 3--5 of \citet{lind03}.  Furthermore,
depending on the nature of the time variation, the region of interest
may conceivably include redshifts greater than $2$.  In addition,
although 5 bursts in Set A have $z>2$, the mean redshift in the sample
is $\bar{z}\sim 1.3$, and with {\it Swift}, it is likely that GRBs
will dominate the $1<z<2$ region --- as compared to SNe Ia --- several
years before the launch of {\it SNAP} ($\sim$2010).  Indeed there are
13 GRBs in our sample in the redshift range $0.65 < z < 2$ \ (9 in the
range $0.9 < z < 2$), which is {\it already} comparable to the number
of high-z SNe Ia so far discovered with {\it HST} (Table 3 of
\citealt{ries04b}).  This intermediate-to-high redshift regime is
clearly important for more precisely constraining $\Omega_M$,
$\Omega_{\Lambda}$, $w$, its possible time variation, and the
transition redshift to the epoch of deceleration
\citep{ries04a,ries04b}.

Thus, as also stressed by \citet{ghir04b}, what may evolve from this
work is a combination of GRBs and SNe Ia, where SNe Ia are primarily
useful for determining $\Omega_{\Lambda}$ and $w$ at low $z$, and GRBs
serve to provide independent, and potentially more accurate,
constraints on $\Omega_M$ (without many low-$z$ bursts, GRBs alone are
essentially insensitive to $\Omega_{\Lambda}$). GRBs could ultimately
serve as an independent cross check to the systematic errors that
would plague a SNe Ia sample with relatively sparse coverage in the
$1<z<2$ region, as outlined in \citet{lind03}.  This is in addition to
the orthogonality of GRBs to the systematic errors that are
potentially the most problematic for SNe Ia, e.g., dust,
$K$-corrections, and evolution.

\subsection{Using the Standardized Energy for Cosmography}
\label{sec:cosmo}

Granting that GRBs are cosmographically useful in principle, we can
test this in practice for the current sample, noting of course that
the sample is small (19 bursts), depends on the typically unknown
external density, and is not well sampled at low redshift.  Since the
theoretical distance modulus $DM_{\rm th}$ and apparent GRB distance
modulus $DM_{\gamma}$ are functions which have complex, but different,
dependences on the cosmological parameters $\Omega_{M}$, and
$\Omega_{\Lambda}$, a minimization of the scatter in the residuals
$DM_{\rm th} - DM_{\gamma} \approx (10/3)\log\left(A_{\gamma}/1\right)$
(in the small angle jet limit) can in principle be a useful tool to
probe the geometry of the universe.

We first stress the need to re-calibrate the slope of the relation for
different cosmologies.  To quantify this, although $\eta$ changes by
no more than $\sim$25\% across the full grid, this variation, as well
as the error in determining the slope for each cosmology ($\sim$5\%)
must be self-consistently taken into account in the fit to the GRB
Hubble diagram.  Even changes of $\gtrsim 5\%$ in $\eta$ (and thus,
$\kappa$ and $E^{*}$, since $\kappa/(E^{*})^\eta$ and $\eta$ are the
two fundamental parameters in the fit) affect the apparent GRB
luminosity distance sensitively as $Dl_{\gamma} \propto
({E^{*}})^{2/3}(E_p/\kappa)^{2/3\eta}$ in the small angle limit
(eq.~\ref{eq:Dl_gamma}).  Ultimately, without a low redshift training
set to calibrate $\eta$ or an {\it a priori} value of $\eta$ from
physics, assuming a value of $\eta$ {\it derived in a given cosmology}
will effectively input prior information about that cosmology into the
analysis.  As shown in \S~\ref{sec:compare}, this concern affects the
analysis of \citet{dai04}.

\begin{figure*}[htp] 
\centerline{\psfig{file=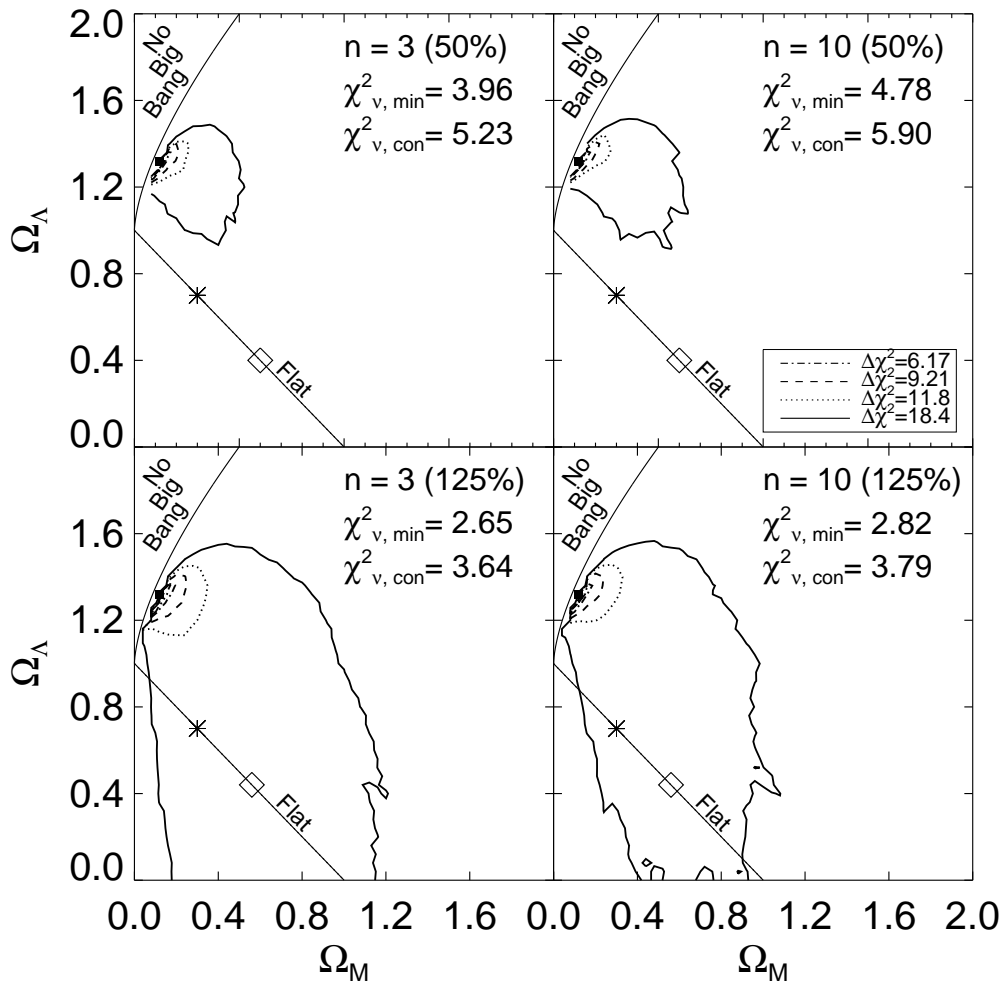,width=5.5in}}
\caption{\small Plotted are $\chi^2$ contours for the GRB Hubble
diagram for Set A (17 dof), for a range of assumed densities
[cm$^{-3}$] and fractional errors.  All other assumptions are as in
the text.  The jagged contours are an artifact of finite grid
resolution, as is the discrepancy $\chi^2_{\nu,{\rm con}}=5.90$
(here), vs. $6.04$ (in Fig.~\ref{fig3}) for the standard
($\Omega_M$,$\Omega_{\Lambda}$,$h$) = (0.3,0.7,0.7), concordance
cosmology (e.g., ``con'').  With the steep shape of the surface, the
outermost contour corresponds to $\Delta\chi^2=18.4$ --- nominally
99.99\% confidence for 2 parameters.  As with the $E_p$--$E_{\gamma}$
relation itself, the goodness of fit at the minimum of the surface
($\chi^2_{\nu,{\rm min}}$) in all panels is clearly sensitive to
density and its error.  However, independent of these assumptions, the
best fit ($\Omega_M$, $\Omega_{\Lambda}$)=($0.12$, $1.32$) cosmology
(black filled square) lies abutting the cosmic loitering line, which
borders the region in the $\Omega_M$--$\Omega_{\Lambda}$ plane for
which there is no Big Bang. Also over-plotted are the standard
concordance cosmology (asterisk) and the best fit cosmology assuming
flatness (diamond). However, in each panel, the data yield a poor fit
for any cosmology in our grid, precluding the use of the $\chi^2$
contours shown for meaningful cosmological parameter determination.
One can recover a good fit in the Hubble diagram, for example, by
increasing the density error arbitrarily, although for errors
$\lesssim$125\% (assumed in \citealt{ghir04b}), the peculiar cosmology
favored by GRBs remains essentially {\it invariant} to these choices.
This cosmology is inconsistent with flatness and is close to
conflicting with a Big Bang.  Although not shown here, even when
removing individual bursts --- which sensitively changes the global
shape of the surface via small number statistics --- the favored
loitering cosmology persists (also see \citealt{ghir04b,firm05}).  In
light of the independent evidence from SNe Ia, LSS, and the CMB, this
result strongly supports the idea that --- at least for the current
data --- GRBs are simply not useful for cosmography (although see
\citealt{firm05}, who claim a Bayesian rather than $\chi^2$ analysis
removes the loitering problem).}
\label{fig4}
\end{figure*}

Although the intrinsic scatter (and sensitivity to input assumptions)
in the Ghirlanda relation limits the precision of this cosmographic
method, one can still apply a self-consistent approach to the current
sample of GRBs with the required spectral and afterglow data and
confirmed spectroscopic redshifts. First, for a given cosmology, we
determine $E_{\gamma}$, $\sigma_{E_{\gamma}}$ for all GRBs of
interest, assuming values for the $\gamma$--ray efficiency
$\xi$, the external density $n$, its error, and other data
where appropriate.  We then re-fit the $E_{p}$--$E_{\gamma}$
correlation to find $\eta$, ($\sigma_{\eta}$), $\kappa$,
($\sigma_{\kappa}$) for that cosmology. After fitting for $\eta$, we
determine the value of the normalization $E^{*}$ for that cosmology
that minimizes the covariance between $\eta$ and $\kappa$ in order to
eliminate the related terms from the error analysis.  We then
determine $DM_{\gamma}$ and $\sigma_{DM_\gamma}$ for all GRBs in the
set.  We repeat this for a grid of cosmologies spanning the range [$0
\le \Omega_M$, $\Omega_{\Lambda} \le 2$].  For each cosmology we then
compute
$$
\chi^2(\Omega_M,\Omega_\Lambda,\eta,\kappa) = \nonumber
$$
\begin{equation}
\sum_{i=1}^{N_{\rm GRB}} \left(\frac{DM_\gamma(z_i;\Omega_M,\Omega_\Lambda,\eta,\kappa) - DM_{\rm th}(z_i;\Omega_M,\Omega_\Lambda)}{\sigma_{DM_\gamma}(\Omega_M,\Omega_\Lambda,\eta,\kappa)}\right)^2,
\label{eq:chi2}
\end{equation}
\noindent where $N_{\rm GRB}$ is the number of GRBs. We do this for
all cosmologies in our grid (with maximum resolution 51 x 51) and
construct a $\chi^2$ surface, shown in Figure~\ref{fig4} (for Set A)
for a range of assumptions for density and its error.  In principle,
the minimum $\chi^2$ should then correspond to the favored
($\Omega_M$,$\Omega_{\Lambda}$) cosmology. Equivalently, the cosmology
can be parameterized in terms of ($\Omega_M$,$w$), as in
\citet{ries04b}, but the sample requires a substantial fraction of
low-$z$ bursts (which our sample is lacking) for optimal sensitivity
to $w$.  Again, note that there is no need to marginalize over $E^{*}$
(implicit in the the zero point ${\rm zp}$ for $DM_{\gamma}$;
eq.~\ref{eq:DM_gamma}) because the parameters $\kappa/(E^{*})^\eta$
and $\eta$ are fit directly from the data for each cosmology.
Similarly, there is no need to marginalize over the Hubble constant
because its effect cancels in eq.~\ref{eq:chi2}; the $5{\rm log}(h)$
implicit each term of the numerator cancels and the denominator (a log
space error) is a fractional error in real space, which is independent
of $h$.

Although the method is, in principle, sound, the current data provide
essentially {\it no meaningful constraints} on the cosmological
parameters because the shape and normalization of the $\chi^2$
surfaces is highly sensitive to input assumptions and data references
for individual bursts (which are outliers to the Ghirlanda relation
for some input data and not for others).  Under our assumptions, the
data do not give a good fit for the Hubble diagram in {\it any}
cosmology in our grid [$0 \le \Omega_M $,$\Omega_{\Lambda} \le 2$],
with a minimum $\chi^2_{\nu} = 4.78$ for the ($\Omega_M$,
$\Omega_{\Lambda}$) = ($0.12$, $1.32$) cosmology (see
Figure~\ref{fig4}).  Although not shown here, for our assumptions, we
find 2$\chi^2_{\nu} \gtrsim 2$ (a poor fit), at the minimum of the
Hubble diagram surface for each of the data sub sets A, G, and D.
This is not surprising since the Ghirlanda relation itself --- the
basis for the cosmographic standard candle assumption --- is not well
fit by a power law under our input assumptions in any reasonable
cosmology for any of the data sets.

As with the Ghirlanda relation, by changing the input assumptions,
once can improve the Hubble diagram fit. However, an interesting ---
but somewhat anticlimactic --- feature emerges.  As shown in
Figure~\ref{fig4}, for our Set A, the peculiar, GRB-favored loitering
cosmology ($\Omega_M$, $\Omega_{\Lambda}$) = ($0.12$, $1.32$) remains
essentially invariant over a range of input assumptions for density
and its error.  Although not shown herein, we have confirmed that this
strange attractor-like behavior (at the surface minimum) remains for
our data Set G (less so for Set D), although the shape and minimum
$\chi^2_{\nu}$ of the surface do change sensitively due to small
number statistics.  This is not surprising, as \citet{ghir04b} find a
similar best fit cosmology, ($\Omega_M$, $\Omega_{\Lambda}$) =
($0.07$, $1.2$), for their data Set G$^{*}$, although this point is
overshadowed as they present their fit jointly with SNe Ia data (see
\S~\ref{sec:compare}).  Extending upon the work of \citet{ghir04b},
\citet{firm05} also note the appearance of ``mathematically
undesirable attractors'' near the loitering region, claiming that they
are mathematical artifacts which can be removed with a new Bayesian
approach. The \citet{firm05} method does not use the traditional
goodness of fit from a $\chi^2$ analysis, and although it probably
deserves further study, it is unclear if it is warranted given the
data and sensitivity to input assumptions.  At least for our data and
assumptions, the best fit parameters and errors are only meaningful if
the fit is implicitly good, which is not the case for all density
assumptions with fractional errors $\lesssim 125\%$ (the value assumed
in \citealt{ghir04b}).  This is illustrated in Figure~\ref{fig4}.
Thus, on statistical grounds, we are not entitled to believe the best
fit loitering cosmology currently favored by GRBs, relieving us of the
burden of explaining a cosmology inconsistent with flatness which
comes close to seriously challenging the Big Bang model.  All told,
the results herein indicate that, when considering the full data set
for a range of input assumptions, GRBs are simply not yet useful for
cosmography.

\section{Cosmography Comparisons}
\label{sec:compare}

Since there are a host of potential uncertainties in this nascent
approach to GRB cosmography, at present, we focus on constraining
$\Omega_M$ and $\Omega_{\Lambda}$ using GRBs {\it alone}.  While
\citet{dai04} and \citet{ghir04b} have attempted to constrain $w$, the
former with GRBs alone, and the latter using a combined fit with SNe
Ia, we consider this well-motivated but likely premature, due both to
the presence of many unaddressed and potentially problematic
systematic errors (which we attempt to address in \S~\ref{sec:sys}),
along with the aforementioned sensitivity to input assumptions, data
set, and the relatively small GRB sample compared to SNe Ia.

\subsection{Addressing the Dai et al.\ Cosmographic Analysis}
\label{sec:dai}

\citet{dai04} have made use of the $E_p$--$E_{\gamma}$ relation to form
a more standard candle and test cosmological models. They report
remarkably tight constraints on $\Omega_M = 0.35^{+0.15}_{-0.15}$
(68.3\% confidence assuming flatness).  Yet, there are a number of
reasons why we believe this work has significantly overstated the
cosmographic power of GRBs. First, the \citeauthor{dai04} sample
contains 7 fewer bursts (12 vs.\ 19; $>$ 50\%) than our sample.  As
seen in Table~\ref{table2} and graphically in Figure~\ref{fig1}, two
of the absent bursts (GRBs 990510 and 030226) are the two most extreme
outliers in $A_{\gamma}$. GRB\, 990510 remains an outlier independent
of density assumptions as it has a density constraint \citep{pana02},
while GRB\, 030226 is an outlier under either set of assumptions,
worsening for the \citet{dai04} density assumption relative to
ours. \citeauthor{dai04} do offer some justification to exclude these
two bursts, but clearly these exclusions --- which we feel are
unwarranted --- help to significantly tighten the scatter and improve
the cosmology statistics.

Second, the authors did not perform the fit of the Ghirlanda relation
self-consistently but instead assumed the slope of the relation to be
fixed for all cosmologies. The value $\eta^{-1} = 1.5 \pm 0.08$
derived in \citet{dai04} assumes a ($\Omega_M$, $\Omega_{\Lambda}$,
$h$) = (0.27, 0.73, 0.71) cosmology.  The authors treat their fit as a
rough confirmation of the $\eta = 0.706 \pm 0.047$ slope found in
\citet{ghir04} for a slightly different ($\Omega_M$,
$\Omega_{\Lambda}$, $h$) = (0.3, 0.7, 0.7) cosmology, and fix
$\eta^{-1} \equiv 1.5$, thereafter, neglecting the derived
uncertainty. \citet{dai04} do attempt to justify this and note ``this
power to be insensitive to $\Omega_M$'' for their data set.  However,
as we have shown, while the particular value of the slope does not
vary dramatically, even for a wide range of cosmologies, one can not
ignore even this small cosmology dependence in the context of
self-consistent cosmography. By fixing $\eta^{-1} \equiv 1.5$, the
\citet{dai04} analysis ignores the fact that the value of $\eta$ is
not known {\it a priori}, but instead is a simple empirical (bad) fit
to noisy data.  \citet{ghir04b} also express similar concerns in their
discussion of the \citet{dai04} analysis.

As with \citet{ghir04} and \citet{ghir04b}, \citet{dai04} assumed a
density ($n=3$ cm$^{-3}$) which improves the fit relative to our
assumption of $n=10$ cm$^{-3}$.  \citet{dai04} also assume the small
angle approximation, which, as mentioned, is accurate for
$E_{\gamma}$, $Dl_{\gamma}$, and $\sigma_{E_{\gamma}}$, but {\it
overestimates} the error in $Dl_{\gamma}$, which can be derived
analytically (see \S~\ref{sec:error}).  Alone, overestimating the
errors $\sigma_{Dl_{\gamma}}$ improves the fit to the GRB Hubble
diagram.  However, this is compensated for in eq.\ 5 of \citet{dai04}
relative to our eq.~\ref{eq:Dl_gamma_err}, since they assume a smaller
error on the density ($11\%$ v.s. $50\%$), and their eq.\ 5 neglects
the error terms we include involving $\eta$, $\kappa$, and the
$k$-correction.  These competing effects lead us to derive similar
typical errors on the distance modulus, where we find $\sim 0.42$ mag,
vs. $\sim 0.45$ mag in \citet{dai04}. However, it is hard to perform a
direct comparison since the authors do not report a goodness of fit
for their favored cosmology, whereas we find a minimum $\chi^2/$dof $=
4.78$, (17 dof) for the GRB Hubble diagram under our assumptions for
our Set A.  Furthermore, \citeauthor{dai04}\ present constraints on
the equation of state parameter $w$ given priors on flatness and
$\Omega_M$, which is an interesting potential application of GRB
cosmology, but may be premature given the small dataset, the large
dependence upon the outliers, and the strong sensitivity to input
assumptions.

\subsection{Addressing the Ghirlanda et al.~(2004b) Analysis}
\label{sec:ghir04b}

\citet{ghir04b} have taken a number of steps to improve upon the Dai
et al.~analysis. They have rightfully acknowledged that the
$E_p$--$E_{\gamma}$ correlation should be re-calibrated for each
cosmology and should include the uncertainty in the slope $\eta$ when
performing a cosmographic analysis. They too, independent of our work,
have noted that GRBs alone are insensitive to the measurement of
$\Omega_{\Lambda}$ (we specifically note that this insensitivity is
directly attributable to the lack of low-redshift bursts, although
\citealt{ghir04b} do suggest the need for more lower-$z$ bursts). The
\citealt{ghir04b} analysis does not include 4 bursts (these bursts
actually slightly {\it improve} the goodness of fit of the relation to
a power law). \citeauthor{ghir04b} also avoid using the small angle
approximation to calculate $E_{\gamma}$ in practice, although they do
not present the equations for the error analysis explicitly.

Both our fit to the $E_p$--$E_{\gamma}$ relation and the
\citeauthor{ghir04b} fit --- with $\chi^2_{\nu}=3.71$ (17 dof) and
$\chi^2_{\nu}=1.27$ (13 dof), respectively --- follow from our
different input assumptions and data selection choices. This
highlights the extreme sensitivity to the input assumptions
(especially density), uncovered here when trying to reconcile the
differences between our respective works.

Our original disagreements stemmed from the difficulty involved in
interpreting the cosmographic method of analysis in \citet{ghir04b},
which, in contrast to \citet{dai04}, is presented in words but not
explicitly formulated in equations. As mentioned, from \citet{ghir04b}
alone, it is not clear that when they ``allow $n$ to cover the full
[1-10] cm$^{-3}$ range'', this means $n=3^{+7}_{-2}$ cm$^{-3}$
$\rightarrow \sigma_{n} \approx \sqrt{7\times2} = \sqrt{14} = 3.74$
cm$^{-3}$ (roughly 125\% error), which is required to reproduce
$\chi^2_{\nu}=1.27$ for the fit to the $E_p$--$E_{\gamma}$ relation
from their data.  This turns out to be crucial, because without this
extra information, it is not possible to compare or even reproduce
their results for the $E_p$--$E_{\gamma}$ relation from
\citet{ghir04b} and \citet{ghir04} alone.  Ultimately, however,
investigation of this elucidated the sensitivity to density.

Rather than focusing on the cosmology selected by GRBs alone, the
authors report a joint fit with SNe Ia.  By including a set of 15 GRBs
(with large errors) along with 156 (better constrained) SNe Ia data
points (the ``Gold'' sample of \citealt{ries04b}), it is clear that
the joint fit presented in \citet{ghir04b} is dominated by the
supernovae, which already are consistent with today's favored
cosmology concordance model derived from CMB data \citep{sper03}, and
large-scale structure \citep{tegm04}.  \citet{ghir04b} argue that SNe
Ia themselves are only marginally consistent with {\it WMAP}, whereas
the combined SNe Ia + GRB fit results in contours that are more
consistent with a flat, $\Lambda$--dominated universe. However, this
line of reasoning ignores the fact that GRBs alone are strikingly {\it
inconsistent} with {\it WMAP} or flatness, where the best fit found in
\citet{ghir04b} straddles the cosmic loitering line, which borders the
region in the $\Omega_M$--$\Omega_{\Lambda}$ plane for which there is
{\it no Big Bang} (although see \citealt{firm05}).  While it is
certainly reasonable to assume flatness as a prior and explore the
outcome, we feel it is important to stress the cosmographic potential
of GRBs alone, and first determine whether GRB cosmography is robust
and comparable to cosmography with SNe Ia before attempting to combine
them.  Ultimately, the sensitivity to input assumptions and data
selection we have found here makes it currently inappropriate to use
GRBs for cosmography, let alone combine with other better understood
standard candles.

\section{Potential Biases for Future GRB Cosmography}
\label{sec:sys}

Here, we briefly identify some major potential systematic errors
concerning GRB cosmography. The list is not meant to be comprehensive,
but to serve as the starting point for future work. We do not discuss
possible selection effects on the sample (e.g., Malquist bias), but see
\citet{band05} which considers selection effects in testing the
consistency of a large sample of {\it BATSE} bursts with the Ghirlanda
relation, extending upon similar work for the Amati relation
\citep{naka04b}. Although \citet{band05} conclude that as many as
$\sim33\%$ of the bursts in their sample may not be consistent with
the Ghirlanda relation, this depends sensitively on the assumed
distribution for $f_b$. Under the least model-dependent assumption
which only requires $E_{\gamma} \leq E_{\rm iso}$ for all bursts
(e.g., $f_b \le 1$), \citet{band05} estimate that only 1.6\% of their
sample is inconsistent with the Ghirlanda relation.  In any case, the
\citet{band05} analysis raises the possibility that the Ghirlanda
relation itself may merely reflect observational selection effects,
which, if true, would fundamentally undermine any cosmographic use of
the relation.

\subsection{Cosmological $k$-correction}
\label{sec:kcor}
The choice of rest frame bolometric bandpass for the cosmological
$k$-correction \citep{bloo01b} $[E_1,E_2]$ is implicit in the
definition of, and any interpretation of, $E_{\gamma}$
(eq.~\ref{eq:egamma}).  If any bursts have $E_p < E_1$ (or $E_p >
E_2$) keV then we may be systematically underestimating the fluence
and energy outside the bandpass.  In our Set A, however, the lowest
$E_p$ bursts -- (030329: $79$ keV, 021211: $91$ keV, 041006: $109$
keV, and XRF 030429: $128$ keV) all have $E_p > 20$ keV by at least a
factor of $\sim$4. XRFs 020903 and 030723 have only upper limits $E_p
< 10$ keV and $E_p < 30$ keV, respectively, so are not included in Set
A.  There is one burst, however, with $E_p > 2000$ keV, 990123: $E_p =
2030$ keV (the second closest is 000911: $E_p = 1192$ keV).  Thus, for
some bursts, we slightly underestimate the energy.  As such, a
bandpass of $[1,10^{4}]$ keV
\citep{bloo01b,amat02,dai04,ghir04,ghir04b}, may be more appropriate
than the traditional {\it BATSE} bandpass, although this choice has a
much smaller effect than the sensitivity to input assumptions, at
least for the current sample.  For future samples, with several XRFs
(or GRBs) with low $E_p$, it may be more appropriate to choose $E_1 <
1$ keV (also stressed by G. Ghirlanda --- private communication).  In
contrast, there are diminishing returns for increasing $E_2$
arbitrarily, as the typical fractional error on the $k$-correction
increases from $\sim$11\% for $[1,10^{4}]$ keV, to $\sim$ 25\% for the
$[1,10^{5}]$ keV bandpass, with the typical $k$-correction only
increasing from $\sim$1.5 to $\sim$2. Furthermore, for $E_2 > 10^{4}$
keV (10 MeV), we are surpassing the limit beyond which we have strong
evidence to believe in our extrapolation of the Band spectrum.

\subsection{Covariance Between Observables}
\label{sec:covar}
Ignoring covariance where it exists will systematically underestimate
the error on the GRB distance modulus. However, as shown in earlier
error analysis, even assuming maximal covariance --- which we argue is
unlikely --- leads to at most a factor of $\lesssim2$ underestimate of
the errors in $E_{\gamma}$, $A_{\gamma}$, or $Dl_{\gamma}$,
respectively.

\citet{bloo03} discuss possible covariances between $S_{\gamma}$ and
the inference of $\theta_{\rm jet}$ (or $f_b$) arguing that the
effects should be small as the two quantities are determined from the
observationally distinct measurements of different phenomena - i.e.,
the prompt emission and the afterglow.  \citet{bloo03} also argue
that, despite both being derived from broad band afterglow modeling,
$t_{\rm jet}$ and $n$ should have small covariance, because $t_{\rm
jet}$ is usually determined from early optical/IR afterglow data
whereas $n$ --- in the rare cases where it is estimated --- is best
constrained by late time radio data (see their footnote 6).
\citet{bloo01b} also argue that the possible covariance between
$S_{\gamma}$ and $k$ is small, introducing at most a factor of $\sim$2
uncertainty into the error on $k$ (see their \S 2.1).

Because of the $k$-correction, $E_{\gamma}$ = $E_{\gamma}[{ k}(E_p)]$,
and thus $E_p$ and $E_{\gamma}$ are not completely independent
variables.  As such, there is certainly some covariance, but it should
be small in practice, because $k$ and $\sigma_{k}$ are only slowly
varying functions of their inputs and depend most on the choice of
rest frame bolometric bandpass $[E_1,E_2]$ keV.  Although the goodness
of fit to the Ghirlanda relation worsens (under our assumptions) if
one ignores the k-correction (i.e., by assuming $k=1$ for all bursts),
the value of $E_{\gamma}$ itself depends on a combination of
observables with no relation to $E_p$ (e.g., $t_{\rm jet}$, $n$,
etc...), implying that the $E_p$--$E_{\gamma}$ relation itself is not
in doubt on these grounds.  As such, there is also certainly an
intrinsic correlation between $E_p$ and $E_{\gamma}$, but unlike the
covariance above, which describes a mathematical dependence affecting
the correlated {\it measurement} of $E_{\gamma}$ and $E_p$, the {\it
intrinsic} correlation is presumably based on local GRB physics, and
is therefore not reflective of observations with correlated errors
(although, again, see \citealt{band05}).

Finally, a judicious choice of $E^{*}$ can minimize the covariance
between the measurements of the parameters $\eta$ and $\kappa$
(i.e., the off-diagonal elements in the covariance matrix of the
$E_p$--$E_{\gamma}$ fit $\rightarrow$ 0), thus eliminating covariance
terms from the Ghirlanda parameters in the cosmography error analysis.
A different choice of $E^{*}$ would not affect the value of
$Dl_{\gamma}$ or $A_{\gamma}$, since the value of $\kappa$ in the fit
to the Ghirlanda relation would change to compensate, scaling as
$\kappa \propto (E^{*})^{\eta}$.

\subsection{Gravitational Lensing}
\label{sec:lens}

Gravitational lensing is not likely to dominate the systematics,
though higher redshift bursts are more likely to be lensed than lower
redshift SNe Ia. \citet{bloo03c} has argued, based on beaming, that
the probability of detection for a high-redshift GRB is largely
unaffected by Malquist bias (but see also \citealt{balt05}); so the
principal concern is whether the inferred values of $E_{\gamma}$ will be
systematically skewed for bursts at higher redshift. The probabilities
of strong lensing or micro-lensing the GRB are small, $< 5 \times
10^{-3}$ ($z_{\rm GRB} < 5$; \citealt{porc01}) and $\le$0.01
\citep{nemi98}, respectively.  Here we disregard the higher
probability of micro-lensing of the afterglow, since afterglow fluxes
are not used to derive $E_{\gamma}$, although, clearly, a micro-lensed
afterglow could confound the measurement of $t_{\rm jet}$. Still,
strongly-lensed GRBs should be more recognizable as such by the
observations of strong foreground absorption in the early-afterglow
spectra and/or the presence of a galaxy near the burst line-of-sight
in late-time imaging. Weak lensing, with a broad probability of
amplification between 0.8 and 1.2, is expected at $z>3$ in a
$\Lambda$CDM model \citep{wangy02} but, since there is roughly an
equal probability of amplification and de-amplification, weak lensing
biases are systematically suppressed with a larger sample size.

\subsection{Wind-Blown Circumburst Environment}
\label{sec:wind}

If GRB progenitors are massive Wolf-Rayet type stars as in the popular
collapsar model \citep{woos93} or the hypernova model \citep{pacz98},
one naturally expects at least some bursts to go off in the presence
of a wind-blown environment (WIND) where the radial density profile
varies as the inverse square of the radial distance
\citep{chev99,chev00,liz03}.  Following equation (31) of
\citet{chev00}, a WIND modifies our equation~\ref{eq:theta_jet} for
the jet opening angle, and in general, $E_{\gamma}$ will be smaller
when inferred for the WIND case for the same value of $t_{\rm jet}$
and typical density scalings (e.g., $A_{*}$=1, defined in
\citealt{chev00}).  Thus, in the context of the fit to the Ghirlanda
relation, a WIND will help an outlier burst to fall on the relation
only if $E_{\gamma}$, calculated assuming an ISM, was too large (i.e.,
the data point has excess energy on the x-axis in Figure~\ref{fig1}
relative to the best fit line).  Of the bursts that are outliers in
this sense (970508, 011211, 020124, and 020813 -- with 011121 and
010921 as lower and upper limit outliers respectively), only for
011121 is there strong support for a WIND \citep{pric02b}.  For GRB
970508, the analysis of \citet{frai00} claims to rule out a WIND,
whereas \citet{chev00} and \citet{pana02} claim support for a WIND.
For the remaining bursts in our sample Set A where a WIND has been
supported by at least some analyses: 980703; \citet{pana01c}, 991216;
\citet{pana01c} and \citet{pana02}, 021004; \citet{liz03} (but see
\citealt{pand03}), 030226; \citet{dai03}, the WIND would tend to lower
the energy in the x-axis of Figure~\ref{fig1}, making them greater
outliers.  Generally, there is a lack of strong evidence for a WIND
for most bursts.  Furthermore, WIND interaction with the ambient
medium (termination shock) may still lead to a roughly constant
density (ISM) profile beyond some radius \citep{rami01}.  As such, the
ISM assumption is reasonable and does not lead to a major systematic
error relative to the WIND case.

\subsection{Density Assumptions}
\label{sec:n_dist}

The assumption of the {\it same} density for all bursts lacking
constraints leads to a potentially major systematic error. From the
set of 12 bursts with the best constrained densities in
Table~\ref{table1}, estimates range from $0.29$--$30$ cm$^{-3}$ with a
mean of 16.5 cm$^{-3}$ and a standard deviation of 12.7 cm$^{-3}$.
This gives some justification to our earlier order of magnitude
assumption of $n=10$ cm$^{-3}$, but highlights the large uncertainty
in assuming the {\it same} density for all unknown bursts, which, in
nature will be drawn from a wider distribution.  Current constraints
limit density roughly to the $0.1$--$100$ cm$^{-3}$ range or
greater\footnote{\citet{pana02} report a very low density of
$1.9^{+0.5}_{-1.5} \times 10^{-3}$ cm$^{-3}$ for GRB 990123 (see their
Table 2), although this estimate has been superseded by more recent
analyses -- e.g., \citet{pana04}, where the authors report considerably
higher densities in the range $0.1$--$1$ cm$^{-3}$ (see their Figure 1).}
(see \citealt{pana02}).  In addition, even these constraints are
highly uncertain as density is not measured directly but requires
detailed broadband afterglow modeling, where in most cases, the fit
parameters are under-constrained by the sparse data and the model
uncertainties may be {\it much greater} than the reported statistical
uncertainties. All this indicates that, at the very least, a more
conservative error assumption is appropriate for the density. This, of
course, would naturally improve the fit to the Ghirlanda relation.

Despite the uncertainties, we have shown that good fits are {\it
possible} simply by changing the unknown density (and error) for all
bursts (Figure~\ref{fig2}).  However, granting that the true
densities likely follow some wide distribution (rather than the
effective $\delta$-function we have been assuming), one can allow the
assumed densities of individual bursts to vary, drawing them from this
distribution.  As a simple exercise, we use a Komolgorov-Smirnoff
(K-S) test to determine whether the distribution of known densities is
consistent with the distribution of densities tuned to make {\it all}
the nominal outlier bursts fall on the relation.  For simplicity, we
fix $\xi=0.2$ for the exercise.  From the set of the 12 most
reliable density estimates, only 6 are also in Set A, leaving 13 of 19
bursts with no density constraints.  Fitting the $E_p$--$E_{\gamma}$
relation only with those 6 bursts, we find $\eta = 0.806 \pm 0.074$,
$\chi^2_{\nu}=3.96$ (4 dof).  Using this as a baseline fit, we solve
for the individual densities necessary to make all remaining 13 bursts
fall on the relation.  Comparing the set of 12 known and 13 tuned
densities via a K--S test indicates an acceptable consistency with a
K--S probability of 11\%, which is meaningful if there are at least 4
bursts in each set \citep{pres92_c}.\footnote{One can extend this to a
larger distribution including tentative density estimates (see
Appendix~\ref{sec:data_selection}).  This yields a set with 11
additional bursts for a total of 23 with known + tentative density
estimates.  This set is consistent with the set of 13 bursts tuned to
fall on the relation with K--S probability of 32\%.  Furthermore, if
we relax the constraints, and do not require {\it all} outlier bursts
to be tuned to fall on the relation, we can achieve even greater
levels of consistency, although the K--S test itself is only a
consistency test, not a measure of the goodness of fit
\citep{pres92_c}.}  If the density distributions had been obviously
inconsistent, say with K--S probabilities $\ll5\%$, then the relation
would remain a poor fit {\it independent of any assumptions for the
density}.  In fact, we find that there are reasonable density choices
consistent with the distribution of known densities which lead to a
good fit for the relation.\footnote{This, for example, is not true for
the Amati relation for which depends on assumptions concerning the
rest frame $k$-correction bandpass $[E_1,E_2]$, but not on
$\xi$ or $n$.  As such, the goodness of fit of that relation
can not be made to approach unity by changing the assumptions.}.  This
leaves some hope that a sample of bursts with well-constrained
densities (and efficiencies) in the {\it Swift} era may reveal an
underlying good of fit for the relation, which is a prerequisite for
any cosmographic utility. More detailed analysis would require
marginalizing over, or sampling statistically from, the assumed
density distribution (e.g., a Monte Carlo simulation), which is beyond
the scope of this work.

\subsection{Assuming a $\gamma$--ray Production Efficiency}
\label{sec:eta_gamma}

As with density, assuming the {\it same} efficiency for all bursts
represents a potential systematic error, although, as discussed, it is
likely to be a much weaker effect.  Following, \citet{frai01}, we
assume a $\gamma$--ray production efficiency of $\xi=0.2$
(20\%) for all bursts in our sample.  In the context of the internal
shock model, this is consistent with the range of theoretically
predicted efficiencies: $\lesssim 1\%-90\%$
(\citealt{koba97,kuma99,lazz99,belo00,guet01,koba01}). Furthermore,
$\xi$ has been reported for individual bursts for the
$[20,2000]$ keV bandpass, as determined with radio fireball
calorimetry, X-Ray modeling of the afterglow kinetic energy, or other
estimates of the total energy of the fireball
\citep{pana01c,pana02,yost03,berg03,berg04a,lloy04b}.  Estimates for
individual bursts range from $\sim$3\% (970508:
\citealt{yost03,berg04a}), to as high as $\sim$88\% (991208:
\citealt{pana02}). Although the various techniques used to estimate
$\xi$ are highly uncertain, clearly the assumption of a
constant efficiency for each burst is suspect. In light of the
uncertainty involved, it may be appropriate to at least assume some
error on $\xi$, for example a 50--100\% error, in future
work.  As with density, a potential future approach involves assuming
a distribution (e.g., a Gaussian), and marginalizing over it, or
sampling from it statistically.

\section{Discussion}
\label{sec:dis}

We have shown that, given our set of assumptions, the fit to the
Ghirlanda relation remains poor in the standard cosmology
($\chi^2_{\nu} > 3$), across the entire grid of cosmologies, and for
all data subsets that we consider.  As was the case for the standard
cosmology, $\chi^2_{\nu}$ can only be made acceptable at the minimum
of these surfaces by changing the input assumptions or by choosing
different parameter references for individual bursts.  Although this
casts doubt on the current cosmographic utility of the relation, it
must be stressed that it is {\it possible} to obtain good fits simply
by changing the density (and/or efficiency) --- to otherwise
reasonable values --- for bursts without reliable constraints.

The value of $\eta$ (and $\kappa$) might be determined {\it a priori}
from {\bf i)} a well motivated theoretical model or {\bf ii)} by
measuring a sample of GRBs at low redshift, where the observed GRB
properties are essentially independent of the cosmology
\citep{ghir04b}.  This low-$z$ population would represent a ``training
set'', analogous to the training set of low-$z$ SNe Ia used to
calibrate the various light curve shape corrections to Type Ia SNe
magnitudes in a cosmology independent way.  With GRBs, at present, we
do not have the luxury of such low redshift calibration, so the best
one can do is use the {\it data itself} as the training set, and
calibrate the relation separately for each cosmology.  As also
realized by \citet{ghir04b} (in contrast to \citealt{dai04}), this is
currently the only self-consistent way to do cosmology with GRBs.

Procuring such a training set may be feasible in practice.  In the
current sample of $39$ bursts with known redshifts
(Table~\ref{table1}), GRBs 980425, 030329, and 031203 have $z < 0.17$.
With a similar detection ratio, {\it Swift} may find $\sim 5$--$10$
such low-$z$ bursts/year, and with its higher sensitivity, possibly an
even greater number if the intrinsic population has been
underestimated.  This would conservatively provide a reasonable
training set of $\sim 10$--$20$ objects within only 2 years.  It is
noteworthy, however, that only 1 burst {\it in our sample thus far}
(GRB 030329, $z$=0.1685) falls into this potential training set class.
As seen in Fig.~\ref{fig1}, 030329 is remarkable because it it highly
discrepant from the mean in energy, yet falls extremely close to the
best fit line for the $E_p$--$E_{\gamma}$ relation. As a single burst,
it is the low redshift anchor of the relation and comes as close as
possible to acting as a cosmology independent calibrator.  However,
one needs more than a single point to constrain a slope, and although
more low-$z$ bursts are expected with {\it Swift}, it remains to be
seen whether they will {\it actually fall on the relation}.  In fact,
as mentioned, the two lowest redshift GRBs (980425 and 031203) are the
two {\it most striking outliers} to the relation regardless of any
assumptions about the value of the ambient density.  Rather than a
failure of the $E_p$--$E_{\gamma}$ relation, such outliers might serve
as a diagnostics for identifying different progenitor classes, going
beyond the recognition of purely sub-energetic bursts, which are now
quite common.

In fact, with the relatively recent discoveries of GRBs 030329 and
031203 and XRFs 020903 and 030723, the existence of true outliers in
the $E_{\gamma}$ distribution became incontrovertible; GRB\,980425 is
not simply a singular anomaly in prompt-burst energy release. Without
compelling reason to exclude these outlier bursts on
energy-independent grounds, what was once a promising prospect, the
$E_{\gamma}$ distribution (e.g., $\epsilon_{\gamma}$), is clearly a
poor standard candle.  Even if there exists a standard reservoir of
energy in GRB explosions, on conceptual grounds, it is entirely
plausible that $E_{\gamma}$ should differ from burst to burst, sensitive
to the variation in $\gamma$--ray efficiency.  The energy channeled
into gravitational-radiation, neutrinos, and the supernova explosion
are also likely offer significant contributions to the total energy
budget.
 
Furthermore, \citet{berg03c} have shown that the kinetic energy
($E_k$) in relativistic ejecta (as proxied by the radio and X-ray
afterglow) may be comparable (if not greater) than the $E_{\gamma}$.  Of
course, this hypothesis, in concert with the Ghirlanda power-law,
implies a trivial connection of $E_k$ upon $\eta$ and $E_p$: $E_k =
E_{\rm tot} - (E_p/\kappa)^{1/\eta}$, with $E_{\rm tot} \approx 5$ foe
$= 5 \times 10^{51}$ erg \citep{berg03c}. In this context, although in
their fit to GRBs with X-ray afterglow \citet{lloy04} do not find a
constant $E_{\rm tot}$, it is curious to note that those authors do
find that $E_k \propto E_p^{1.5 \pm 0.5}$, a power-law consistent with
$1/\eta \sim 3/2$. Perhaps more interesting, if $E_{\rm tot}$ is
indeed constant then the efficiency of shock conversion to
$\gamma$-rays, $\xi$, should be $\xi \approx 0.2
(E_p/\kappa)^{1/\eta}$ (as opposed to $\xi \propto E_p^{0.4
\pm 0.1}$ found by \citealt{lloy04}), suggesting XRFs are
lower-efficiency shocks, rather than off-axis GRBs. If there are
multiple jet components \citepeg{berg03c}, the value of $\xi$
is even less than implied by the relation. Also, such a connection
between $\xi$ and $E_p$ would imply that the inference of
$\theta_{\rm jet}$ inheres an implicit dependence upon $E_p$,
requiring a reformulation of $E_{\gamma}$ and thus the Ghirlanda
relation.

Inherent in the reconstruction of $E_{\gamma}$ is also a systematic
uncertainty in the jet structure and diversity of the bursts.  We have
cast the correction formalism in the context of the top-hat model,
where the energy per solid angle remains constant across the cone of
the jet, independent of the observer's viewing angle relative to the
central beaming axis.  If instead, all GRBs jets are universal with
the energy per steradian falling as the square of the azimuthal angle
\citep{ross02,zhanb02a}, or a Gaussian profile
\citep{zhanb04a,lloy04}, a similar {\it spread} of the resultant
$E_{\gamma}$ distributions is inferred. In such alternative jet
prescriptions, we have confirmed that $E_{\gamma}$ still correlates
with $E_p$\footnote{Applying more realistic jet models
\citep{ross02,zhanb02a,zhanb04a,lloy04} comes at the cost of
introducing additional free parameters, which we have no way of
simultaneously determining {\it a priori}.  As such, we must make
assumptions about them when analyzing different jet models, even
though they may not be constant from burst to burst.} and so we argue
that the need to specify a particular jet model is {\it obviated}: all
that is required is the existence of an empirical correlation between
$E_{p}$ and some function of observables, which may happen to be
interpreted as $E_{\gamma}$ in some particular jet model.  Although
the Ghirlanda relation has been interpreted in the context of a top
hat jet model, it is ultimately derived empirically from observables.
 
Although there is still some uncertainty surrounding the physical
basis for the SN Ia light curve peak luminosity-decline rate
correlations \citep{mazz01,timm03,ropk04}, the basic mechanism
involving sensitivity to $^{56}$Ni production is fairly well
understood \citep{pint01}.  In contrast, the physics that gives rise
to the intrinsic correlation between $E_p$ and $E_{\gamma}$ is not
well-understood (although see~\citealt{rees04} and~\citet{eich04} with
the latter concerning the related $E_p$--$E_{\rm iso}$
correlation). While the choice of jet model is irrelevant if one is
interested only in an empirical correlation, it is {\it highly
relevant} if one is seeking a meaningful physical explanation. In
particular, understanding $E_{\gamma}$ {\it alone} requires a more
physical jet model than a simple top hat, as $E_{\gamma}$ has a clear
physical interpretation as the total beaming-corrected $\gamma$--ray
energy, which is computed differently between jet models.  Indeed, a
structured jet, with more energy on axis, finds natural support in
numerical simulations of the ``collapsar model'' \citep{macf01}.  At
the very least, physical jets are likely to have an energy profile
much more complicated that some simple analytic function -- for
example, a highly variable jet core with ``wings''
\citep{rami_rome04}. If future $E_p-E_{\gamma}$ data provide better
support for a single power law, for example, the slope $\eta$ may
contain information about the underlying physics, and might be useful
in actually {\it constraining} jet models, as the intrinsic value of
$E_{\gamma}$ (and probably $E_p$) clearly depend on the jet structure
of the burst.  The physical origin of $E_p$ itself is even less
understood (although, again, see \citealt{rees04}).

Ultimately, as stressed by \citet{ghir04}, the relation clearly has
great promise to lend insight into GRB radiation mechanisms, and is
likely more fundamental that the long discussed $E_p$--$E_{\rm iso}$
relation \citep{naka04b}.  As discussed, a theoretical {\it
cosmologically-independent} explanation for the relation would help
reduce the uncertainties in the determination of $A_{\gamma}$, the
$C_\gamma$ correction term for each burst, and $DM_{\gamma}$ by
effectively reducing $\sigma_{\eta}$ ($\sigma_{\kappa}$) to nil (this,
too, has been noted by \citealt{ghir04b}).  However, better
understanding of the underlying density distribution is required in
either case.

\section{Conclusion}
\label{sec:concl}

Regardless of the physical basis for the Ghirlanda relation, we have
shown that the $C_\gamma$ correction provides a significant
improvement to the standard candle. Further improvements to the
$C_\gamma$ corrections should be possible with more detailed
observations: more early-time measurements of GRB afterglows should
help constrain the density of the circumburst medium, along with its
radial dependence (which may arise from a stellar wind;
\citealt{chev00}), testing our assumption of a constant-density
medium. In addition, the value of the conversion efficiency to
gamma-rays ($\xi$) may not be constant and may indeed be a measurable
quantity for each burst \citepeg{pana02,yost03,berg04a,lloy04b}.  Even
with incomplete density (efficiency) data, a more detailed analysis
can be completed in future work by assuming probability distributions
for $n$ and $\xi$, and marginalizing over them, and/or sampling from
them in a statistical (Monte Carlo) fashion.

The existence of a relationship between $E_{\gamma}$ and another
intrinsic property of the GRB mechanism also augers well for the
potential refinement of the standard energy with additional
relations. For example, correlations between $E_{\gamma}$ and/or $E_p$
and GRB temporal profiles (e.g., variability)
\citep{feni00a,reic01,lloy02b,scha03a} and/or spectral evolution
(e.g., spectral lags) \citep{norr00,scha01,norr02} might prove useful
in reducing the scatter of the dimensionless GRB standard candle
$A_\gamma$. That is, the Ghirlanda relation may prove to be a
projection from a higher-dimension ``fundamental plane'' involving
additional observables.

If, with an expanded dataset and additional refinements to $C_\gamma$,
GRBs prove to be standardizable candles, tests of cosmological models
could be performed to redshifts $z \sim 10$ or higher
\citep{lamb03,brom02}, a lever arm where Hubble diagrams diverge most
which could help pin down {\bf i)} the matter density to higher
precision, {\bf ii)} the redshift of the transition to the epoch of
deceleration, and {\bf iii)} systematics of the dark energy and its
time variation \citep{lind03}, complementary to Type Ia SNe
\citep{ries04b}. Such redshifts are higher than Type Ia supernovae
could ever reach ($z_{\rm max} \sim 1.7$), even in the best of all
possible Type Ia detection scenarios ({\it JWST} notwithstanding),
with a sample essentially free of reddening/extinction by dust, and
with potentially less systematically biased $k$-corrections, several
years before the expected launch of the {\it SNAP} satellite
\citep{scha03a,lind04}.  Ultimately, if the dark energy shows exotic
time variation, ultra high redshift cosmology (e.g., $z>2$) may prove
quite interesting, lending insight into much more than the matter
density.

Also of great interest, an expanded set (with better constrained
densities) will allow for tests of the evolution of the GRB standard
candle $A_\gamma$ with redshift --- clearly a crucial insight if
high-redshift bursts are to be used for cosmography. With the current
sample, no evolution in the corrected energies is apparent, from
redshifts of 0.1 to 4.5, a difference in look back time that is
$\sim$80\% the age of the universe (see Fig~\ref{fig3}), although, of
course, this depends on density assumptions for individual bursts
which could conceivably be tuned to mimic evolution.  Even so, any
systematic evolutionary effects (which must occur at some limiting
redshift when the GRB progenitors become Population III stars;
\citealt{bark01}) are bound to be different than those for Type Ia
supernovae, providing a complementary, independent check.

While indeed more promising than $E_{\gamma}$ (see Fig.~\ref{fig3}) or
the $E_p$--$E_{\rm iso}$ relation (which can be used to construct a
corrected standard candle roughly intermediate in accuracy between
$\epsilon_{\gamma}$ and $A_{\gamma}$ since the Amati relation is
implicit in the Ghirlanda relation), in strong contrast to the
conclusions of \citet{dai04} and \citet{ghir04b}, we have found this
new GRB standard candle $A_{\gamma}$ provides essentially no
meaningful constraints on $\Omega_M$ and $\Omega_\Lambda$ with the
current, small sample of less than 20 events, most notably due to the
sensitivity to data selection choices and assumptions for the
unknown density (efficiency). 

Still, despite the current uncertainties and rather strong dependence
on input assumptions and data selection, we believe the
standardization of GRB energetics holds promise, thanks to the
discovery of the $E_p$--$E_{\gamma}$ relation. SNe Ia data and, by
extension, GRB data probe an orthogonal region in the parameter space.
Whereas CMB power spectrum measurements are sensitive most to
$\Omega_{M}h^2$, $\Omega_{b}h^2$, and $\Omega_{\rm tot}= \Omega_{M} +
\Omega_{\Lambda}$, SNe Ia measurements, and hence GRB measurements are
sensitive essentially to the difference $\Omega_{M} -
\Omega_{\Lambda}$, with GRBs being most sensitive to $\Omega_M$.
Aside from providing more bursts for statistics, with accurate and
homogeneously-determined GRB and afterglow parameters, we expect that
the {\it Swift} satellite will yield further refinements towards a
standardizable GRB energy.

To that end, we stress the importance of early-time broad band ground
based follow up observations to help constrain the ambient density
(efficiency) of future bursts (also of independent interest for
constraining the progenitors).  We also highlight the continued
relevance of the {\it HETE II} satellite (with its 30-400 keV
bandpass) concerning {\it all} applications of the $E_p$--$E_{\gamma}$
relation, as the spectral coverage of the BAT detector on {\it Swift}
is limited largely to the narrow 15-150 keV range \citep{gehr04}.  As
such, the current work strengthens the science case for the ongoing
symbiosis of {\it HETE II} and {\it Swift}.  By further exploring the
$E_p$--$E_{\gamma}$ relation in this manner, we may potentially lend
insight towards both our understanding of GRBs and to the expansion
history of the universe.

{\bf Addendum:} Recent work \citep{firm05} extends upon the work of
\citet{ghir04b}, while \citet{xu05} extends upon the work of
\citet{dai04}.  The new work, however, does not take into account the
sensitivity to input assumptions involving density, $\gamma$--ray
efficiency, etc...  As such, the major criticisms presented herein
still extend to that new work.

\acknowledgements 

A.S.F. acknowledges support from a National Science Foundation
Graduate Research Fellowship and the Harvard University Department of
Astronomy.  J.S.B. gratefully acknowledges a fellowship from the
Harvard Society of Fellows and the generous research support from the
Harvard-Smithsonian Center For Astrophysics.  We thank R.\ Narayan,
R.\ Kirshner, A.\ Filippenko, and K.\ Stanek for helpful discussions
and comments.  The suggestions and criticisms of the anonymous referee
were thoughtful and thorough, leading to a considerably more
comprehensive paper compared to our original submission. We thank
G. Ghirlanda, D.~Lazzati and collaborators for patience and care in
explaining to us the details of their energetics and cosmological
analysis.

\appendix

\section{Data Selection}
\label{sec:data_selection}

Expanding upon \S~\ref{sec:data_selection2}, we discuss data selection
concerns for the observables of interest defined in the text ($E_p$,
$z$, $S_{\gamma}$, $t_{\rm jet}$, $n$, $\xi$, $\alpha$, and
$\beta$), detailing specific cases for individual bursts as is
relevant.

Peak energy measurements are occasionally inconsistent from different
satellites (e.g.,\ GRB 970508), and in these
cases, we choose the bursts that have spectra which are well described
by the Band model and we preferentially choose $E^{\rm obs}_p$
measurements with reported error bars.  For 970508, \citet{jime01}
report $E^{\rm obs}_p=389$ along with $\alpha=-1.191$, $\beta=-1.831$
(all without error bars), whereas \citet{amat02} report $E^{\rm
obs}_p=79 \pm 23$ along with $\alpha=-1.71 \pm 0.1$, $\beta=-2.2 \pm
0.25$.  The \citet{jime01} data for GRB 970508 have no reported error
bars and have $\beta > -2$, which can not be interpreted in the
context of the Band Model.  As such, we use the
\citet{amat02} reference.  

In the absence of reported values of $\alpha$ or $\beta$ (there are no
cases of both missing in our sample), we choose values consistent with
those measured in the sample, although this choice is not critical in
the analysis. In our sample there are 29 bursts with measured
redshifts, peak energies, and $\alpha$, and 20 bursts with measured
redshifts, peak energies, and $\beta$ (for recent bursts observed by
{\it HETE II} in the $[30,400]$ keV bandpass, it is often difficult to
constrain the high energy spectral slope $\beta$).  For the first set,
we find $\bar{\alpha} = -1.11$ with a standard deviation of 0.36 and
for the second set we find mean values of $\bar{\beta} = -2.30$ with a
standard deviation of 0.29.  These values are also representative of
those found for a large sample of bright {\it BATSE} bursts \citep{pree00}.
Thus, in the absence of constraints, we set $\alpha=-1$ (choosing
$\alpha=-1.1$ would not affect the analysis) and $\beta=-2.3$ where
appropriate (the latter is also assumed in \citealt{atte03}, and similar
averages are used in
\citealt{ghir04}).

Occasionally, measurements of fluence $S_{\gamma}$ from different
satellites are inconsistent, but more often than not, we can not
determine whether two independent measurements are inconsistent if
either one or both do not report 1-$\sigma$ error bars. As noted, in
these cases, we use input fluence measurements with reported errors
with priority over fluence measurements in wider bandpasses.  For
example, for pre-{\it HETE II} bursts, we generally will choose a {\it
BeppoSAX} burst measured in the $[40,700]$ keV bandpass with reported
fluence errors over a {\it BATSE} bursts measured in the larger
$[20,2000]$ keV bandpass when the latter does not have reported
fluence errors.

In the case of competing $t_{\rm jet}$ measurements, we choose the
best sampled light curve with the smallest errors on the best fit
value of $t_{\rm jet}$, preferring early time optical data where
available.  However, there may be problems with any measurement that
reports $t_{\rm jet}$ errors of smaller than 10\%, due to intractable
uncertainties in the afterglow modeling (D. Lazzati -- private
communication).  As such, there is reason to consider a lower limit
criteria for fractional errors of 10\% on the jet-break time.
Although we do not modify the reported measurement errors for any
bursts in this fashion, if we did, it would affect the following
bursts: [GRB/XRF: $t_{\rm jet}$ [days]; reference] $\rightarrow$
[011211: $1.56 \pm 0.02$; \citealt{jako03}], [990510: $1.2 \pm 0.08$;
\citealt{harr01}, $1.57 \pm 0.03$; \citealt{stan99}, $1.6 \pm 0.2$;
\citealt{isra99} (we reference the latter)], [021004: $6.5 \pm 0.2$;
\citealt{pand03}], and [030329: $0.481 \pm 0.033$; \citealt{pric03b}].
In addition, [GRB 000926: $t_{\rm jet} = 1.8 \pm 0.1$ \citep{harr01}]
also has a reported jet break error of less than 10\%, but it is not
included in our sample because $E^{\rm obs}_p$ is not found in the
literature.  Again, we do not alter any reported errors, but as an
example, \citet{ghir04} do change the reported error for 011211 from
$t_{\rm jet} = 1.56 \pm 0.02$ to $t_{\rm jet} = 1.56 \pm 0.15$
(e.g., 10\%).

Density measurements require detailed broad band afterglow modeling
(see \citealt{pana02,yost03}), and are generally unknown for most
bursts, requiring us to assume a value.  Of the 52 bursts listed in
Table~\ref{table1}, only 12 have reliable density estimates which are
listed here.  However, at least an additional 11 bursts have densities
reported in the literature: [GRB/XRF: $n$ [cm$^{-3}$]; reference]
$\rightarrow$ [980519: $0.14^{+0.32}_{-0.03}$; \citealt{pana02}],
[990123: $1.9^{+0.5}_{-1.5} \times 10^{-3}$, $0.1-1$;
\citealt{pana02,pana04}, 000911: $0.07$; \citealt{pric02a}], [020124:
$1$; \citealt{berg02b}], [020405: $0.08$, \citealt{berg03d}], [020427:
$1$; \citealt{amat04b}], [020903: $100$; \citealt{sode04}], [021211:
$<1$, $>30$; \citealt{kuma03,pana04}], [030226: $100$;
\citealt{dai03}], [030723: $1$; \citealt{huan04}], and [040924:
$0.01$; \citealt{fan04}], but we do not list them in
Table~\ref{table1} because either {\bf i)} the densities are from
estimates other than broadband afterglow modeling, {\bf ii)} the
estimate assumed a redshift (e.g., 980519, $z=1$ assumed in
\citealt{pana02}), {\bf iii)} the estimates had been contradicted by
further analyses of the same data (e.g., 990123;
\citealt{pana02,pana04}), or {\bf iv)} the densities are unreliable
for some other reason, such as being distinctly presented as tentative
by the authors (e.g., \citealt{pric02a}).

The question of data selection is relatively moot for redshift
measurements as they are the most accurate (usually confirmed by
several follow up spectra) and precise (negligible errors) of our
input observables.  In any case, it is clear that spectroscopic
redshifts are preferred over photometric redshifts, with no preference
between emission or absorption redshifts.  Ultimately, in the case of
spectroscopic redshifts with multiple independent confirmations we
take the measurement with the highest precision, although the results
are rather insensitive to whether the redshift is known to, 3 or 6
decimal places, for example.

\section{Data Comparison and Potential Outliers}
\label{sec:ghir_comp}

Here we note burst by burst differences between our references and
those used in other work \citep{ghir04,ghir04b,dai04}, emphasizing its
effect on the outliers status of individual bursts. Again, references
to sets A, G, and D only refer to the burst names in those subsets,
not to individual data references.

Our data selection differs from \citet{ghir04} mostly from updates to
$S_{\gamma}$ and $E^{\rm obs}_p$ taken from \citet{saka04b}, which was
recently added to the literature, superseding the analysis of
\citet{barr03}, reported in \citet{ghir04}, as the new work now
incorporates a joint fit with the {\it WXM} X-Ray data.  This affects
bursts: 020124, 020813, 030226, and 030328 most significantly for
$E^{\rm obs}_p$ and $S_{\gamma}$.  Additionally, for 011211,
$S_{\gamma}$ is not listed in \citet{ghir04}, although the burst is
used in their analysis, and probably also uses $S_{\gamma} = 5\times
10^{-6}$ erg cm$^{-2}$ \citep{holl02}, which we reference.  Other
minor differences include slightly different references for $t_{\rm
jet}$ and $n$ for 030329, although this makes little difference in the
analysis.

In comparing the outliers between Sets A and G, as noted, using
$t_{\rm jet} = 15 \pm 5$ days (i.e., $t_{\rm jet} = 10$--$20$ days;
\citealt{berg02b}), GRB 020124 is an outlier, although it is not an
outlier with $t_{\rm jet} = 3 \pm 0.4$ days as reported in
\citet{ghir04}, also citing \citep{berg02b} along with
\citet{goro_gcn_1224} and \citet{bloo03} for the same burst in their
Table 2, although we believe the reference group itself is specious.
GRB 021004 ($z = 2.332$) was a significant outlier if we take $E^{\rm
obs}_p = 1080$ keV ($E_p = 3600$ keV) \citep{barr03}.  However, it is
no longer an outlier using $E^{\rm obs}_p = 79.79$ keV, updated from
\citep{saka04b}, a more current analysis of {\it HETE II} burst
spectra.  The only outlier that we include in Set A that is not also
in Set G is 970508, which \citet{ghir04} left out of their sample due
to conflicting $E^{\rm obs}_p$ reports from \citet{amat02} and
\citet{jime01}, where we use the former reference herein.  As noted,
\citet{dai04} do not include 970508, along with the major outliers
990510 and 030226, which they argue should be left out on grounds,
which are, at best, controversial.

Other bursts not in Set A are also minor (1-$\sigma$) to major
($2$--$3 \ \sigma$) outliers in $A_{\gamma}$ depending on the
assumptions regarding $t_{\rm jet}$, $E_p$, and $z$: i) GRB 010222,
with $E_p > 887$ keV \citep{amat02} is a major 3-$\sigma$ outlier.
ii) GRB 010921 falls significantly off the relation if one assumes
$t_{\rm jet} = 33 \pm 6.5$ days \citep{pric02a}.  It is consistent
with the relation if we interpret this jet break as an upper limit, as
we do here, conservatively, following \citet{ghir04}.  \citet{pric02c}
had previously noted the afterglow light curve to also be consistent
with an early jet break $t_{\rm jet} < 1$ day, which would still make
010921 a minor outlier, although in the opposite sense. iii) GRB
011121 ($t_{\rm jet} > 7$ days; \citealt{pric02b}) is an outlier if we
assume $E_p = 295 \pm 35$ keV as reported in \citet{amat04}, although
it is consistent with the relation if we assume $E_p > 952$ keV (Piro
et. al 04, in preparation, as cited by \citet{ghir04}.  iv) GRB 000911
is a major outlier if one assumes $t_{\rm jet} = 0.6$ days, or the
firmer upper limit of $t_{\rm jet} < 1.5$ days from \citet{pric02a},
along with $n=10$ cm$^{-3}$.  \citet{pric02a} also tentatively suggest
a largely uncertain broadband afterglow fit of $n= 0.07$ cm$^{-3}$,
but this would only make 000911 more of an outlier.  The recently
discovered GRB 040924 is also a major outlier under the assumptions
made here.

Several bursts with uncertain redshift also are outliers under
reasonable assumptions.  GRB 980326 has a redshift suggestion of $z
\sim 1.0$ \citep{bloo99} and $t_{\rm jet} < 0.4$ days \citep{groo98},
which make it a 3-$\sigma$ outlier from $A_{\gamma}=1$.  There is some
indication that GRB 980519 has $z\sim 1.5$ \citep{bloo03}, which would
make it a 3-$\sigma$ outlier.  Furthermore, the recently discovered
GRB 030528 with $E^{\rm obs}_p = 32$ keV, $0.4 < t_{\rm jet} < 4$ days
(i.e., $t_{\rm jet} = 2.2 \pm 1.8$ days), and $z<1$ tentatively
reported by \citet{rau04}, also falls off the relation.  Bursts that
are 1--3 $\sigma$ outliers regardless of membership in Set A are
indicated in the leftmost column of Table~\ref{table2}.


\textwidth=6.9in 
\textheight=9.1in
\columnsep=5.0mm 
\parindent=6.0mm
\voffset=-5mm
\hoffset=-5mm

\begin{deluxetable}{rlccccccccc}
\rotate
\tabletypesize{\scriptsize}
\tablewidth{9.2in}
\tablecaption{Compilation of Spectra and Energetics Input Data\label{table1}}
\tablecolumns{11}
\tablehead{
\colhead{GRB} 		& \colhead{$z$} 			&\colhead{$S_{\gamma}$} 		&\colhead{Bandpass} 	 		& \colhead{$t_{\rm jet}$} 	& \colhead{$n$}			&\colhead{$\alpha$} 		&\colhead{$\beta$} 		 		& \colhead{$E^{\rm obs}_p$} 	& \colhead{$E_p$} 	 	& \colhead{References}\\
\colhead{/XRF} 			& \colhead{} 				&\colhead{[$10^{-6}$ {\tiny erg cm$^{-2}$}]} 	&\colhead{[keV]} 	 		& \colhead{[days]} 		& \colhead{[cm$^{-3}$]}			&\colhead{} 			&\colhead{} 			 & \colhead{[kev]} 		& \colhead{[kev]}  		& \colhead{($z$, $S_{\gamma}=S$, $t_{\rm jet}=t$, $n$, $\alpha$, $\beta$, $E_p$)}\\
\colhead{\tablenotemark{a}} 	& \colhead{\tablenotemark{b}} 		&\colhead{\tablenotemark{c}} 		&\colhead{} 		 		& \colhead{\tablenotemark{d}} 		& \colhead{\tablenotemark{e}}		&\colhead{\tablenotemark{f}} 	&\colhead{\tablenotemark{g}} 	& \colhead{\tablenotemark{h}} 	& \colhead{\tablenotemark{i}} 	 & \colhead{}}
\startdata
970228  &        0.6950         &        11.00 (1.00)   &        40, 700        &        \nodata        &        10.00  (5.00) $*$      &        -1.54 (0.08)   &        -2.50 (0.40)   &        115 (38)       &        195 (64)       &        $z$: 1, $S$: 2, $\alpha$: 2, $\beta$: 2, $E_p$: 2    \\
970508  &        0.8349         &         1.80 (0.30)   &        40, 700        &        25.00 (5.00)   &         1.00  (0.50) $**$     &        -1.71 (0.10)   &        -2.20 (0.25)   &        79 (23)        &        145 (42)       &        $z$: 3, $S$: 2, $t$: 4, $n$: 4, $\alpha$: 2, $\beta$: 2, $E_p$: 2    \\
970828  &        0.9578         &        96.00 (9.60)$*$        &        20, 2000       &         2.20 (0.40)   &        10.00  (5.00) $*$      &        -0.70 (0.08)   &        -2.07 (0.37)   &        298 (60)       &        583 (117)      &        $z$: 5, $S$: 6, $t$: 5, $\alpha$: 7, $\beta$: 6, $E_p$: 6    \\
971214  &        3.4180         &         8.80 (0.90)   &        40, 700        &        $>$  2.50      &        10.00  (5.00) $*$      &        -0.76 (0.17)   &        -2.70 (1.10)   &        155 (30)       &        685 (133)      &        $z$: 8, $S$: 2, $t$: 8, $\alpha$: 2, $\beta$: 2, $E_p$: 2    \\
980326  &        [1.00]$*$      &         0.75 (0.15)   &        40, 700        &        $<$  0.40      &        10.00  (5.00) $*$      &        -1.23 (0.21)   &        -2.48 (0.31)   &        47 (5)         &        [94] (10)      &        $z$: 9, $S$: 2, $t$: 10, $\alpha$: 2, $\beta$: 2, $E_p$: 10  \\
980329  &        [2.95]$*$      &        65.00 (5.00)   &        40, 700        &         0.29 (0.20)   &        29.00  (10.00)         &        -0.64 (0.14)   &        -2.20 (0.80)   &        237 (38)       &        [936] (150)    &        $z$: 11, $S$: 2, $t$: 12, $n$: 12, $\alpha$: 2, $\beta$: 2, $E_p$: 2         \\
980425  &        0.0085         &         3.87 (0.39)$*$        &        20, 2000       &        \nodata        &        10.00  (5.00) $*$      &        -1.27 (0.25)   &        -2.30 (0.46)$*$        &        118 (24)       &        119 (24)       &        $z$: 13, $S$: 6, $\alpha$: 6, $E_p$: 6       \\
980519  &        [2.50]         &        10.30 (1.03)$*$        &        20, 2000       &         0.55 (0.17)   &        10.00  (5.00) $*$      &        -1.35 (0.27)   &        -2.30 (0.46)$*$        &        205 (41)       &        [718] (144)    &        $S$: 6, $t$: 14, $\alpha$: 6, $E_p$: 6       \\
980613  &        1.0969         &         1.00 (0.20)   &        40, 700        &        $>$  3.10      &        10.00  (5.00) $*$      &        -1.43 (0.24)   &        -2.70 (0.60)   &        93 (43)        &        195 (90)       &        $z$: 15, $S$: 2, $t$: 16, $\alpha$: 2, $\beta$: 2, $E_p$: 2  \\
980703  &        0.9662         &        22.60 (2.26)$*$        &        20, 2000       &         3.40 (0.50)   &        28.00  (10.00)         &        -1.31 (0.26)   &        -2.40 (0.26)   &        254 (51)       &        499 (100)      &        $z$: 17, $S$: 6, $t$: 18, $n$: 18, $\alpha$: 7, $\beta$: 6, $E_p$: 6         \\
981226  &        [1.50]         &         0.40 (0.10)   &        40, 700        &        $>$  5.00      &        10.00  (5.00) $*$      &        -1.25 (0.05)   &        -2.60 (0.70)   &        61 (15)        &        [153] (38)     &        $S$: 19, $t$: 20, $\alpha$: 19, $\beta$: 19, $E_p$: 19       \\
990123  &        1.6004         &        300.00 (40.00)         &        40, 700        &         2.04 (0.46)   &        10.00  (5.00) $*$      &        -0.89 (0.08)   &        -2.45 (0.97)   &        781 (62)       &        2031 (161)     &        $z$: 21, $S$: 2, $t$: 21, $\alpha$: 2, $\beta$: 2, $E_p$: 2  \\
990506  &        1.3066         &        194.00 (19.40)$*$      &        20, 2000       &        \nodata        &        10.00  (5.00) $*$      &        -1.37 (0.28)   &        -2.15 (0.43)   &        283 (57)       &        653 (131)      &        $z$: 22, $S$: 6, $\alpha$: 7, $\beta$: 6, $E_p$: 6   \\
990510  &        1.6187         &        19.00 (2.00)   &        40, 700        &         1.57 (0.03)   &         0.29 $^{+0.11}_{-0.15}$       &        -1.23 (0.05)   &        -2.70 (0.40)   &        163 (16)       &        427 (42)       &        $z$: 23, $S$: 2, $t$: 24, $n$: 25, $\alpha$: 2, $\beta$: 2, $E_p$: 2         \\
990705  &        0.8424         &        75.00 (8.00)   &        40, 700        &         1.00 (0.20)   &        10.00  (5.00) $*$      &        -1.05 (0.21)   &        -2.20 (0.10)   &        189 (15)       &        348 (28)       &        $z$: 26, $S$: 2, $t$: 27, $\alpha$: 2, $\beta$: 2, $E_p$: 2  \\
990712  &        0.4331         &        11.00 (0.30)   &        2, 700         &         1.60 (0.20)   &        10.00  (5.00) $*$      &        -1.88 (0.07)   &        -2.48 (0.56)   &        65 (11)        &        93 (16)        &        $z$: 23, $S$: 28, $t$: 29, $\alpha$: 2, $\beta$: 2, $E_p$: 2         \\
991208  &        0.7055         &        100.00 (10.00)         &        25, 1000       &        $<$  2.10      &        18.00 $^{+22.00}_{-6.00}$      &        \nodata        &        \nodata        &        \nodata        &        \nodata        &        $z$: 30, $S$: 31, $t$: 32, $n$: 25   \\
991216  &        1.0200         &        194.00 (19.40)$*$      &        20, 2000       &         1.20 (0.40)   &         4.70 $^{+6.80}_{-1.80}$       &        -1.23 (0.25)   &        -2.18 (0.39)   &        318 (64)       &        642 (128)      &        $z$: 33, $S$: 6, $t$: 34, $n$: 25, $\alpha$: 7, $\beta$: 6, $E_p$: 6         \\
000131  &        4.5000         &        35.10 (8.00)   &        26, 1800       &        $<$  3.50      &        10.00  (5.00) $*$      &        -1.20 (0.10)   &        -2.40 (0.10)   &        163 (13)       &        897 (72)       &        $z$: 35, $S$: 35, $t$: 35, $\alpha$: 35, $\beta$: 35, $E_p$: 35      \\
000210  &        0.8463         &        61.00 (2.00)   &        40, 700        &        $>$  0.88      &        10.00  (5.00) $*$      &        \nodata        &        \nodata        &        \nodata        &        \nodata        &        $z$: 36, $S$: 36, $t$: 36   \\
000214  &        [0.42]$*$      &         1.42 (0.40)   &        40, 700        &        \nodata        &        10.00  (5.00) $*$      &        -1.62 (0.13)   &        -2.10 (0.42)   &        $>$ 82         &        $>$ 116        &        $z$: 37, $S$: 2, $\alpha$: 2, $\beta$: 2, $E_p$: 2   \\
000301C         &        2.0335         &         2.00 (0.60)   &        150, 1000      &         7.30 (0.50)   &        26.00  (12.00)         &        \nodata        &        \nodata        &        \nodata        &        \nodata        &        $z$: 38, $S$: 39, $t$: 40, $n$: 41   \\
000418  &        1.1182         &        20.00 (2.00)$*$        &        15, 1000       &        25.70 (5.10)   &        27.00 $^{+250.00}_{-14.00}$    &        \nodata        &        \nodata        &        \nodata        &        \nodata        &        $z$: 22, $S$: 42, $t$: 42, $n$: 25   \\
000630  &        [1.50]         &         2.00 (0.20)$*$        &        25, 100        &        $>$  4.00      &        10.00  (5.00) $*$      &        \nodata        &        \nodata        &        \nodata        &        \nodata        &        $S$: 43, $t$: 44    \\
000911  &        1.0585         &        230.00 (23.00)$*$      &        15, 8000       &        $<$  1.50      &        10.00  (5.00) $*$      &        -1.11 (0.12)   &        -2.32 (0.41)   &        579 (116)      &        1192 (239)     &        $z$: 45, $S$: 45, $t$: 45, $\alpha$: 45, $\beta$: 45, $E_p$: 45      \\
000926  &        2.0369         &         6.20 (0.62)$*$        &        25, 100        &         1.80 (0.10)   &        27.00  (3.00)  &        \nodata        &        \nodata        &        \nodata        &        \nodata        &        $z$: 46, $S$: 47, $t$: 48, $n$: 48   \\
010222  &        1.4769         &        120.00 (3.00)  &        2, 700         &         0.93 $^{+0.15}_{-0.06}$       &         1.70  (0.85) $**$     &        -1.35 (0.19)   &        -1.64 (0.02)   &        $>$ 358        &        $>$ 887        &        $z$: 49, $S$: 50, $t$: 51, $n$: 25, $\alpha$: 2, $\beta$: 2, $E_p$: 2        \\
010921  &        0.4509         &        18.42 $^{+ 0.97}_{- 0.95}$     &        2, 400         &        $<$ 33.00      &        10.00  (5.00) $*$      &        -1.55 (0.08)   &        -2.30 (0.46)   &        89 (17)        &        129 (25)       &        $z$: 52, $S$: 53, $t$: 54, $\alpha$: 53, $\beta$: 55, $E_p$: 53      \\
011121  &        0.3620         &        24.00 (2.40)$*$        &        25, 100        &        $>$  7.00      &        10.00  (5.00) $*$      &        -1.42 (0.14)   &        -2.30 (0.46)$*$        &        217 (26)       &        296 (35)       &        $z$: 56, $S$: 57, $t$: 57, $\alpha$: 7, $E_p$: 7     \\
011211  &        2.1400         &         5.00 (0.50)$*$        &        40, 700        &         1.56 (0.02)   &        10.00  (5.00) $*$      &        -0.84 (0.09)   &        -2.30 (0.46)$*$        &        59 (7)         &        185 (22)       &        $z$: 58, $S$: 58, $t$: 59, $\alpha$: 7, $E_p$: 7     \\
020124  &        3.1980         &         8.10 $^{+ 0.89}_{- 0.77}$     &        2, 400         &        15.00 (5.00)   &        10.00  (5.00) $*$      &        -0.79 (0.15)   &        -2.30 (0.46)   &        87 (15)        &        365 (63)       &        $z$: 60, $S$: 53, $t$: 61, $\alpha$: 53, $\beta$: 55, $E_p$: 53      \\
020331  &        [1.50]         &         0.69 $^{+ 0.84}_{- 0.74}$     &        2, 400         &        \nodata        &        10.00  (5.00) $*$      &        -0.79 (0.13)   &        -2.30 (0.46)   &        92 (17)        &        [229] (43)     &        $S$: 53, $\alpha$: 53, $\beta$: 55, $E_p$: 53        \\
020405  &        0.6899         &        74.00 (0.70)$*$        &        15, 2000       &         1.67 (0.52)   &        10.00  (5.00) $*$      &         0.00 (0.25)   &        -1.87 (0.23)   &        364 (73)       &        615 (123)      &        $z$: 62, $S$: 62, $t$: 62, $\alpha$: 62, $\beta$: 62, $E_p$: 63      \\
020427  &         $<$ 2.30      &         0.58 (0.04)   &        2, 28  &        $>$ 17.00      &        10.00  (5.00) $*$      &        -1.00 (0.20)   &        -2.10 (0.26)   &        3 (3)  &        $<$ 9  &        $z$: 64, $S$: 65, $t$: 65, $\alpha$: 65, $\beta$: 7, $E_p$: 65       \\
020813  &        1.2540         &        97.87 $^{+ 1.27}_{- 1.28}$     &        2, 400         &         0.43 (0.06)   &        10.00  (5.00) $*$      &        -0.94 (0.03)   &        -1.57 (0.04)   &        142 (13)       &        320 (30)       &        $z$: 66, $S$: 53, $t$: 66, $\alpha$: 53, $\beta$: 53, $E_p$: 53      \\
020903  &        0.2510         &         0.10 $^{+ 0.06}_{- 0.03}$     &        2, 400         &        \nodata        &        10.00  (5.00) $*$      &        -1.00 (0.20)$*$        &        -2.62 (0.55)   &        3 (1)  &        3 (1)  &        $z$: 67, $S$: 53, $\beta$: 53, $E_p$: 53     \\
021004  &        2.3351         &         2.55 $^{+ 0.69}_{- 0.50}$     &        2, 400         &         6.50 (0.20)   &        30.00 $^{+270.00}_{-27.00}$    &        -1.01 (0.19)   &        -2.30 (0.46)$*$        &        80 (35)        &        266 (117)      &     $z$: 68, $S$: 53, $t$: 69, $n$: 70, $\alpha$: 53, $E_p$: 53     \\
021211  &        1.0060         &         3.53 $^{+ 0.21}_{- 0.21}$     &        2, 400         &         1.40 (0.50)   &        10.00  (5.00) $*$      &        -0.86 (0.10)   &        -2.18 (0.25)   &        46 (7)         &        91 (14)        &        $z$: 71, $S$: 53, $t$: 72, $\alpha$: 53, $\beta$: 53, $E_p$: 53      \\
030115  &        [2.20]     &         2.31 $^{+ 0.40}_{- 0.32}$     &        2, 400         &        \nodata        &        10.00  (5.00) $*$      &        -1.28 (0.14)   &        -2.30 (0.46)$*$        &        83 (34)        &        [265] (110)    &        $z$: 73, $S$: 53, $\alpha$: 53, $E_p$: 53    \\
030226  &        1.9860         &         5.61 $^{+ 0.69}_{- 0.61}$     &        2, 400         &         0.83 (0.10)   &        10.00  (5.00) $*$      &        -0.89 (0.17)   &        -2.30 (0.46)   &        97 (21)        &        290 (64)       &        $z$: 74, $S$: 53, $t$: 75, $\alpha$: 53, $\beta$: 55, $E_p$: 53      \\
030323  &        3.3718         &         1.23 $^{+ 0.37}_{- 0.34}$     &        2, 400         &        $>$  1.40      &        10.00  (5.00) $*$      &        -1.62 (0.25)   &        -2.30 (0.46)$*$        &        \nodata        &        \nodata        &        $z$: 76, $S$: 53, $t$: 76, $\alpha$: 53      \\
030324  &         $<$ 2.70      &         1.82 $^{+ 0.33}_{- 0.30}$     &        2, 400         &        \nodata        &        10.00  (5.00) $*$      &        -1.45 (0.14)   &        -2.30 (0.46)$*$        &        147 (203)      &        $<$ 543        &        $z$: 73, $S$: 53, $\alpha$: 53, $E_p$: 53    \\
030328  &        1.5200         &        36.95 $^{+ 1.40}_{- 1.42}$     &        2, 400         &         0.80 (0.10)   &        10.00  (5.00) $*$      &        -1.14 (0.03)   &        -2.09 (0.40)   &        126 (13)       &        318 (34)       &        $z$: 77, $S$: 53, $t$: 78, $\alpha$: 53, $\beta$: 53, $E_p$: 53      \\
030329  &        0.1685         &        163.00 $^{+ 1.40}_{- 1.30}$    &        2, 400         &         0.48 (0.03)   &         5.50  (2.75) $**$     &        -1.26 (0.02)   &        -2.28 (0.06)   &        68 (2)         &        79 (3)         &        $z$: 79, $S$: 53, $t$: 63, $n$: 63, $\alpha$: 53, $\beta$: 53, $E_p$: 53     \\
030429  &        2.6564         &         0.85 $^{+ 0.15}_{- 0.13}$     &        2, 400         &         1.77 (1.00)   &        10.00  (5.00) $*$      &        -1.12 (0.25)   &        -2.30 (0.46)$*$        &        35 (10)        &        128 (35)       &        $z$: 80, $S$: 53, $t$: 81, $\alpha$: 53, $E_p$: 53   \\
030528  &         $<$ 1.00      &        11.90 $^{+ 0.76}_{- 0.78}$     &        2, 400         &         2.20 (1.80)   &        10.00  (5.00) $*$      &        -1.33 (0.15)   &        -2.65 (0.98)   &        32 (5)         &        $<$ 64         &        $z$: 82, $S$: 53, $t$: 82, $\alpha$: 53, $\beta$: 53, $E_p$: 53      \\
030723  &         $<$ 2.10      &         0.03 $^{+ 0.06}_{- 0.01}$     &        2, 400         &         1.67 (0.30)   &        10.00  (5.00) $*$      &        -1.00 (0.20)$*$        &        -1.90 (0.20)   &        $<$ 9  &        $<$ 28         &        $z$: 83, $S$: 53, $t$: 84, $\beta$: 53, $E_p$: 53    \\
031203  &        0.1055         &         1.20 (0.12)$*$        &        20, 2000       &        \nodata        &        10.00  (5.00) $*$      &        -1.00 (0.20)$*$        &        -2.30 (0.46)$*$        &        $>$ 190        &        $>$ 210        &        $z$: 85, $S$: 86, $E_p$: 87  \\
040511  &        [1.50]         &        10.00 (1.00)$*$        &        30, 400        &         1.20 (0.40)   &        10.00  (5.00) $*$      &        -0.67 (0.07)   &        -2.30 (0.46)$*$        &        131 (26)       &        [328] (65)     &        $z$: 88, $S$: 89, $t$: 90, $\alpha$: 91, $E_p$: 73   \\
040701  &        0.2146         &         0.45 (0.08)   &        2, 25  &        \nodata        &        10.00  (5.00) $*$      &        \nodata        &        \nodata        &        \nodata        &        \nodata        &        $z$: 92, $S$: 93       \\
040924  &        0.8590         &         2.73 (0.12)   &        20, 500        &        $<$  1.00      &        10.00  (5.00) $*$      &        -1.17 (0.23)   &        -2.30 (0.46)$*$        &        67 (6)         &        125 (11)       &        $z$: 94, $S$: 95, $t$: 94, $\alpha$: 73, $E_p$: 95   \\
041006  &        0.7160         &         7.00 (0.70)$*$        &        30, 400        &         1.10 (0.60)   &        10.00  (5.00) $*$      &        -1.37 (0.27)   &        -2.30 (0.46)$*$        &        63 (13)        &        109 (22)       &        $z$: 96, $S$: 97, $t$: 98, $\alpha$: 73, $E_p$: 73   \\
\enddata

\tablenotetext{a} {Upper/lower limit data are indicated with $<$ and
$>$ respectively. $1$-$\sigma$ errors are indicated to the right of
data values in parentheses ( ). References are given in order for
redshift (``$z$''), fluence (``$S$''), jet-break time (``$t$''),
density (``n''), low energy band spectral slope (``$\alpha$''), high
energy band spectral slope (``$\beta$''), and spectral peak energy
(``$E_p$'') (See \S~\ref{sec:data_selection2},
Appendix~\ref{sec:data_selection}).}

\tablenotetext{b} {Spectroscopic redshift $z$. GRBs marked with $*$
have upper {\it and} lower limits, where the $z$ indicated is the
mean.  For GRBs 980519, 000630, 020331, 030115, and 040511, we assume
the redshift in square brackets to calculate the data in the table.}

\tablenotetext{c} {GRB fluence $S_{\gamma}$ calculated in the observed
bandpass [$e_1$,$e_2$] keV.  Fluence errors are assumed to be 10\%
when not reported in the literature (marked with $*$).  When multiple
fluence references are available, we choose measurements prioritized
according to those with reported fluence errors, then those with the
widest observed bandpass, preferring published papers over GCNs.  When
asymmetric fluence errors are reported in the literature
(e.g., \citet{saka04b}), we use the geometric mean to construct
approximate symmetric errors, i.e., $\sigma_{S_{\gamma}} \approx
\sqrt{\sigma_{S_{\gamma}}^{+}\sigma_{S_{\gamma}}^{-}}$.}

\tablenotetext{d} { Afterglow jet-break time $t_{\rm jet}$.  When
multiple references are available, we choose those reporting early
time optical data.}
  
\tablenotetext{e} { Ambient density $n$ inferred from broadband
afterglow modeling assuming a constant density ISM.  We assume ${\rm
n} = 10 \pm 5$ cm$^{-3}$ (50\% error) in the absence of constraints
from broadband afterglow modeling (marked with $*$), and also 50\%
error when the error is not reported (marked with $**$).}

\tablenotetext{f} {\ Low energy ``Band'' spectral slope $\alpha$.
When $\beta$ is reported in the literature but $\alpha$ is not, we set
$\alpha=-1$ (marked with $*$; See Appendix~\ref{sec:data_selection}).
When multiple references are available
(i.e., \citealt{jime01,amat02,amat04}), we list the values with
reported errors and assume an error of 20\% otherwise.}

\tablenotetext{g} { High energy ``Band'' spectral slope $\beta$.
Following \citet{atte03}, when $\alpha$ is reported in the literature
and $\beta$ is not, we fix $\beta$=-2.3 (marked with $*$; Also see
Appendix~\ref{sec:data_selection}).  We assume an error of 20\% when
not reported.}

\tablenotetext{h} { Observed spectral peak energy $E^{\rm obs}_p =
E^{\rm obs}_o(2+\alpha)$.  When asymmetric peak energy errors for
$E^{\rm obs}_p$ are reported in the literature [e.g., \citet{saka04b}],
we assume $\sigma_{E^{\rm obs}_p} \approx \sqrt{\sigma_{E^{\rm
obs}_p}^{+}\sigma_{E^{\rm obs}_p}^{-}}$ (i.e the geometric mean), to
calculate the approximate symmetric errors reported in this table.}
  
\tablenotetext{i} { Rest frame spectral peak energy $E_p = E^{\rm
obs}_p(1+z) = E^{\rm obs}_o(2+\alpha)(1+z) = E_o(2+\alpha)$.  $E_p$
values calculated for uncertain redshifts are marked with square
brackets [ ].}

\tablerefs{ \tiny
1. \citet{bloo01a}; 
2. \citet{amat02}; 
3. \citet{bloo98}; 
4. \citet{frai00}; 
5. \citet{djor01a}; 
6. \citet{jime01}; 
7. \citet{amat04}; 
8. \citet{kulk98}; 
9. \citet{bloo99}; 
10. \citet{groo98}; 
11. \citet{lamb99}; 
12. \citet{yost02}; 
13. \citet{tin_iauc_6896}; 
14. \citet{jaun01}; 
15. \citet{djor03c}; 
16. \citet{halp_gcn_134}; 
17. \citet{djor98}; 
18. \citet{frai03}; 
19. \citet{fron00b}; 
20. \citet{frai99}; 
21. \citet{kulk99}; 
22. \citet{bloo03b}; 
23. \citet{vree01}; 
24. \citet{stan99}; 
25. \citet{pana02}; 
26. \citet{lefl02}; 
27. \citet{mase00}; 
28. \citet{fron01}; 
29. \citet{bjor01}; 
30. \citet{djor_gcn_481}; 
31. \citet{hurl00}; 
32. \citet{saga00}; 
33. \citet{djor_gcn_510}; 
34. \citet{halp00}; 
35. \citet{ande00}; 
36. \citet{piro02}; 
37. \citet{anto00}; 
38. \citet{cast_gcn_605}; 
39. \citet{jens01}; 
40. \citet{berg00}; 
41. \citet{pana01d}; 
42. \citet{berg01}; 
43. \citet{hurl_gcn_736}; 
44. \citet{fynb01}; 
45. \citet{pric02a}; 
46. \citet{cast_gcn_851}; 
47. \citet{pric01}; 
48. \citet{harr01}; 
49. \citet{mira02}; 
50. \citet{intz01}; 
51. \citet{gala03b}; 
52. \citet{pric02c}; 
53. \citet{saka04b}; 
54. \citet{pric03a}; 
55. \citet{atte03}; 
56. \citet{garn03}; 
57. \citet{pric02b}; 
58. \citet{holl02}; 
59. \citet{jako03}; 
60. \citet{hjor03b}; 
61. \citet{berg02b}; 
62. \citet{pric03c}; 
63. \citet{pric03b}; 
64. \citet{vand_gcn_2380}; 
65. \citet{amat04b}; 
66. \citet{bart03}; 
67. \citet{sode04}; 
68. \citet{moll02}; 
69. \citet{pand03}; 
70. \citet{scha03c}; 
71. \citet{vree_gcn_1785}; 
72. \citet{holl04}; 
73. \citet{hete2}; 
74. \citet{grei_gcn_1886}; 
75. \citet{klos04}; 
76. \citet{vree04}; 
77. \citet{mart_gcn_1980}; 
78. \citet{ande_gcn_1993}; 
79. \citet{bloo_gcn_2212}; 
80. \citet{weid_gcn_2215}; 
81. \citet{jako04}; 
82. \citet{rau04}; 
83. \citet{fynb_gcn_2327}; 
84. \citet{dull_gcn_2336}; 
85. \citet{proc_gcn_2482}; 
86. \citet{wats04}; 
87. \citet{sazo04}; 
88. \citet{berg_gcn_2610}; 
89. \citet{dull_gcn_2588}; 
90. \citet{bers_gcn_2602}; 
91. \citet{ghir04}; 
92. \citet{kels_gcn_2627}; 
93. \citet{barr_gcn_2620}; 
94. \citet{wier_gcn_2800}; 
95. \citet{gole_gcn_2754}; 
96. \citet{pric_gcn_2791}; 
97. \citet{gala_gcn_2770}; 
98. \citet{dava_gcn_2788} 
}

\end{deluxetable}

\newpage

\textwidth=6.6in 
\textheight=9.2in
\columnsep=5.0mm 
\parindent=6.0mm
\hoffset=00mm

\begin{deluxetable}{rlcccccccccl}
\rotate
\tabletypesize{\scriptsize}
\tablewidth{9.6in}
\tablecaption{Derived Energetics Parameters \label{table2}}
\tablecolumns{12}
\tablehead{
\colhead{Data} 			& \colhead{GRB} 		& \colhead{z} 			& \colhead{$k$}			&\colhead{$\log(E_{iso})$} 	&\colhead{$\theta_{\rm jet}$} 	& \colhead{$\log(E_{\gamma})$} 	& \colhead{$\epsilon_{\gamma}$} 	& \colhead{$A_{\gamma}$} 	 & \colhead{$C_{\gamma}$} 	& \colhead{$DM_{\gamma},{\rm unc}$} 	&\colhead{$DM_{\gamma}$} \\
\colhead{Set(s)} 		& \colhead{/XRF} 		& \colhead{}  			& \colhead{}			& \colhead{[erg]} 		& \colhead{[deg]} 	      	& \colhead{[erg]}        	& \colhead{}          		& \colhead{} 			 & \colhead{[mag]}  		& \colhead{[mag]} 			& \colhead{[mag]}\\ 
\colhead{\tablenotemark{a}} 	& \colhead{\tablenotemark{b}}  	& \colhead{\tablenotemark{c}} 	& \colhead{\tablenotemark{d}} 	& \colhead{\tablenotemark{e}} 	& \colhead{\tablenotemark{f}}  	& \colhead{\tablenotemark{g}}   & \colhead{h} 			& \colhead{\tablenotemark{i}} 	 & \colhead{\tablenotemark{j}}  & \colhead{\tablenotemark{k}} 		& \colhead{\tablenotemark{l}} 	 }
\startdata
\nodata         &        970228         &        0.6950         &         1.44 (0.07)   &        52.30 (0.05)   &        \nodata        &        $<$ 52.30 $\ddagger$   &        $<$ 25.05 $\ddagger$   &        $<$ 17.70 $\ddagger$   &        \nodata        &        \nodata        &        \nodata        \\
A       &        970508$\dagger$        &        0.8349         &         1.55 (0.08)   &        51.71 (0.08)   &        21.83 (2.18)   &        50.56 (0.10)   &         0.46 (0.11)   &         2.00 (1.00)   &        -1.19 (0.64)   &        44.74 (0.48)   &        42.59 (0.72)   \\
A, G, D         &        970828         &        0.9578         &         0.82 (0.08)   &        53.28 (0.06)   &         7.26 (0.68) $*$       &        51.18 (0.09)   &         1.91 (0.40)   &         1.04 (0.39)   &         1.82 (0.45)   &        43.05 (0.43)   &        43.92 (0.55)   \\
\nodata         &        971214$^{*}$   &        3.4180         &         1.09 (0.14)   &        53.36 (0.07)   &        $>$  5.48      &        $>$ 51.02      &        $>$  1.32 (0.25)       &        $>$  0.57 (0.20)       &        $<$  2.17 (0.44)       &        $<$ 46.97 (0.39)       &        $<$ 48.18 (0.52)   \\
\nodata         &        980326$^{*}$$\dagger$$\dagger$$\dagger$        &        [1.00]$*$      &        [1.65] (0.21)  &        51.51 (0.10)   &        $<$  6.33      &        $<$ 49.29      &        $<$ [0.02] (0.01)      &        $<$ [0.21] (0.06)      &        $>$ [-2.13] (0.27)     &        $>$ [49.46] (0.47) & $>$ [46.38] (0.43)     \\
\nodata         &        980329         &        [2.95]$*$      &        [0.97] (0.09)  &        54.07 (0.05)   &        \nodata        &        \nodata        &        \nodata        &        \nodata        &        \nodata        &        \nodata        &        \nodata        \\
\nodata         &        980425         &        0.0085         &         1.00 (0.00)   &        47.79 (0.04)   &        \nodata        &        $<$ 47.79 $\ddagger$   &        $<$ 0.00077 $\ddagger$         &        $<$   0.028 $\ddagger$         &        \nodata        &        \nodata        &        \nodata    \\
\nodata         &        980519$\dagger$$\dagger$$\dagger$      &        [2.50]         &        [0.86] (0.09) $*$      &        53.10 (0.06)   &        [3.65] (0.43)  &        [50.41] (0.11)         &        [0.32] (0.08)  &        [0.13] (0.05)  &        [2.27] (0.46)  &        [48.19] (0.52)         &        [49.50] (0.59)      \\
\nodata         &        980613$^{*}$   &        1.0969         &         1.47 (0.24)   &        51.66 (0.11)   &        $>$ 12.82      &        $>$ 50.06      &        $>$  0.14 (0.03)       &        $>$  0.40 (0.30)       &        $<$ -0.55 (1.01)       &        $<$ 47.16 (0.49)       &        $<$ 45.65 (1.06)   \\
A, G, D         &        980703         &        0.9662         &         0.94 (0.08)   &        52.71 (0.06)   &        11.42 (0.83)   &        51.01 (0.07)   &         1.29 (0.22)   &         0.89 (0.31)   &         1.48 (0.45)   &        43.64 (0.35)   &        44.17 (0.51)   \\
\nodata         &        981226$^{*}$   &        [1.50]         &         1.58 (0.18)   &        51.56 (0.12)   &        $>$ 14.80      &        $>$ 50.08      &        $>$ [0.15] (0.04)      &        $>$ [0.61] (0.28)      &        $<$ [-1.08] (0.54)     &        $<$ [47.93] (0.52)     &        $<$ [45.90] (0.65) \\
A, G, D         &        990123         &        1.6004         &         1.13 (0.01)   &        54.34 (0.06)   &         4.68 (0.50)   &        51.86 (0.10)   &         9.09 (2.11)   &         0.77 (0.24)   &         4.52 (0.30)   &        42.17 (0.48)   &        45.74 (0.45)   \\
\nodata         &        990506         &        1.3066         &         0.87 (0.10)   &        53.87 (0.07)   &        \nodata        &        $<$ 53.87 $\ddagger$   &        $<$  933.40 $\ddagger$         &        $<$   59.46 $\ddagger$         &        \nodata        &        \nodata        &        \nodata    \\
A, G    &        990510$\dagger$$\dagger$$\dagger$      &        1.6187         &         1.29 (0.03)   &        53.20 (0.05)   &         3.77 (0.22)   &        50.54 (0.06)   &         0.43 (0.06)   &         0.38 (0.08)   &         1.14 (0.24)   &        46.61 (0.28)   &        46.80 (0.31)   \\
A, G, D         &        990705         &        0.8424         &         1.30 (0.05)   &        53.26 (0.05)   &         5.56 (0.55) $*$       &        50.93 (0.09)   &         1.07 (0.23)   &         1.27 (0.32)   &         0.70 (0.20)   &        43.55 (0.44)   &        43.30 (0.37)   \\
A, G, D         &        990712         &        0.4331         &         0.74 (0.08)   &        51.59 (0.05)   &        11.78 (0.93) $*$       &        49.91 (0.08)   &         0.10 (0.02)   &         0.87 (0.28)   &        -2.15 (0.39)   &        45.18 (0.36)   &        42.08 (0.47)   \\
\nodata         &        991208$^{*}$   &        0.7055         &         1.09 (0.03) $*$       &        53.15 (0.05)   &        $<$  8.39      &        $<$ 51.18      &        $<$  1.90 (0.37)       &        \nodata        &        \nodata        &        $>$ 42.24 (0.39)       &        \nodata        \\
A, G, D         &        991216         &        1.0200         &         0.88 (0.09)   &        53.66 (0.06)   &         4.66 (0.73)   &        51.18 (0.14)   &         1.91 (0.63)   &         0.91 (0.41)   &         2.03 (0.45)   &        43.22 (0.67)   &        44.30 (0.66)   \\
\nodata         &        000131$^{*}$   &        4.5000         &         0.85 (0.07)   &        54.04 (0.11)   &        $<$  4.71      &        $<$ 51.57      &        $<$  4.67 (1.09)       &        $<$  1.34 (0.39)       &        $>$  2.75 (0.24)       &        $>$ 45.85 (0.48)       &        $>$ 47.65 (0.42)   \\
\nodata         &        000210$^{*}$   &        0.8463         &         1.28 (0.10) $*$       &        53.16 (0.04)   &        $>$  5.43      &        $>$ 50.82      &        $>$  0.82 (0.13)       &        \nodata        &        \nodata        &        $<$ 43.94 (0.32)       &        \nodata        \\
\nodata         &        000214$^{**}$  &        [0.42]$*$      &        [1.39] (0.13)  &        50.94 (0.13)   &        \nodata        &        \nodata        &        $<$  1.11 $\ddagger$   &        $<$  3.67 $\ddagger$   &        \nodata        &        \nodata        &        \nodata        \\
\nodata         &        000301C        &        2.0335         &         1.37 (0.36) $*$       &        52.44 (0.17)   &        13.88 (1.12)   &        50.90 (0.14)   &         1.00 (0.32)   &        \nodata        &        \nodata        &        46.00 (0.66)   &        \nodata        \\
\nodata         &        000418         &        1.1182         &         1.00 (0.02) $*$       &        52.81 (0.04)   &        22.95 (6.52)   &        51.71 (0.25)   &         6.45 (3.72)   &        \nodata        &        \nodata        &        41.69 (1.17)   &        \nodata        \\
\nodata         &        000630$^{*}$   &        [1.50]         &        [4.21] (1.56) $*$      &        52.68 (0.17)   &        $>$  9.85      &        $>$ 50.85      &        $>$ [0.89] (0.29)      &        \nodata        &        \nodata        &        $<$ [45.37] (0.66)     &        \nodata        \\
\nodata         &        000911$^{*}$$\dagger$$\dagger$         &        1.0585         &         0.63 (0.12)   &        53.63 (0.10)   &        $<$  5.58      &        $<$ 51.30      &        $<$  2.53 (0.55)       &        $<$  0.47 (0.19)       &        $>$  3.37 (0.48)       &        $>$ 42.92 (0.45)       &    $>$ 45.33 (0.57)        \\
\nodata         &        000926         &        2.0369         &         3.91 (1.33) $*$       &        53.38 (0.15)   &         6.28 (0.32)   &        51.16 (0.12)   &         1.82 (0.49)   &        \nodata        &        \nodata        &        45.15 (0.55)   &        \nodata        \\
\nodata         &        010222$^{**}$$\dagger$$\dagger$$\dagger$       &        1.4769         &         1.03 (0.04) $*$       &        53.84 (0.02)   &         3.29 (0.24)   &        51.05 (0.07)   &         1.42 (0.21)   &        $<$  0.41 (0.15)       &        $>$  2.73 (0.46)       &        44.65 (0.31)   &    $>$ 46.42 (0.51)        \\
\nodata         &        010921$^{*}$   &        0.4509         &         0.97 (0.10)   &        51.96 (0.05)   &        $<$ 32.76      &        $<$ 51.16      &        $<$  1.84 (0.40)       &        $<$  9.61 (3.53)       &        $>$ -1.45 (0.44)       &        $>$ 41.08 (0.43)       &        $>$ 38.68 (0.54)   \\
\nodata         &        011121$^{*}$$\dagger$  &        0.3620         &         3.70 (0.63) $*$       &        52.46 (0.09)   &        $>$ 16.24      &        $>$ 51.06      &        $>$  1.46 (0.30)       &        $>$  2.20 (0.62)       &        $<$  0.35 (0.28)       &        $<$ 40.88 (0.43)       &        $<$ 40.28 (0.41)    \\
A, G, D         &        011211$\dagger$        &        2.1400         &         1.43 (0.11)   &        52.89 (0.06)   &         5.98 (0.39) $*$       &        50.63 (0.07)   &         0.53 (0.08)   &         1.61 (0.40)   &        -0.66 (0.28)   &        47.06 (0.32)   &        45.45 (0.36)   \\
A, G, D         &        020124$\dagger$$\dagger$       &        3.1980         &         1.02 (0.02)   &        53.25 (0.04)   &        11.30 (1.59)   &        51.54 (0.13)   &         4.33 (1.25)   &         4.77 (1.88)   &         0.81 (0.39)   &        45.07 (0.59)   &        44.93 (0.57)   \\
\nodata         &        020331         &        [1.50]         &        [1.11] (0.06) $*$      &        51.64 (0.50)   &        \nodata        &        $<$ [51.64] $\ddagger$         &        $<$  5.52 $\ddagger$   &        $<$  5.48 $\ddagger$   &        \nodata        &        \nodata        &        \nodata    \\
A, G, D         &        020405$\dagger$$\dagger$       &        0.6899         &         0.90 (0.07) $*$       &        52.92 (0.03)   &         7.68 (1.02)   &        50.87 (0.12)   &         0.93 (0.25)   &         0.47 (0.19)   &         1.93 (0.45)   &        43.22 (0.55)   &        44.20 (0.60)   \\
\nodata         &        020427$^{*}$   &         $<$ 2.30      &         1.43 (0.72)   &        $>$52.01       &        $>$ 18.52      &        $>$ 50.72      &        $>$ [0.67] (0.27)      &        \nodata  &        $<$ [-7.15] (2.16)         &        $<$ [46.91] (0.83)     &        $<$ [38.81] (2.24)      \\
A, G, D         &        020813$\dagger$        &        1.2540         &         1.50 (0.03)   &        53.77 (0.01)   &         3.24 (0.26) $*$       &        50.98 (0.07)   &         1.19 (0.19)   &         1.60 (0.36)   &         0.52 (0.23)   &        44.46 (0.34)   &        44.03 (0.33)   \\
\nodata         &        020903         &        0.2510         &         0.28 (0.28)   &        48.62 (0.48)   &        \nodata        &        $<$ 48.62 $\ddagger$   &        $<$  0.0052 $\ddagger$         &        $<$  3.64 $\ddagger$   &        \nodata        &        \nodata        &        \nodata        \\
A       &        021004         &        2.3351         &         1.04 (0.06) $*$       &        52.52 (0.10)   &        12.73 (4.55)   &        50.91 (0.32)   &         1.03 (0.76)   &         1.82 (1.79)   &         0.12 (0.95)   &        46.33 (1.50)   &        45.49 (1.43)   \\
A       &        021211         &        1.0060         &         1.07 (0.11)   &        52.00 (0.05)   &         8.78 (1.30)   &        50.07 (0.13)   &         0.15 (0.05)   &         1.28 (0.51)   &        -2.19 (0.36)   &        46.90 (0.63)   &        43.75 (0.57)   \\
\nodata         &        030115         &        [2.20]         &        [1.01] (0.06) $*$      &        52.42 (0.07)   &        \nodata        &        \nodata        &        $<$ 33.34 $\ddagger$   &        $<$ 15.77 $\ddagger$   &        \nodata        &        \nodata        &        \nodata        \\
A, G    &        030226$\dagger$$\dagger$$\dagger$      &        1.9860         &         1.08 (0.05)   &        52.76 (0.05)   &         4.99 (0.39)   &        50.34 (0.08)   &         0.27 (0.05)   &         0.43 (0.16)   &         0.31 (0.49)   &        47.81 (0.37)   &        47.17 (0.55)   \\
\nodata         &        030323$^{*}$   &        3.3718         &         1.05 (0.03) $*$       &        52.48 (0.13)   &        $>$  5.71      &        $>$ 50.18      &        $>$  0.19 (0.05)       &        \nodata        &        \nodata        &        $<$ 49.75 (0.54)       &        \nodata        \\
\nodata         &        030324         &         $<$ 2.70      &        [1.00] (0.09) $*$      &        $>$52.47       &        \nodata        &        \nodata        &        $<$ 37.29 $\ddagger$   &        $<$  8.29 $\ddagger$   &        \nodata        &        \nodata        &        \nodata        \\
A, G, D         &        030328         &        1.5200         &         1.15 (0.11)   &        53.39 (0.04)   &         4.37 (0.35) $*$       &        50.86 (0.08)   &         0.90 (0.16)   &         1.22 (0.30)   &         0.51 (0.25)   &        45.38 (0.35)   &        44.93 (0.35)   \\
A, G, D         &        030329         &        0.1685         &         1.01 (0.03)   &        52.04 (0.01)   &         6.60 (0.45)   &        49.86 (0.06)   &         0.09 (0.01)   &         0.99 (0.18)   &        -2.50 (0.17)   &        43.01 (0.28)   &        39.56 (0.26)   \\
A, G    &        030429         &        2.6564         &         0.96 (0.06) $*$       &        52.11 (0.08)   &         7.41 (1.64) $*$       &        50.03 (0.20)   &         0.14 (0.06)   &         0.71 (0.45)   &        -1.46 (0.61)   &        49.60 (0.94)   &        47.19 (0.90)   \\
\nodata         &        030528$\dagger$        &         $<$ 1.00      &        [0.82] (0.17) $*$      &        $>$52.40       &         $>$  9.27 $*$         &        $>$50.52       &        $>$ [0.41] (0.27)      &        $>$ [6.22] (4.32)      &        $<$ [-2.97] (0.37)     &        $<$ [45.38] (1.32)     &    $<$ [41.45] (1.01)      \\
\nodata         &        030723$^{**}$  &         $<$ 2.10      &        [1.19] (0.13) $*$      &        $>$50.61       &         $>$ 11.89     &        $>$48.94       &        $>$ [0.01] (0.01)      &        $>$ [0.57] (0.36)      &        $<$ [-4.78] (0.51)     &        $<$ [52.63] (1.08)     &        $<$ [46.89] (0.92)  \\
\nodata         &        031203$^{**}$  &        0.1055         &         0.99 (0.02) $*$       &        49.48 (0.04)   &        \nodata        &        $<$ 49.48 $\ddagger$   &        $<$  0.04 $\ddagger$   &        $<$  0.21 $\ddagger$   &        \nodata        &        \nodata        &        \nodata        \\
\nodata         &        040511$\dagger$        &        [1.50]         &        [1.36] (0.05) $*$      &        52.89 (0.05)   &        [5.91] (0.83) $*$      &        [50.61] (0.13)         &        [0.51] (0.15)  &        [0.67] (0.28)  &        [0.57] (0.44)  &        [46.16] (0.60)         &        [45.77] (0.61)      \\
\nodata         &        040701         &        0.2146         &        21.61 (15.12)  &        51.03 (0.31)   &        \nodata        &        $<$ 51.03 $\ddagger$   &        \nodata        &        \nodata        &        \nodata        &        \nodata        &        \nodata        \\
\nodata         &        040924$^{*}$$\dagger$$\dagger$$\dagger$        &        0.8590         &         1.28 (0.05)   &        51.83 (0.03)   &        $<$  8.36      &        $<$ 49.86      &        $<$  0.09 (0.01)       &        $<$  0.49 (0.11)       &        $>$ -1.52 (0.23)       &        $>$ 47.18 (0.31)   & $>$ 44.71 (0.32)       \\
A       &        041006         &        0.7160         &         1.56 (0.09)   &        52.16 (0.05)   &         8.11 (1.74) $*$       &        50.16 (0.19)   &         0.18 (0.08)   &         1.23 (0.66)   &        -1.81 (0.45)   &        45.67 (0.89)   &        42.90 (0.78)   \\
\enddata

\tablenotetext{a} {Data Sets A, G, and D are as described in
\S~\ref{sec:prev}. Data calculated for uncertain redshifts is marked
with square brackets [ ]. Values are calculated assuming a cosmology
of ($\Omega_M$,$\Omega_\Lambda$,$h$)=(0.3,0.7,0.7).  Set A is used
to fit the Ghirlanda relation.  Set E (not marked) consists of the 23
bursts with $z$ and $t_{\rm jet}$ with no upper/lower limits, and is
used to calculate the median energy $\bar{E_{\gamma}}$.  Standard
candle variables computed for bursts not in Sets A or E assume the
parameters calculated from the fits to the standard sets,
e.g., ($\eta$, $\kappa$, $\bar{E_{\gamma}}$).}

\tablenotetext{b} {GRB names with $^*$ have upper/lower limits on
$t_{\rm jet}$, and $^{**}$ have upper/lower limits on $E^{\rm obs}_p$.
See Table~\ref{table1} for input data, references.  Bursts that have
$A_{\gamma} \sim 1$ have the least scatter about the Hubble
diagram. Bursts marked with $\dagger$, $\dagger\dagger$, and
$\dagger\dagger\dagger$ are 1-$\sigma$, 2-$\sigma$, and 3-$\sigma$
outliers in $A_{\gamma}$, respectively.}

\tablenotetext{c} { See Table~\ref{table1} for redshift references.}

\tablenotetext{d} {Cosmological $k$-correction calculated for a
rest frame ``bolometric'' gamma-ray bandpass of [20,2000] keV.  Values
of $k$ marked with $*$ are calculated for bursts with no
spectral information via the template spectra method of
\citet{bloo01b}, which has been adapted to incorporate upper/lower
limit information on $E^{\rm obs}_p/E^{\rm obs}_o$. \citet{bloo01b}
also describes the formalism for calculating $k$ and its error,
given known ``Band'' spectral parameters $\alpha$, $\beta$, $E_p$.
See Table~\ref{table1} for ``Band'' parameters, references, which are
inputs to the $k$-correction.}

\tablenotetext{e} { Isotropic-equivalent gamma-ray energy $E_{\rm
iso}$ is calculated via eq.~\ref{eq:egamma} for a rest frame
``bolometric'' bandpass of [20,2000] keV.}

\tablenotetext{f} { Top hat jet half-opening angle $\theta_{\rm jet}$
is calculated via eq.~\ref{eq:theta_jet}. Values marked with $*$
assume a constant ISM density of $n = 10 \pm 5$ in the absence of
constraints from broadband afterglow modeling.  Upper limits on
$\theta_{\rm jet}$ come from upper limits on the jet break time,
$t_{\rm jet}$.  Lower limits on $\theta_{\rm jet}$ come from either
lower limits on $t_{\rm jet}$ or upper limits on $z$ and a measured
$t_{jet}$ (i.e., GRB 030528, XRF 030723 for the latter).  See
Table~\ref{table1} for input densities, jet break times, references.}

\tablenotetext{g} { Beaming corrected top hat gamma-ray energy
$E_{\gamma} = E_{\rm iso}[1-cos(\theta_{\rm jet})]$
(eq.~\ref{eq:egamma}), for a rest frame ``bolometric'' bandpass of
[20,2000] keV. When jet break times (and hence opening angles) are not
available in the literature, we indicate the upper limit $E_{\gamma}$
with $\ddagger$.  In other words, $E_{\gamma}=E_{\rm iso} f_b \leq
E_{\rm iso}$ is always true since $f_b \leq 1$, and $E_{\gamma} =
E_{\rm iso}$ only in the limit of isotropy ($f_b = 1$).}

\tablenotetext{h} {$\epsilon_{\gamma}=E_{\gamma}/\bar{E_{\gamma}}$,
the ``uncorrected'' standard candle. The error in
$\epsilon_{\gamma}=E_{\gamma}/\bar{E_{\gamma}}$ is given by
$\sigma_{\epsilon_{\gamma}} = \sigma_{E_{\gamma}}/\bar{E_{\gamma}}$.}
  
\tablenotetext{i} {$A_{\gamma}$ the dimensionless ``corrected'' GRB
standard candle (defined in eq.~\ref{eq:agamma}) has a spread of no
more than a factor of $\sim$2-3, as compared to the distribution of
$\epsilon_{\gamma}$, which spans several orders of magnitude.}

\tablenotetext{j} {$C_{\gamma}$ is the ``GRB energy correction'' term
that helps standardize the energetics. Due to the Ghirlanda relation,
large, positive $C_\gamma$ values correspond to bursts with $E_{\gamma}
> \bar{E}_{\gamma}$ and vice versa for bursts with $C_\gamma < 0$.  As
noted, the maximal spread in $C_{\gamma}$$\sim8$ magnitudes, reflects
the large underlying spread in $\epsilon_{\gamma}$. }

\tablenotetext{k} {$DM_{\gamma,{\rm unc}}$, the apparent GRB distance
modulus [mag] derived assuming $\epsilon_{\gamma} \equiv 1$. }

\tablenotetext{l} {$DM_{\gamma}$ the apparent GRB distance modulus [mag]
derived assuming $A_{\gamma} \equiv 1$. Note that $C_{\gamma}\approx
DM_{\gamma} - DM_{\gamma,{\rm unc}} + (10/3){\rm
log}(\bar{E_{\gamma}}/E^{*})$. For reference, $(10/3){\rm
log}(\bar{E_{\gamma}}/E^{*}) =0.95$ mag in the standard cosmology. }

\end{deluxetable}


\newpage 

\begin{thebibliography}{196}
\expandafter\ifx\csname natexlab\endcsname\relax\def\natexlab#1{#1}\fi

\bibitem[{{Amati}(2004)}]{amat04}
{Amati}, L. 2004, {astro-ph/0405318}

\bibitem[{{Amati} {et~al.}(2004){Amati}, {Frontera}, {in't Zand}, {Capalbi},
  {Landi}, {Soffitta}, {Vetere}, {Antonelli}, {Costa}, {Del Sordo}, {Feroci},
  {Guidorzi}, {Heise}, {Masetti}, {Montanari}, {Nicastro}, {Palazzi}, \&
  {Piro}}]{amat04b}
{Amati}, L. {et~al.} 2004, \aap, 426, 415

\bibitem[{{Amati} {et~al.}(2002){Amati}, {Frontera}, {Tavani}, {in't Zand},
  {Antonelli}, {Costa}, {Feroci}, {Guidorzi}, {Heise}, {Masetti}, {Montanari},
  {Nicastro}, {Palazzi}, {Pian}, {Piro}, \& {Soffitta}}]{amat02}
---. 2002, \aap, 390, 81

\bibitem[{{Andersen} {et~al.}(2000){Andersen}, {Hjorth}, {Pedersen}, {Jensen},
  {Hunt}, {Gorosabel}, {M{\o}ller}, {Fynbo}, {Kippen}, {Thomsen}, {Olsen},
  {Christensen}, {Vestergaard}, {Masetti}, {Palazzi}, {Hurley}, {Cline},
  {Kaper}, \& {Jaunsen}}]{ande00}
{Andersen}, M.~I. {et~al.} 2000, \aap, 364, L54

\bibitem[{{Andersen} {et~al.}(2003){Andersen}, {Masi}, {Chile}, {Jensen}, \&
  {Hjorth}}]{ande_gcn_1993}
{Andersen}, M.~I., {Masi}, G., {Chile}, E., {Jensen}, B.~L., \& {Hjorth}, J.
  2003, {GCN Report} 1993

\bibitem[{{Antonelli} {et~al.}(2000){Antonelli}, {Piro}, {Vietri}, {Costa},
  {Soffitta}, {Feroci}, {Amati}, {Frontera}, {Pian}, {Zand}, {Heise},
  {Kuulkers}, {Nicastro}, {Butler}, {Stella}, \& {Perola}}]{anto00}
{Antonelli}, L.~A. {et~al.} 2000, \apjl, 545, L39

\bibitem[{{Atteia}(2003)}]{atte03}
{Atteia}, J.-L. 2003, \aap, 407, L1

\bibitem[{{Baltz} \& {Hui}(2005)}]{balt05}
{Baltz}, E.~A., \& {Hui}, L. 2005, \apj, 618, 403

\bibitem[{{Band} {et~al.}(1993){Band}, {Matteson}, {Ford}, {Schaefer},
  {Palmer}, {Teegarden}, {Cline}, {Briggs}, {Paciesas}, {Pendleton}, {Fishman},
  {Kouveliotou}, {Meegan}, {Wilson}, \& {Lestrade}}]{band93}
{Band}, D. {et~al.} 1993, \apj, 413, 281

\bibitem[{{Band} \& {Preece}(2005)}]{band05}
{Band}, D., \& {Preece}, R. 2005, {astro-ph/0501559}

\bibitem[{{Barkana} \& {Loeb}(2001)}]{bark01}
{Barkana}, R., \& {Loeb}, A. 2001, \physrep, 349, 125

\bibitem[{{Barraud} {et~al.}(2003){Barraud}, {Olive}, {Lestrade}, {Atteia},
  {Hurley}, {Ricker}, {Lamb}, {Kawai}, {Boer}, {Dezalay}, {Pizzichini},
  {Vanderspek}, {Crew}, {Doty}, {Monnelly}, {Villasenor}, {Butler}, {Levine},
  {Yoshida}, {Shirasaki}, {Sakamoto}, {Tamagawa}, {Torii}, {Matsuoka},
  {Fenimore}, {Galassi}, {Tavenner}, {Donaghy}, {Graziani}, \&
  {Jernigan}}]{barr03}
{Barraud}, C. {et~al.} 2003, \aap, 400, 1021

\bibitem[{{Barraud} {et~al.}(2004){Barraud}, {Ricker}, {Atteia}, {Kawai},
  {Lamb}, {Woosley}, {Donaghy}, {Fenimore}, {Galassi}, {Graziani}, {Matsuoka},
  {Nakagawa}, {Sakamoto}, {Sato}, {Shirasaki}, {Suzuki}, {Tamagawa}, {Torii},
  {Urata}, {Yamazaki}, {Yamamoto}, {Yoshida}, {Butler}, {Crew}, {Doty},
  {Dullighan}, {Prigozhin}, {Vanderspek}, {Villasenor}, {Jernigan}, {Levine},
  {Azzibrouck}, {Braga}, {Manchanda}, {Pizzichini}, {Boer}, {Olive}, {Dezalay},
  \& {Hurley}}]{barr_gcn_2620}
---. 2004, {GCN Report} 2620

\bibitem[{{Barth} {et~al.}(2003){Barth}, {Sari}, {Cohen}, {Goodrich}, {Price},
  {Fox}, {Bloom}, {Soderberg}, \& {Kulkarni}}]{bart03}
{Barth}, A.~J. {et~al.} 2003, \apjl, 584, L47

\bibitem[{{Beloborodov}(2000)}]{belo00}
{Beloborodov}, A.~M. 2000, \apjl, 539, L25

\bibitem[{{Berger}(2004{\natexlab{a}})}]{berg_gcn_2610}
{Berger}, E. 2004{\natexlab{a}}, {GCN Report} 2610

\bibitem[{{Berger}(2004{\natexlab{b}})}]{berger04}
---. 2004{\natexlab{b}}, {Presentation, INT Workshop on the GRB--SN connection,
  Seattle, WA, July 12-24, 2004}

\bibitem[{{Berger} {et~al.}(2001){Berger}, {Diercks}, {Frail}, {Kulkarni},
  {Bloom}, {Sari}, {Halpern}, {Mirabal}, {Taylor}, {Hurley}, {Pooley},
  {Becker}, {Wagner}, {Terndrup}, {Statler}, {Wik}, {Mazets}, \&
  {Cline}}]{berg01}
{Berger}, E. {et~al.} 2001, \apj, 556, 556

\bibitem[{{Berger} {et~al.}(2002{\natexlab{a}}){Berger}, {Kulkarni}, {Bloom},
  {Price}, {Fox}, {Frail}, {Axelrod}, {Chevalier}, {Colbert}, {Costa},
  {Djorgovski}, {Frontera}, {Galama}, {Halpern}, {Harrison}, {Holtzman},
  {Hurley}, {Kimble}, {McCarthy}, {Piro}, {Reichart}, {Ricker}, {Sari},
  {Schmidt}, {Wheeler}, {Vanderppek}, \& {Yost}}]{berg02b}
---. 2002{\natexlab{a}}, \apj, 581, 981

\bibitem[{{Berger} {et~al.}(2002{\natexlab{b}}){Berger}, {Kulkarni}, \&
  {Frail}}]{berg02}
{Berger}, E., {Kulkarni}, S.~R., \& {Frail}, D.~A. 2002{\natexlab{b}}, American
  Astronomical Society Meeting, 201, 0

\bibitem[{{Berger} {et~al.}(2003{\natexlab{a}}){Berger}, {Kulkarni}, \&
  {Frail}}]{berg03}
---. 2003{\natexlab{a}}, \apj, 590, 379

\bibitem[{{Berger} {et~al.}(2004){Berger}, {Kulkarni}, \& {Frail}}]{berg04a}
---. 2004, \apj, 612, 966

\bibitem[{{Berger} {et~al.}(2003{\natexlab{b}}){Berger}, {Kulkarni}, {Pooley},
  {Frail}, {McIntyre}, {Wark}, {Sari}, {Soderberg}, {Fox}, {Yost}, \&
  {Price}}]{berg03c}
{Berger}, E. {et~al.} 2003{\natexlab{b}}, \nat, 426, 154

\bibitem[{{Berger} {et~al.}(2000){Berger}, {Sari}, {Frail}, {Kulkarni},
  {Bertoldi}, {Peck}, {Menten}, {Shepherd}, {Moriarty-Schieven}, {Pooley},
  {Bloom}, {Diercks}, {Galama}, \& {Hurley}}]{berg00}
---. 2000, \apj, 545, 56

\bibitem[{{Berger} {et~al.}(2003{\natexlab{c}}){Berger}, {Soderberg}, {Frail},
  \& {Kulkarni}}]{berg03d}
{Berger}, E., {Soderberg}, A.~M., {Frail}, D.~A., \& {Kulkarni}, S.~R.
  2003{\natexlab{c}}, \apjl, 587, L5

\bibitem[{{Bersier} {et~al.}(2004){Bersier}, {Castro}, {Rhoads}, {Fruchter},
  {Levan}, {Gorosabel}, {Kouveliotou}, {Patel}, {Merrill}, {Bally},
  {Walawender}, \& {Reipurth}}]{bers_gcn_2602}
{Bersier}, D. {et~al.} 2004, {GCN Report} 2602

\bibitem[{{Bj{\" o}rnsson} {et~al.}(2001){Bj{\" o}rnsson}, {Hjorth},
  {Jakobsson}, {Christensen}, \& {Holland}}]{bjor01}
{Bj{\" o}rnsson}, G., {Hjorth}, J., {Jakobsson}, P., {Christensen}, L., \&
  {Holland}, S. 2001, \apjl, 552, L121

\bibitem[{{Bloom}(2003)}]{bloo03c}
{Bloom}, J.~S. 2003, \aj, 125, 2865

\bibitem[{{Bloom} {et~al.}(2003{\natexlab{a}}){Bloom}, {Berger}, {Kulkarni},
  {Djorgovski}, \& {Frail}}]{bloo03b}
{Bloom}, J.~S., {Berger}, E., {Kulkarni}, S.~R., {Djorgovski}, S.~G., \&
  {Frail}, D.~A. 2003{\natexlab{a}}, \aj, 125, 999

\bibitem[{{Bloom} {et~al.}(2001{\natexlab{a}}){Bloom}, {Djorgovski}, \&
  {Kulkarni}}]{bloo01a}
{Bloom}, J.~S., {Djorgovski}, S.~G., \& {Kulkarni}, S.~R. 2001{\natexlab{a}},
  \apj, 554, 678

\bibitem[{{Bloom} {et~al.}(1998){Bloom}, {Djorgovski}, {Kulkarni}, \&
  {Frail}}]{bloo98}
{Bloom}, J.~S., {Djorgovski}, S.~G., {Kulkarni}, S.~R., \& {Frail}, D.~A. 1998,
  \apjl, 507, L25

\bibitem[{{Bloom} {et~al.}(2003{\natexlab{b}}){Bloom}, {Frail}, \&
  {Kulkarni}}]{bloo03}
{Bloom}, J.~S., {Frail}, D.~A., \& {Kulkarni}, S.~R. 2003{\natexlab{b}}, \apj,
  594, 674

\bibitem[{{Bloom} {et~al.}(2001{\natexlab{b}}){Bloom}, {Frail}, \&
  {Sari}}]{bloo01b}
{Bloom}, J.~S., {Frail}, D.~A., \& {Sari}, R. 2001{\natexlab{b}}, \aj, 121,
  2879

\bibitem[{{Bloom} {et~al.}(1999){Bloom}, {Kulkarni}, {Djorgovski},
  {Eichelberger}, {Cote}, {Blakeslee}, {Odewahn}, {Harrison}, {Frail},
  {Filippenko}, {Leonard}, {Riess}, {Spinrad}, {Stern}, {Bunker}, {Dey},
  {Grossan}, {Perlmutter}, {Knop}, {Hook}, \& {Feroci}}]{bloo99}
{Bloom}, J.~S. {et~al.} 1999, \nat, 401, 453

\bibitem[{{Bloom} {et~al.}(2003{\natexlab{c}}){Bloom}, {Morrell}, \&
  {Mohanty}}]{bloo_gcn_2212}
{Bloom}, J.~S., {Morrell}, N., \& {Mohanty}, S. 2003{\natexlab{c}}, {GCN
  Report} 2212

\bibitem[{{Bromm} \& {Loeb}(2002)}]{brom02}
{Bromm}, V., \& {Loeb}, A. 2002, \apj, 575, 111

\bibitem[{{Castro} {et~al.}(2000{\natexlab{a}}){Castro}, {Diercks},
  {Djorgovski}, {Kulkarni}, {Galama}, {Bloom}, {Harrison}, \&
  {Frail}}]{cast_gcn_605}
{Castro}, S.~M., {Diercks}, A., {Djorgovski}, S.~G., {Kulkarni}, S.~R.,
  {Galama}, T.~J., {Bloom}, J.~S., {Harrison}, F.~A., \& {Frail}, D.~A.
  2000{\natexlab{a}}, {GCN Report} 605

\bibitem[{{Castro} {et~al.}(2000{\natexlab{b}}){Castro}, {Djorgovski},
  {Kulkarni}, {Bloom}, {Galama}, {Harrison}, \& {Frail}}]{cast_gcn_851}
{Castro}, S.~M., {Djorgovski}, S.~G., {Kulkarni}, S.~R., {Bloom}, J.~S.,
  {Galama}, T.~J., {Harrison}, F.~A., \& {Frail}, D.~A. 2000{\natexlab{b}},
  {GCN Report} 851

\bibitem[{{Chevalier} \& {Li}(1999)}]{chev99}
{Chevalier}, R.~A., \& {Li}, Z. 1999, \apjl, 520, L29

\bibitem[{{Chevalier} \& {Li}(2000)}]{chev00}
---. 2000, \apj, 536, 195

\bibitem[{{Cohen} \& {Piran}(1997)}]{cohe97}
{Cohen}, E., \& {Piran}, T. 1997, \apjl, 488, L7

\bibitem[{{Dai} {et~al.}(2004){Dai}, {Liang}, \& {Xu}}]{dai04}
{Dai}, Z.~G., {Liang}, E.~W., \& {Xu}, D. 2004, \apjl, 612, L101

\bibitem[{{Dai} \& {Wu}(2003)}]{dai03}
{Dai}, Z.~G., \& {Wu}, X.~F. 2003, \apjl, 591, L21

\bibitem[{{D'Avanzo} {et~al.}(2004){D'Avanzo}, {Covino}, {Antonelli},
  {Fugazza}, {Malesani}, {Fiore}, {Cocchia}, {Masetti}, {Pian}, {Stella},
  {Lorenzi}, \& {Barrena}}]{dava_gcn_2788}
{D'Avanzo}, P. {et~al.} 2004, {GCN Report} 2788

\bibitem[{{Dermer}(1992)}]{derm92}
{Dermer}, C.~D. 1992, Physical Review Letters, 68, 1799

\bibitem[{{Djorgovski} {et~al.}(2003){Djorgovski}, {Bloom}, \&
  {Kulkarni}}]{djor03c}
{Djorgovski}, S.~G., {Bloom}, J.~S., \& {Kulkarni}, S.~R. 2003, \apjl, 591, L13

\bibitem[{{Djorgovski} {et~al.}(1999{\natexlab{a}}){Djorgovski}, {Dierks},
  {Bloom}, {Kulkarni}, {Filippenko}, {Hillenbrand}, \&
  {Carpenter}}]{djor_gcn_481}
{Djorgovski}, S.~G., {Dierks}, A., {Bloom}, J.~S., {Kulkarni}, S.~R.,
  {Filippenko}, A.~V., {Hillenbrand}, L.~A., \& {Carpenter}, J.
  1999{\natexlab{a}}, {GCN Report} 481

\bibitem[{{Djorgovski} {et~al.}(2001){Djorgovski}, {Frail}, {Kulkarni},
  {Bloom}, {Odewahn}, \& {Diercks}}]{djor01a}
{Djorgovski}, S.~G., {Frail}, D.~A., {Kulkarni}, S.~R., {Bloom}, J.~S.,
  {Odewahn}, S.~C., \& {Diercks}, A. 2001, \apj, 562, 654

\bibitem[{{Djorgovski} {et~al.}(1999{\natexlab{b}}){Djorgovski}, {Goodrich},
  {Kulkarni}, {Bloom}, {Dierks}, {Harrison}, \& {Frail}}]{djor_gcn_510}
{Djorgovski}, S.~G., {Goodrich}, R., {Kulkarni}, S.~R., {Bloom}, J.~S.,
  {Dierks}, A., {Harrison}, F., \& {Frail}, D.~A. 1999{\natexlab{b}}, {GCN
  Report} 510

\bibitem[{{Djorgovski} {et~al.}(1998){Djorgovski}, {Kulkarni}, {Bloom},
  {Goodrich}, {Frail}, {Piro}, \& {Palazzi}}]{djor98}
{Djorgovski}, S.~G., {Kulkarni}, S.~R., {Bloom}, J.~S., {Goodrich}, R.,
  {Frail}, D.~A., {Piro}, L., \& {Palazzi}, E. 1998, \apjl, 508, L17

\bibitem[{{Dullighan} {et~al.}(2003){Dullighan}, {Butler}, {Vanderspek},
  {Villasenor}, \& {Ricker}}]{dull_gcn_2336}
{Dullighan}, A., {Butler}, N., {Vanderspek}, R., {Villasenor}, J., \& {Ricker},
  G. 2003, {GCN Report} 2336

\bibitem[{{Dullighan} {et~al.}(2004){Dullighan}, {Ricker}, {Atteia}, {Kawai},
  {Lamb}, {Woosley}, {Donaghy}, {Fenimore}, {Galassi}, {Graziani}, {Matsuoka},
  {Nakagawa}, {Sakamoto}, {Sato}, {Shirasaki}, {Suzuki}, {Tamagawa}, {Torii},
  {Urata}, {Yamazaki}, {Yamamoto}, {Yoshida}, {Butler}, {Crew}, {Doty},
  {Prigozhin}, {Vanderspek}, {Villasenor}, {Jernigan}, {Levine}, {Azzibrouck},
  {Braga}, {Manchanda}, {Pizzichini}, {Boer}, {Olive}, {Dezalay}, {Barraud}, \&
  {Hurley}}]{dull_gcn_2588}
{Dullighan}, A. {et~al.} 2004, {GCN Report} 2588

\bibitem[{{Eichler} \& {Levinson}(2004)}]{eich04}
{Eichler}, D., \& {Levinson}, A. 2004, \apjl, 614, L13

\bibitem[{{Fan} {et~al.}(2004){Fan}, {Zhang}, {Kobayashi}, \&
  {Meszaros}}]{fan04}
{Fan}, Y.~Z., {Zhang}, B., {Kobayashi}, S., \& {Meszaros}, P. 2004,
  {astro-ph/0410060}

\bibitem[{{Fenimore} \& {Ramirez-Ruiz}(2000)}]{feni00a}
{Fenimore}, E., \& {Ramirez-Ruiz}, E. 2000, {astro-ph/0004176}

\bibitem[{{Firmani} {et~al.}(2005){Firmani}, {Ghisellini}, {Ghirlanda}, \&
  {Avila-Reese}}]{firm05}
{Firmani}, C., {Ghisellini}, G., {Ghirlanda}, G., \& {Avila-Reese}, V. 2005,
  {astro-ph/0501395}

\bibitem[{{Frail} {et~al.}(1999){Frail}, {Kulkarni}, {Bloom}, {Djorgovski},
  {Gorjian}, {Gal}, {Meltzer}, {Sari}, {Chaffee}, {Goodrich}, {Frontera}, \&
  {Costa}}]{frai99}
{Frail}, D.~A. {et~al.} 1999, \apjl, 525, L81

\bibitem[{{Frail} {et~al.}(2001){Frail}, {Kulkarni}, {Sari}, {Djorgovski},
  {Bloom}, {Galama}, {Reichart}, {Berger}, {Harrison}, {Price}, {Yost},
  {Diercks}, {Goodrich}, \& {Chaffee}}]{frai01}
---. 2001, \apjl, 562, L55

\bibitem[{{Frail} {et~al.}(2000){Frail}, {Waxman}, \& {Kulkarni}}]{frai00}
{Frail}, D.~A., {Waxman}, E., \& {Kulkarni}, S.~R. 2000, \apj, 537, 191

\bibitem[{{Frail} {et~al.}(2003){Frail}, {Yost}, {Berger}, {Harrison}, {Sari},
  {Kulkarni}, {Taylor}, {Bloom}, {Fox}, {Moriarty-Schieven}, \&
  {Price}}]{frai03}
{Frail}, D.~A. {et~al.} 2003, \apj, 590, 992

\bibitem[{{Frontera} {et~al.}(2001){Frontera}, {Amati}, {Vietri}, {Zand},
  {Costa}, {Feroci}, {Heise}, {Masetti}, {Nicastro}, {Orlandini}, {Palazzi},
  {Pian}, {Piro}, \& {Soffitta}}]{fron01}
{Frontera}, F. {et~al.} 2001, \apjl, 550, L47

\bibitem[{{Frontera} {et~al.}(2000){Frontera}, {Antonelli}, {Amati},
  {Montanari}, {Costa}, {Dal Fiume}, {Giommi}, {Feroci}, {Gennaro}, {Heise},
  {Masetti}, {Muller}, {Nicastro}, {Orlandini}, {Palazzi}, {Pian}, {Piro},
  {Soffitta}, {Stornelli}, {in 't Zand}, {Frail}, {Kulkarni}, \&
  {Vietri}}]{fron00b}
---. 2000, \apj, 540, 697

\bibitem[{{Fynbo} {et~al.}(2003){Fynbo}, {Hjorth}, {Gorosabel}, {Vreeswijk}, \&
  {Rhoads}}]{fynb_gcn_2327}
{Fynbo}, J.~P.~U., {Hjorth}, J., {Gorosabel}, J., {Vreeswijk}, P.~M., \&
  {Rhoads}, J.~E. 2003, {GCN Report} 2327

\bibitem[{{Fynbo} {et~al.}(2001){Fynbo}, {Jensen}, {Gorosabel}, {Hjorth},
  {Pedersen}, {M{\o}ller}, {Abbott}, {Castro-Tirado}, {Delgado}, {Greiner},
  {Henden}, {Magazz{\` u}}, {Masetti}, {Merlino}, {Masegosa}, {{\O}stensen},
  {Palazzi}, {Pian}, {Schwarz}, {Cline}, {Guidorzi}, {Goldsten}, {Hurley},
  {Mazets}, {McClanahan}, {Montanari}, {Starr}, \& {Trombka}}]{fynb01}
{Fynbo}, J.~U. {et~al.} 2001, \aap, 369, 373

\bibitem[{{Galama} {et~al.}(2003){Galama}, {Reichart}, {Brown}, {Kimble},
  {Price}, {Berger}, {Frail}, {Kulkarni}, {Yost}, {Gal-Yam}, {Bloom},
  {Harrison}, {Sari}, {Fox}, \& {Djorgovski}}]{gala03b}
{Galama}, T.~J. {et~al.} 2003, \apj, 587, 135

\bibitem[{{Galassi} {et~al.}(2004){Galassi}, {Ricker}, {Atteia}, {Kawai},
  {Lamb}, {Woosley}, {Donaghy}, {Fenimore}, {Graziani}, {Matsuoka}, {Nakagawa},
  {Sakamoto}, {Sato}, {Shirasaki}, {Suzuki}, {Tamagawa}, {Urata}, {Yamazaki},
  {Yamamoto}, {Yoshida}, {Butler}, {Crew}, {Doty}, {Dullighan}, {Prigozhin},
  {Vanderspek}, {Villasenor}, {Jernigan}, {Levine}, {Azzibrouck}, {Braga},
  {Manchanda}, {Pizzichini}, {Barraud}, {Boer}, {Olive}, {Dezalay}, \&
  {Hurley}}]{gala_gcn_2770}
{Galassi}, M. {et~al.} 2004, {GCN Report} 2770

\bibitem[{{Garnavich} {et~al.}(2003){Garnavich}, {Stanek}, {Wyrzykowski},
  {Infante}, {Bendek}, {Bersier}, {Holland}, {Jha}, {Matheson}, {Kirshner},
  {Krisciunas}, {Phillips}, \& {Carlberg}}]{garn03}
{Garnavich}, P.~M. {et~al.} 2003, \apj, 582, 924

\bibitem[{{Gehrels} {et~al.}(2004){Gehrels}, {Chincarini}, {Giommi}, {Mason},
  {Nousek}, {Wells}, {White}, {Barthelmy}, {Burrows}, {Cominsky}, {Hurley},
  {Marshall}, {M{\' e}sz{\' a}ros}, {Roming}, {Angelini}, {Barbier}, {Belloni},
  {Campana}, {Caraveo}, {Chester}, {Citterio}, {Cline}, {Cropper}, {Cummings},
  {Dean}, {Feigelson}, {Fenimore}, {Frail}, {Fruchter}, {Garmire}, {Gendreau},
  {Ghisellini}, {Greiner}, {Hill}, {Hunsberger}, {Krimm}, {Kulkarni}, {Kumar},
  {Lebrun}, {Lloyd-Ronning}, {Markwardt}, {Mattson}, {Mushotzky}, {Norris},
  {Osborne}, {Paczynski}, {Palmer}, {Park}, {Parsons}, {Paul}, {Rees},
  {Reynolds}, {Rhoads}, {Sasseen}, {Schaefer}, {Short}, {Smale}, {Smith},
  {Stella}, {Tagliaferri}, {Takahashi}, {Tashiro}, {Townsley}, {Tueller},
  {Turner}, {Vietri}, {Voges}, {Ward}, {Willingale}, {Zerbi}, \&
  {Zhang}}]{gehr04}
{Gehrels}, N. {et~al.} 2004, \apj, 611, 1005

\bibitem[{{Ghirlanda} {et~al.}(2004{\natexlab{a}}){Ghirlanda}, {Ghisellini}, \&
  {Lazzati}}]{ghir04}
{Ghirlanda}, G., {Ghisellini}, G., \& {Lazzati}, D. 2004{\natexlab{a}}, \apj,
  616, 331

\bibitem[{{Ghirlanda} {et~al.}(2004{\natexlab{b}}){Ghirlanda}, {Ghisellini},
  {Lazzati}, \& {Firmani}}]{ghir_rome04}
{Ghirlanda}, G., {Ghisellini}, G., {Lazzati}, D., \& {Firmani}, C.
  2004{\natexlab{b}}, {Presentation, Gamma-Ray Bursts in the Afterglow Era: 4th
  Workshop, Rome, Italy, Oct 18-22, 2004}

\bibitem[{{Ghirlanda} {et~al.}(2004{\natexlab{c}}){Ghirlanda}, {Ghisellini},
  {Lazzati}, \& {Firmani}}]{ghir04b}
---. 2004{\natexlab{c}}, \apjl, 613, L13

\bibitem[{{Golenetskii} {et~al.}(2004){Golenetskii}, {Aptekar}, {Mazets},
  {Pal'shin}, {Frederiks}, \& {Cline}}]{gole_gcn_2754}
{Golenetskii}, S., {Aptekar}, R., {Mazets}, E., {Pal'shin}, V., {Frederiks},
  D., \& {Cline}, T. 2004, {GCN Report} 2754

\bibitem[{{Gorosabel} {et~al.}(2002){Gorosabel}, {Hjorth}, {Pursimo}, {Kaas},
  {Fynbo}, {Moller}, {Jensen}, {Pedersen}, \& {Andersen}}]{goro_gcn_1224}
{Gorosabel}, J. {et~al.} 2002, {GCN Report} 1224

\bibitem[{{Greiner} {et~al.}(2003){Greiner}, {Guenther}, {Klose}, \&
  {Schwarz}}]{grei_gcn_1886}
{Greiner}, J., {Guenther}, E., {Klose}, S., \& {Schwarz}, R. 2003, {GCN Report}
  1886

\bibitem[{{Groot} {et~al.}(1998){Groot}, {Galama}, {Vreeswijk}, {Wijers},
  {Pian}, {Palazzi}, {van Paradijs}, {Kouveliotou}, {in 't Zand}, {Heise},
  {Robinson}, {Tanvir}, {Lidman}, {Tinney}, {Keane}, {Briggs}, {Hurley},
  {Gonzalez}, {Hall}, {Smith}, {Covarrubias}, {Jonker}, {Casares}, {Frontera},
  {Feroci}, {Piro}, {Costa}, {Smith}, {Jones}, {Windridge}, {Bland-Hawthorn},
  {Veilleux}, {Garcia}, {Brown}, {Stanek}, {Castro-Tirado}, {Gorosabel},
  {Greiner}, {Jaeger}, {Bohm}, \& {Fricke}}]{groo98}
{Groot}, P.~J. {et~al.} 1998, \apjl, 502, L123

\bibitem[{{Guetta} {et~al.}(2001){Guetta}, {Spada}, \& {Waxman}}]{guet01}
{Guetta}, D., {Spada}, M., \& {Waxman}, E. 2001, \apj, 557, 399

\bibitem[{{Halpern} \& {Fesen}(1998)}]{halp_gcn_134}
{Halpern}, J.~P., \& {Fesen}, R. 1998, {GCN Report} 134

\bibitem[{{Halpern} {et~al.}(2000){Halpern}, {Uglesich}, {Mirabal}, {Kassin},
  {Thorstensen}, {Keel}, {Diercks}, {Bloom}, {Harrison}, {Mattox}, \&
  {Eracleous}}]{halp00}
{Halpern}, J.~P. {et~al.} 2000, \apj, 543, 697

\bibitem[{{Hamuy} {et~al.}(1995){Hamuy}, {Phillips}, {Maza}, {Suntzeff},
  {Schommer}, \& {Aviles}}]{hamu95}
{Hamuy}, M., {Phillips}, M.~M., {Maza}, J., {Suntzeff}, N.~B., {Schommer},
  R.~A., \& {Aviles}, R. 1995, \aj, 109, 1

\bibitem[{{Hamuy} {et~al.}(1996){Hamuy}, {Phillips}, {Suntzeff}, {Schommer},
  {Maza}, {Smith}, {Lira}, \& {Aviles}}]{hamu96}
{Hamuy}, M., {Phillips}, M.~M., {Suntzeff}, N.~B., {Schommer}, R.~A., {Maza},
  J., {Smith}, R.~C., {Lira}, P., \& {Aviles}, R. 1996, \aj, 112, 2438

\bibitem[{{Harrison} {et~al.}(1999){Harrison}, {Bloom}, {Frail}, {Sari},
  {Kulkarni}, {Djorgovski}, {Axelrod}, {Mould}, {Schmidt}, {Wieringa}, {Wark},
  {Subrahmanyan}, {McConnell}, {McCarthy}, {Schaefer}, {McMahon}, {Markze},
  {Firth}, {Soffitta}, \& {Amati}}]{harr99}
{Harrison}, F.~A. {et~al.} 1999, \apjl, 523, L121

\bibitem[{{Harrison} {et~al.}(2001){Harrison}, {Yost}, {Sari}, {Berger},
  {Galama}, {Holtzman}, {Axelrod}, {Bloom}, {Chevalier}, {Costa}, {Diercks},
  {Djorgovski}, {Frail}, {Frontera}, {Hurley}, {Kulkarni}, {McCarthy}, {Piro},
  {Pooley}, {Price}, {Reichart}, {Ricker}, {Shepherd}, {Schmidt}, {Walter}, \&
  {Wheeler}}]{harr01}
---. 2001, \apj, 559, 123

\bibitem[{{Heise} {et~al.}(2001){Heise}, {in't Zand}, {Kippen}, \&
  {Woods}}]{heis01}
{Heise}, J., {in't Zand}, J., {Kippen}, R.~M., \& {Woods}, P.~M. 2001, in
  Gamma-ray Bursts in the Afterglow Era, 16

\bibitem[{{Hjorth} {et~al.}(2003){Hjorth}, {M{\o}ller}, {Gorosabel}, {Fynbo},
  {Toft}, {Jaunsen}, {Kaas}, {Pursimo}, {Torii}, {Kato}, {Yamaoka}, {Yoshida},
  {Thomsen}, {Andersen}, {Burud}, {Castro Cer{\' o}n}, {Castro-Tirado},
  {Fruchter}, {Kaper}, {Kouveliotou}, {Masetti}, {Palazzi}, {Pedersen}, {Pian},
  {Rhoads}, {Rol}, {Tanvir}, {Vreeswijk}, {Wijers}, \& {van den
  Heuvel}}]{hjor03b}
{Hjorth}, J. {et~al.} 2003, \apj, 597, 699

\bibitem[{{Holland} {et~al.}(2004){Holland}, {Bersier}, {Bloom}, {Garnavich},
  {Caldwell}, {Challis}, {Kirshner}, {Luhman}, {McLeod}, \& {Stanek}}]{holl04}
{Holland}, S.~T. {et~al.} 2004, \aj, 128, 1955

\bibitem[{{Holland} {et~al.}(2002){Holland}, {Soszy{\' n}ski}, {Gladders},
  {Barrientos}, {Berlind}, {Bersier}, {Garnavich}, {Jha}, \& {Stanek}}]{holl02}
---. 2002, \aj, 124, 639

\bibitem[{{Huang} {et~al.}(2004){Huang}, {Wu}, {Dai}, {Ma}, \& {Lu}}]{huan04}
{Huang}, Y.~F., {Wu}, X.~F., {Dai}, Z.~G., {Ma}, H.~T., \& {Lu}, T. 2004, \apj,
  605, 300

\bibitem[{{Hurley} {et~al.}(2000){Hurley}, {Cline}, {Mazets}, {Aptekar},
  {Golenetskii}, {Frederiks}, {Frail}, {Kulkarni}, {Trombka}, {McClanahan},
  {Starr}, \& {Goldsten}}]{hurl00}
{Hurley}, K. {et~al.} 2000, \apjl, 534, L23

\bibitem[{{Hurley} {et~al.}(2001){Hurley}, {Cline}, {Mazets}, {Golenetskii},
  {Guidorzi}, {Montanari}, \& {Frontera}}]{hurl_gcn_736}
{Hurley}, K., {Cline}, T., {Mazets}, E., {Golenetskii}, S., {Guidorzi}, C.,
  {Montanari}, E., \& {Frontera}, F. 2001, {GCN Report} 736

\bibitem[{{Inoue}(2004)}]{inou04}
{Inoue}, S. 2004, \mnras, 348, 999

\bibitem[{{in't Zand} {et~al.}(2001){in't Zand}, {Kuiper}, {Amati},
  {Antonelli}, {Butler}, {Costa}, {Feroci}, {Frontera}, {Gandolfi}, {Guidorzi},
  {Heise}, {Kaptein}, {Kuulkers}, {Nicastro}, {Piro}, {Soffitta}, \&
  {Tavani}}]{intz01}
{in't Zand}, J.~J.~M. {et~al.} 2001, \apj, 559, 710

\bibitem[{{Israel} {et~al.}(1999){Israel}, {Marconi}, {Covino}, {Lazzati},
  {Ghisellini}, {Campana}, {Guzzo}, {Guerrero}, \& {Stella}}]{isra99}
{Israel}, G.~L. {et~al.} 1999, \aap, 348, L5

\bibitem[{{Jakobsson} {et~al.}(2003){Jakobsson}, {Hjorth}, {Fynbo},
  {Gorosabel}, {Pedersen}, {Burud}, {Levan}, {Kouveliotou}, {Tanvir},
  {Fruchter}, {Rhoads}, {Grav}, {Hansen}, {Michelsen}, {Andersen}, {Jensen},
  {Pedersen}, {Thomsen}, {Weidinger}, {Bhargavi}, {Cowsik}, \&
  {Pandey}}]{jako03}
{Jakobsson}, P. {et~al.} 2003, \aap, 408, 941

\bibitem[{{Jakobsson} {et~al.}(2004){Jakobsson}, {Hjorth}, {Fynbo},
  {Weidinger}, {Gorosabel}, {Ledoux}, {Watson}, {Bj{\" o}rnsson},
  {Gudmundsson}, {Wijers}, {M{\" o}ller}, {Pedersen}, {Sollerman}, {Henden},
  {Jensen}, {Gilmore}, {Kilmartin}, {Levan}, {Castro Cer{\' o}n},
  {Castro-Tirado}, {Fruchter}, {Kouveliotou}, {Masetti}, \& {Tanvir}}]{jako04}
---. 2004, \aap, 427, 785

\bibitem[{{Jaunsen} {et~al.}(2001){Jaunsen}, {Hjorth}, {Bj{\" o}rnsson},
  {Andersen}, {Pedersen}, {Kjernsmo}, {Korhonen}, {S{\o}rensen}, \&
  {Palazzi}}]{jaun01}
{Jaunsen}, A.~O. {et~al.} 2001, \apj, 546, 127

\bibitem[{{Jensen} {et~al.}(2001){Jensen}, {Fynbo}, {Gorosabel}, {Hjorth},
  {Holland}, {M{\" o}ller}, {Thomsen}, {Bj{\" o}rnsson}, {Pedersen}, {Burud},
  {Henden}, {Tanvir}, {Davis}, {Vreeswijk}, {Rol}, {Hurley}, {Cline},
  {Trombka}, {McClanahan}, {Starr}, {Goldsten}, {Castro-Tirado}, {Greiner},
  {Bailer-Jones}, {K{\" u}mmel}, \& {Mundt}}]{jens01}
{Jensen}, B.~L. {et~al.} 2001, \aap, 370, 909

\bibitem[{{Jimenez} {et~al.}(2001){Jimenez}, {Band}, \& {Piran}}]{jime01}
{Jimenez}, R., {Band}, D., \& {Piran}, T. 2001, \apj, 561, 171

\bibitem[{{Kelson} {et~al.}(2004){Kelson}, {Koviak}, {Berger}, \&
  {Fox}}]{kels_gcn_2627}
{Kelson}, D.~D., {Koviak}, K., {Berger}, w.~E., \& {Fox}, D.~B. 2004, {GCN
  Report} 2627

\bibitem[{{Klose} {et~al.}(2004){Klose}, {Greiner}, {Rau}, {Henden},
  {Hartmann}, {Zeh}, {Ries}, {Masetti}, {Malesani}, {Guenther}, {Gorosabel},
  {Stecklum}, {Antonelli}, {Brinkworth}, {Cer{\' o}n}, {Castro-Tirado},
  {Covino}, {Fruchter}, {Fynbo}, {Ghisellini}, {Hjorth}, {Hudec},
  {Jel{\'{\i}}nek}, {Kaper}, {Kouveliotou}, {Lindsay}, {Maiorano}, {Mannucci},
  {Nysewander}, {Palazzi}, {Pedersen}, {Pian}, {Reichart}, {Rhoads}, {Rol},
  {Smail}, {Tanvir}, {de Ugarte Postigo}, {Vreeswijk}, {Wijers}, \& {van den
  Heuvel}}]{klos04}
{Klose}, S. {et~al.} 2004, \aj, 128, 1942

\bibitem[{{Kobayashi} {et~al.}(1997){Kobayashi}, {Piran}, \& {Sari}}]{koba97}
{Kobayashi}, S., {Piran}, T., \& {Sari}, R. 1997, \apj, 490, 92

\bibitem[{{Kobayashi} \& {Sari}(2001)}]{koba01}
{Kobayashi}, S., \& {Sari}, R. 2001, \apj, 551, 934

\bibitem[{{Kulkarni} {et~al.}(1998){Kulkarni}, {Djorgoski}, {Ramaprakash},
  {Goodrich}, {Bloom}, {Adelberger}, {Kundic}, {Lubin}, {Frail}, {Frontera},
  {Feroci}, {Nicastro}, {Barth}, {Davis}, {Filippenko}, \& {Newman}}]{kulk98}
{Kulkarni}, S.~R. {et~al.} 1998, \nat, 393, 35

\bibitem[{{Kulkarni} {et~al.}(1999){Kulkarni}, {Djorgovski}, {Odewahn},
  {Bloom}, {Gal}, {Koresko}, {Harrison}, {Lubin}, {Armus}, {Sari},
  {Illingworth}, {Kelson}, {Magee}, {van Dokkum}, {Frail}, {Mulchaey},
  {Malkan}, {McClean}, {Teplitz}, {Koerner}, {Kirkpatrick}, {Kobayashi},
  {Yadigaroglu}, {Halpern}, {Piran}, {Goodrich}, {Chaffee}, {Feroci}, \&
  {Costa}}]{kulk99}
---. 1999, \nat, 398, 389

\bibitem[{{Kumar}(1999)}]{kuma99}
{Kumar}, P. 1999, \apjl, 523, L113

\bibitem[{{Kumar} \& {Granot}(2003)}]{kuma03}
{Kumar}, P., \& {Granot}, J. 2003, \apj, 591, 1075

\bibitem[{{Lamb}(2003)}]{lamb03}
{Lamb}, D.~Q. 2003, in AIP Conf. Proc. 662: Gamma-Ray Burst and Afterglow
  Astronomy 2001: A Workshop Celebrating the First Year of the HETE Mission,
  433--437

\bibitem[{{Lamb} {et~al.}(1999){Lamb}, {Castander}, \& {Reichart}}]{lamb99}
{Lamb}, D.~Q., {Castander}, F.~J., \& {Reichart}, D.~E. 1999, \aaps, 138, 479

\bibitem[{{Lazzati} {et~al.}(1999){Lazzati}, {Ghisellini}, \&
  {Celotti}}]{lazz99}
{Lazzati}, D., {Ghisellini}, G., \& {Celotti}, A. 1999, \mnras, 309, L13

\bibitem[{{Le Floc'h} {et~al.}(2002){Le Floc'h}, {Duc}, {Mirabel}, {Sanders},
  {Bosch}, {Rodrigues}, {Courvoisier}, {Mereghetti}, \& {Melnick}}]{lefl02}
{Le Floc'h}, E. {et~al.} 2002, \apjl, 581, L81

\bibitem[{{Li} \& {Chevalier}(2003)}]{liz03}
{Li}, Z., \& {Chevalier}, R.~A. 2003, \apjl, 589, L69

\bibitem[{{Linder} \& {Collaboration}(2004)}]{lind04}
{Linder}, E.~V., \& {Collaboration}, f.~t.~S. 2004, {astro-ph/0406186}

\bibitem[{{Linder} \& {Huterer}(2003)}]{lind03}
{Linder}, E.~V., \& {Huterer}, D. 2003, \prd, 67, 081303

\bibitem[{{Lloyd-Ronning} {et~al.}(2004){Lloyd-Ronning}, {Dai}, \&
  {Zhang}}]{lloy04}
{Lloyd-Ronning}, N.~M., {Dai}, X., \& {Zhang}, B. 2004, \apj, 601, 371

\bibitem[{{Lloyd-Ronning} \& {Ramirez-Ruiz}(2002)}]{lloy02b}
{Lloyd-Ronning}, N.~M., \& {Ramirez-Ruiz}, E. 2002, {\apj}, 576, 101

\bibitem[{{Lloyd-Ronning} \& {Zhang}(2004)}]{lloy04b}
{Lloyd-Ronning}, N.~M., \& {Zhang}, B. 2004, \apj, 613, 477

\bibitem[{{Loeb} \& {Barkana}(2001)}]{loeb01}
{Loeb}, A., \& {Barkana}, R. 2001, \araa, 39, 19

\bibitem[{{M{\' e}sz{\' a}ros} \& {Rees}(2003)}]{mesz03}
{M{\' e}sz{\' a}ros}, P., \& {Rees}, M.~J. 2003, \apjl, 591, L91

\bibitem[{{M{\" o}ller} {et~al.}(2002){M{\" o}ller}, {Fynbo}, {Hjorth},
  {Thomsen}, {Egholm}, {Andersen}, {Gorosabel}, {Holland}, {Jakobsson},
  {Jensen}, {Pedersen}, {Pedersen}, \& {Weidinger}}]{moll02}
{M{\" o}ller}, P. {et~al.} 2002, \aap, 396, L21

\bibitem[{{MacFadyen} {et~al.}(2001){MacFadyen}, {Woosley}, \&
  {Heger}}]{macf01}
{MacFadyen}, A.~I., {Woosley}, S.~E., \& {Heger}, A. 2001, \apj, 550, 410

\bibitem[{{Martini} {et~al.}(2003){Martini}, {Garnavich}, \&
  {Stanek}}]{mart_gcn_1980}
{Martini}, P., {Garnavich}, P., \& {Stanek}, K.~Z. 2003, {GCN Report} 1980

\bibitem[{{Masetti} {et~al.}(2000){Masetti}, {Palazzi}, {Pian}, {Hunt}, {M{\'
  e}ndez}, {Frontera}, {Amati}, {Vreeswijk}, {Rol}, {Galama}, {van Paradijs},
  {Antonelli}, {Nicastro}, {Feroci}, {Marconi}, {Piro}, {Costa}, {Kouveliotou},
  {Castro-Tirado}, {Falomo}, {Augusteijn}, {B{\" o}hnhardt}, {Lidman}, {Vanzi},
  {Merrill}, {Kaminsky}, {van der Klis}, {Heemskerk}, {van der Hooft},
  {Kuulkers}, {Pedersen}, \& {Benetti}}]{mase00}
{Masetti}, N. {et~al.} 2000, \aap, 354, 473

\bibitem[{{Mazzali} {et~al.}(2001){Mazzali}, {Nomoto}, {Cappellaro},
  {Nakamura}, {Umeda}, \& {Iwamoto}}]{mazz01}
{Mazzali}, P.~A., {Nomoto}, K., {Cappellaro}, E., {Nakamura}, T., {Umeda}, H.,
  \& {Iwamoto}, K. 2001, \apj, 547, 988

\bibitem[{{Mirabal} {et~al.}(2002){Mirabal}, {Halpern}, {Kulkarni}, {Castro},
  {Bloom}, {Djorgovski}, {Galama}, {Harrison}, {Frail}, {Price}, {Reichart},
  {Ebeling}, {Bunker}, {Dawson}, {Dey}, {Spinrad}, \& {Stern}}]{mira02}
{Mirabal}, N. {et~al.} 2002, \apj, 578, 818

\bibitem[{{Nakar} \& {Piran}(2004)}]{naka04b}
{Nakar}, E., \& {Piran}, T. 2004, {astro-ph/0412232}

\bibitem[{{Nemiroff} {et~al.}(1998){Nemiroff}, {Norris}, {Bonnell}, \&
  {Marani}}]{nemi98}
{Nemiroff}, R.~J., {Norris}, J.~P., {Bonnell}, J.~T., \& {Marani}, G.~F. 1998,
  \apjl, 494, L173

\bibitem[{{Norris}(2002)}]{norr02}
{Norris}, J.~P. 2002, \apj, 579, 386

\bibitem[{{Norris} {et~al.}(2000){Norris}, {Marani}, \& {Bonnell}}]{norr00}
{Norris}, J.~P., {Marani}, G.~F., \& {Bonnell}, J.~T. 2000, \apj, 534, 248

\bibitem[{{Paczynski}(1998)}]{pacz98}
{Paczynski}, B. 1998, \apjl, 494, L45

\bibitem[{{Panaitescu}(2001)}]{pana01d}
{Panaitescu}, A. 2001, \apj, 556, 1002

\bibitem[{{Panaitescu} \& {Kumar}(2001)}]{pana01c}
{Panaitescu}, A., \& {Kumar}, P. 2001, \apj, 554, 667

\bibitem[{{Panaitescu} \& {Kumar}(2002)}]{pana02}
---. 2002, \apj, 571, 779

\bibitem[{{Panaitescu} \& {Kumar}(2004)}]{pana04}
---. 2004, \mnras, 353, 511

\bibitem[{{Pandey} {et~al.}(2003){Pandey}, {Sahu}, {Resmi}, {Sagar}, {Anupama},
  {Bhattacharya}, {Mohan}, {Prabhu}, {Bhatt}, {Pandey}, {Parihar}, \&
  {Castro-Tirado}}]{pand03}
{Pandey}, S.~B. {et~al.} 2003, Bulletin of the Astronomical Society of India,
  31, 19

\bibitem[{{Perlmutter} {et~al.}(1997){Perlmutter}, {Gabi}, {Goldhaber},
  {Goobar}, {Groom}, {Hook}, {Kim}, {Kim}, {Lee}, {Pain}, {Pennypacker},
  {Small}, {Ellis}, {McMahon}, {Boyle}, {Bunclark}, {Carter}, {Irwin},
  {Glazebrook}, {Newberg}, {Filippenko}, {Matheson}, {Dopita}, {Couch}, \& {The
  Supernova Cosmology Project}}]{perl97}
{Perlmutter}, S. {et~al.} 1997, \apj, 483, 565

\bibitem[{{Phillips}(1993)}]{phil93}
{Phillips}, M.~M. 1993, \apjl, 413, L105

\bibitem[{{Pinto} \& {Eastman}(2001)}]{pint01}
{Pinto}, P.~A., \& {Eastman}, R.~G. 2001, New Astronomy, 6, 307

\bibitem[{{Piran} {et~al.}(2001){Piran}, {Kumar}, {Panaitescu}, \&
  {Piro}}]{pira01}
{Piran}, T., {Kumar}, P., {Panaitescu}, A., \& {Piro}, L. 2001, \apjl, 560,
  L167

\bibitem[{{Piro} {et~al.}(2002){Piro}, {Frail}, {Gorosabel}, {Garmire},
  {Soffitta}, {Amati}, {Andersen}, {Antonelli}, {Berger}, {Frontera}, {Fynbo},
  {Gandolfi}, {Garcia}, {Hjorth}, {Zand}, {Jensen}, {Masetti}, {M{\o}ller},
  {Pedersen}, {Pian}, \& {Wieringa}}]{piro02}
{Piro}, L. {et~al.} 2002, \apj, 577, 680

\bibitem[{{Porciani} \& {Madau}(2001)}]{porc01}
{Porciani}, C., \& {Madau}, P. 2001, \apj, 548, 522

\bibitem[{{Preece} {et~al.}(2000){Preece}, {Briggs}, {Mallozzi}, {Pendleton},
  {Paciesas}, \& {Band}}]{pree00}
{Preece}, R.~D., {Briggs}, M.~S., {Mallozzi}, R.~S., {Pendleton}, G.~N.,
  {Paciesas}, W.~S., \& {Band}, D.~L. 2000, \apjs, 126, 19

\bibitem[{{Press} {et~al.}(1992){Press}, {Teukolsky}, {Vetterling}, \&
  {Flannery}}]{pres92_c}
{Press}, W.~H., {Teukolsky}, S.~A., {Vetterling}, W.~T., \& {Flannery}, B.~P.
  1992, {Numerical recipes in C. The art of scientific computing} (Cambridge:
  University Press, 1992, 2nd ed.)

\bibitem[{{Price} {et~al.}(2002{\natexlab{a}}){Price}, {Berger}, {Kulkarni},
  {Djorgovski}, {Fox}, {Mahabal}, {Hurley}, {Bloom}, {Frail}, {Galama},
  {Harrison}, {Morrison}, {Reichart}, {Yost}, {Sari}, {Axelrod}, {Cline},
  {Golenetskii}, {Mazets}, {Schmidt}, \& {Trombka}}]{pric02a}
{Price}, P.~A. {et~al.} 2002{\natexlab{a}}, \apj, 573, 85

\bibitem[{{Price} {et~al.}(2002{\natexlab{b}}){Price}, {Berger}, {Reichart},
  {Kulkarni}, {Yost}, {Subrahmanyan}, {Wark}, {Wieringa}, {Frail}, {Bailey},
  {Boyle}, {Corbett}, {Gunn}, {Ryder}, {Seymour}, {Koviak}, {McCarthy},
  {Phillips}, {Axelrod}, {Bloom}, {Djorgovski}, {Fox}, {Galama}, {Harrison},
  {Hurley}, {Sari}, {Schmidt}, {Brown}, {Cline}, {Frontera}, {Guidorzi}, \&
  {Montanari}}]{pric02b}
---. 2002{\natexlab{b}}, \apjl, 572, L51

\bibitem[{{Price} {et~al.}(2003{\natexlab{a}}){Price}, {Fox}, {Kulkarni},
  {Peterson}, {Schmidt}, {Soderberg}, {Yost}, {Berger}, {Djorgovski}, {Frail},
  {Harrison}, {Sari}, {Blain}, \& {Chapman}}]{pric03b}
---. 2003{\natexlab{a}}, \nat, 423, 844

\bibitem[{{Price} {et~al.}(2001){Price}, {Harrison}, {Galama}, {Reichart},
  {Axelrod}, {Berger}, {Bloom}, {Busche}, {Cline}, {Diercks}, {Djorgovski},
  {Frail}, {Gal-Yam}, {Halpern}, {Holtzman}, {Hunt}, {Hurley}, {Jacoby},
  {Kimble}, {Kulkarni}, {Mirabal}, {Morrison}, {Ofek}, {Pevunova}, {Sari},
  {Schmidt}, {Turnshek}, \& {Yost}}]{pric01}
---. 2001, \apjl, 549, L7

\bibitem[{{Price} {et~al.}(2002{\natexlab{c}}){Price}, {Kulkarni}, {Berger},
  {Djorgovski}, {Frail}, {Mahabal}, {Fox}, {Harrison}, {Bloom}, {Yost},
  {Reichart}, {Henden}, {Ricker}, {van der Spek}, {Hurley}, {Atteia}, {Kawai},
  {Fenimore}, \& {Graziani}}]{pric02c}
---. 2002{\natexlab{c}}, \apjl, 571, L121

\bibitem[{{Price} {et~al.}(2003{\natexlab{b}}){Price}, {Kulkarni}, {Berger},
  {Fox}, {Bloom}, {Djorgovski}, {Frail}, {Galama}, {Harrison}, {McCarthy},
  {Reichart}, {Sari}, {Yost}, {Jerjen}, {Flint}, {Phillips}, {Warren},
  {Axelrod}, {Chevalier}, {Holtzman}, {Kimble}, {Schmidt}, {Wheeler},
  {Frontera}, {Costa}, {Piro}, {Hurley}, {Cline}, {Guidorzi}, {Montanari},
  {Mazets}, {Golenetskii}, {Mitrofanov}, {Anfimov}, {Kozyrev}, {Litvak},
  {Sanin}, {Boynton}, {Fellows}, {Harshman}, {Shinohara}, {Gal-Yam}, {Ofek}, \&
  {Lipkin}}]{pric03c}
---. 2003{\natexlab{b}}, \apj, 589, 838

\bibitem[{{Price} {et~al.}(2003{\natexlab{c}}){Price}, {Kulkarni}, {Schmidt},
  {Galama}, {Bloom}, {Berger}, {Frail}, {Djorgovski}, {Fox}, {Henden}, {Klose},
  {Harrison}, {Reichart}, {Sari}, {Yost}, {Axelrod}, {McCarthy}, {Holtzman},
  {Halpern}, {Kimble}, {Wheeler}, {Chevalier}, {Hurley}, {Ricker}, {Costa},
  {Frontera}, \& {Piro}}]{pric03a}
---. 2003{\natexlab{c}}, \apj, 584, 931

\bibitem[{{Price} {et~al.}(2004){Price}, {Roth}, {Rich}, {Schmidt}, {Peterson},
  {Cowie}, {Smith}, \& {Rest}}]{pric_gcn_2791}
{Price}, P.~A., {Roth}, K., {Rich}, J., {Schmidt}, B.~P., {Peterson}, B.~A.,
  {Cowie}, L., {Smith}, C., \& {Rest}, A. 2004, {GCN Report} 2791

\bibitem[{{Prochaska} {et~al.}(2003){Prochaska}, {Bloom}, {Chen}, {Hurley},
  {Dressler}, \& {Osip}}]{proc_gcn_2482}
{Prochaska}, J.~X., {Bloom}, J.~S., {Chen}, H.~W., {Hurley}, K., {Dressler},
  A., \& {Osip}, D. 2003, {GCN Report} 2482

\bibitem[{{R{\" o}pke} \& {Hillebrandt}(2004)}]{ropk04}
{R{\" o}pke}, F.~K., \& {Hillebrandt}, W. 2004, \aap, 420, L1

\bibitem[{{Ramirez-Ruiz}(2004)}]{rami_rome04}
{Ramirez-Ruiz}, E. 2004, {Presentation, Gamma-Ray Bursts in the Afterglow Era:
  4th Workshop, Rome, Italy, Oct 18-22, 2004}

\bibitem[{{Ramirez-Ruiz} {et~al.}(2001){Ramirez-Ruiz}, {Dray}, {Madau}, \&
  {Tout}}]{rami01}
{Ramirez-Ruiz}, E., {Dray}, L.~M., {Madau}, P., \& {Tout}, C.~A. 2001, \mnras,
  327, 829

\bibitem[{{Rau} {et~al.}(2004){Rau}, {Greiner}, {Klose}, {Salvato}, {Castro
  Cer{\' o}n}, {Hartmann}, {Fruchter}, {Levan}, {Tanvir}, {Gorosabel},
  {Hjorth}, {Zeh}, {K{\" u}pc{\" u} Yolda{\c s}}, {Beaulieu}, {Donatowicz},
  {Vinter}, {Castro-Tirado}, {Fynbo}, {Kann}, {Kouveliotou}, {Masetti},
  {M{\o}ller}, {Palazzi}, {Pian}, {Rhoads}, {Wijers}, \& {van den
  Heuvel}}]{rau04}
{Rau}, A. {et~al.} 2004, \aap, 427, 815

\bibitem[{{Rees} \& {Meszaros}(2004)}]{rees04}
{Rees}, M.~J., \& {Meszaros}, P. 2004, {astro-ph/0412702}

\bibitem[{{Reichart} {et~al.}(2001){Reichart}, {Lamb}, {Fenimore},
  {Ramirez-Ruiz}, {Cline}, \& {Hurley}}]{reic01}
{Reichart}, D.~E., {Lamb}, D.~Q., {Fenimore}, E.~E., {Ramirez-Ruiz}, E.,
  {Cline}, T.~L., \& {Hurley}, K. 2001, \apj, 552, 57

\bibitem[{{Rhoads}(1997)}]{rhoa97b}
{Rhoads}, J.~E. 1997, \apjl, 487, L1

\bibitem[{{Riess} {et~al.}(2001){Riess}, {Nugent}, {Gilliland}, {Schmidt},
  {Tonry}, {Dickinson}, {Thompson}, {Budav{\' a}ri}, {Casertano}, {Evans},
  {Filippenko}, {Livio}, {Sanders}, {Shapley}, {Spinrad}, {Steidel}, {Stern},
  {Surace}, \& {Veilleux}}]{ries01}
{Riess}, A.~G. {et~al.} 2001, \apj, 560, 49

\bibitem[{{Riess} {et~al.}(1995){Riess}, {Press}, \& {Kirshner}}]{ries95}
{Riess}, A.~G., {Press}, W.~H., \& {Kirshner}, R.~P. 1995, \apjl, 438, L17

\bibitem[{{Riess} {et~al.}(1996){Riess}, {Press}, \& {Kirshner}}]{ries96}
---. 1996, \apj, 473, 88

\bibitem[{{Riess} {et~al.}(2004{\natexlab{a}}){Riess}, {Strolger}, {Tonry},
  {Casertano}, {Ferguson}, {Mobasher}, {Challis}, {Filippenko}, {Jha}, {Li},
  {Chornock}, {Kirshner}, {Leibundgut}, {Dickinson}, {Livio}, {Giavalisco},
  {Steidel}, {Ben{\'{\i}}tez}, \& {Tsvetanov}}]{ries04b}
{Riess}, A.~G. {et~al.} 2004{\natexlab{a}}, \apj, 607, 665

\bibitem[{{Riess} {et~al.}(2004{\natexlab{b}}){Riess}, {Strolger}, {Tonry},
  {Tsvetanov}, {Casertano}, {Ferguson}, {Mobasher}, {Challis}, {Panagia},
  {Filippenko}, {Li}, {Chornock}, {Kirshner}, {Leibundgut}, {Dickinson},
  {Koekemoer}, {Grogin}, \& {Giavalisco}}]{ries04a}
---. 2004{\natexlab{b}}, \apjl, 600, L163

\bibitem[{{Rossi} {et~al.}(2002){Rossi}, {Lazzati}, \& {Rees}}]{ross02}
{Rossi}, E., {Lazzati}, D., \& {Rees}, M.~J. 2002, \mnras, 332, 945

\bibitem[{{Rutledge} {et~al.}(1995){Rutledge}, {Hui}, \& {Lewin}}]{rutl95}
{Rutledge}, R.~E., {Hui}, L., \& {Lewin}, W.~H.~G. 1995, \mnras, 276, 753

\bibitem[{{Sagar} {et~al.}(2000){Sagar}, {Mohan}, {Pandey}, {Pandey}, \&
  {Castro-Tirado}}]{saga00}
{Sagar}, R., {Mohan}, V., {Pandey}, A.~K., {Pandey}, S.~B., \& {Castro-Tirado},
  A.~J. 2000, Bulletin of the Astronomical Society of India, 28, 15

\bibitem[{{Sakamoto} {et~al.}(2004){Sakamoto}, {Lamb}, {Graziani}, {Donaghy},
  {Suzuki}, {Ricker}, {Atteia}, {Kawai}, {Yoshida}, {Shirasaki}, {Tamagawa},
  {Torii}, {Matsuoka}, {Fenimore}, {Galassi}, {Doty}, {Vanderspek}, {Crew},
  {Villasenor}, {Butler}, {Prigozhin}, {Jernigan}, {Barraud}, {Boer},
  {Dezalay}, {Olive}, {Hurley}, {Levine}, {Monnelly}, {Martel}, {Morgan},
  {Woosley}, {Cline}, {Braga}, {Manchanda}, {Pizzichini}, {Takagishi}, \&
  {Yamauchi}}]{saka04b}
{Sakamoto}, T. {et~al.} 2004, {astro-ph/0409128}

\bibitem[{{Sari} {et~al.}(1999){Sari}, {Piran}, \& {Halpern}}]{sari99}
{Sari}, R., {Piran}, T., \& {Halpern}, J.~P. 1999, \apjl, 519, L17

\bibitem[{{Sazonov} {et~al.}(2004){Sazonov}, {Lutovinov}, \&
  {Sunyaev}}]{sazo04}
{Sazonov}, S.~Y., {Lutovinov}, A.~A., \& {Sunyaev}, R.~A. 2004, \nat, 430, 646

\bibitem[{{Schaefer}(2003)}]{scha03a}
{Schaefer}, B.~E. 2003, \apjl, 583, L67

\bibitem[{{Schaefer} {et~al.}(2001){Schaefer}, {Deng}, \& {Band}}]{scha01}
{Schaefer}, B.~E., {Deng}, M., \& {Band}, D.~L. 2001, \apjl, 563, L123

\bibitem[{{Schaefer} {et~al.}(2003){Schaefer}, {Gerardy}, {H{\" o}flich},
  {Panaitescu}, {Quimby}, {Mader}, {Hill}, {Kumar}, {Wheeler}, {Eracleous},
  {Sigurdsson}, {M{\' e}sz{\' a}ros}, {Zhang}, {Wang}, {Hessman}, \&
  {Petrosian}}]{scha03c}
{Schaefer}, B.~E. {et~al.} 2003, \apj, 588, 387

\bibitem[{{Schmidt}(2001)}]{schmidt01a}
{Schmidt}, M. 2001, \apj, 552, 36

\bibitem[{{Soderberg} {et~al.}(2004){Soderberg}, {Kulkarni}, {Berger}, {Fox},
  {Price}, {Yost}, {Hunt}, {Frail}, {Walker}, {Hamuy}, {Shectman}, {Halpern},
  \& {Mirabal}}]{sode04}
{Soderberg}, A.~M. {et~al.} 2004, \apj, 606, 994

\bibitem[{{Spergel} {et~al.}(2003){Spergel}, {Verde}, {Peiris}, {Komatsu},
  {Nolta}, {Bennett}, {Halpern}, {Hinshaw}, {Jarosik}, {Kogut}, {Limon},
  {Meyer}, {Page}, {Tucker}, {Weiland}, {Wollack}, \& {Wright}}]{sper03}
{Spergel}, D.~N. {et~al.} 2003, \apjs, 148, 175

\bibitem[{{Stanek} {et~al.}(1999){Stanek}, {Garnavich}, {Kaluzny}, {Pych}, \&
  {Thompson}}]{stan99}
{Stanek}, K.~Z., {Garnavich}, P.~M., {Kaluzny}, J., {Pych}, W., \& {Thompson},
  I. 1999, \apjl, 522, L39

\bibitem[{{Takahashi} {et~al.}(2003){Takahashi}, {Oguri}, {Kotake}, \&
  {Ohno}}]{taka03}
{Takahashi}, K., {Oguri}, M., {Kotake}, K., \& {Ohno}, H. 2003,
  {astro-ph/035260}

\bibitem[{{Tegmark} {et~al.}(2004){Tegmark}, {Strauss}, {Blanton}, {Abazajian},
  {Dodelson}, {Sandvik}, {Wang}, {Weinberg}, {Zehavi}, {Bahcall}, {Hoyle},
  {Schlegel}, {Scoccimarro}, {Vogeley}, {Berlind}, {Budavari}, {Connolly},
  {Eisenstein}, {Finkbeiner}, {Frieman}, {Gunn}, {Hui}, {Jain}, {Johnston},
  {Kent}, {Lin}, {Nakajima}, {Nichol}, {Ostriker}, {Pope}, {Scranton},
  {Seljak}, {Sheth}, {Stebbins}, {Szalay}, {Szapudi}, {Xu}, {Annis},
  {Brinkmann}, {Burles}, {Castander}, {Csabai}, {Loveday}, {Doi}, {Fukugita},
  {Gillespie}, {Hennessy}, {Hogg}, {Ivezi{\' c}}, {Knapp}, {Lamb}, {Lee},
  {Lupton}, {McKay}, {Kunszt}, {Munn}, {O'Connell}, {Peoples}, {Pier},
  {Richmond}, {Rockosi}, {Schneider}, {Stoughton}, {Tucker}, {vanden Berk},
  {Yanny}, \& {York}}]{tegm04}
{Tegmark}, M. {et~al.} 2004, \prd, 69, 103501

\bibitem[{{Timmes} {et~al.}(2003){Timmes}, {Brown}, \& {Truran}}]{timm03}
{Timmes}, F.~X., {Brown}, E.~F., \& {Truran}, J.~W. 2003, \apjl, 590, L83

\bibitem[{{Tinney} {et~al.}(1998){Tinney}, {Stathakis}, {Cannon}, \&
  {Galama}}]{tin_iauc_6896}
{Tinney}, C., {Stathakis}, R., {Cannon}, R., \& {Galama}, T.~J. 1998, \iaucirc,
  6896, 1

\bibitem[{{Tonry} {et~al.}(2003){Tonry}, {Schmidt}, {Barris}, {Candia},
  {Challis}, {Clocchiatti}, {Coil}, {Filippenko}, {Garnavich}, {Hogan},
  {Holland}, {Jha}, {Kirshner}, {Krisciunas}, {Leibundgut}, {Li}, {Matheson},
  {Phillips}, {Riess}, {Schommer}, {Smith}, {Sollerman}, {Spyromilio},
  {Stubbs}, \& {Suntzeff}}]{tonr03}
{Tonry}, J.~L. {et~al.} 2003, \apj, 519, 1

\bibitem[{{van Dokkum} \& {Bloom}(2003)}]{vand_gcn_2380}
{van Dokkum}, P.~G., \& {Bloom}, J.~S. 2003, {GCN Report} 2380

\bibitem[{{Vanderspek}(2004)}]{hete2}
{Vanderspek}, R. 2004, {http://space.mit.edu/HETE/Bursts/Data/}

\bibitem[{{Vreeswijk} {et~al.}(2003){Vreeswijk}, {Fruchter}, {Hjorth}, \&
  {Kouveliotou}}]{vree_gcn_1785}
{Vreeswijk}, P., {Fruchter}, A., {Hjorth}, J., \& {Kouveliotou}, C. 2003, {GCN
  Report} 1785

\bibitem[{{Vreeswijk} {et~al.}(2004){Vreeswijk}, {Ellison}, {Ledoux}, {Wijers},
  {Fynbo}, {M{\o}ller}, {Henden}, {Hjorth}, {Masi}, {Rol}, {Jensen}, {Tanvir},
  {Levan}, {Castro Cer{\' o}n}, {Gorosabel}, {Castro-Tirado}, {Fruchter},
  {Kouveliotou}, {Burud}, {Rhoads}, {Masetti}, {Palazzi}, {Pian}, {Pedersen},
  {Kaper}, {Gilmore}, {Kilmartin}, {Buckle}, {Seigar}, {Hartmann}, {Lindsay},
  \& {van den Heuvel}}]{vree04}
{Vreeswijk}, P.~M. {et~al.} 2004, \aap, 419, 927

\bibitem[{{Vreeswijk} {et~al.}(2001){Vreeswijk}, {Fruchter}, {Kaper}, {Rol},
  {Galama}, {van Paradijs}, {Kouveliotou}, {Wijers}, {Pian}, {Palazzi},
  {Masetti}, {Frontera}, {Savaglio}, {Reinsch}, {Hessman}, {Beuermann},
  {Nicklas}, \& {van den Heuvel}}]{vree01}
---. 2001, \apj, 546, 672

\bibitem[{{Wang} \& {Garnavich}(2001)}]{wangy01}
{Wang}, Y., \& {Garnavich}, P.~M. 2001, \apj, 552, 445

\bibitem[{{Wang} {et~al.}(2002){Wang}, {Holz}, \& {Munshi}}]{wangy02}
{Wang}, Y., {Holz}, D.~E., \& {Munshi}, D. 2002, \apjl, 572, L15

\bibitem[{{Watson} {et~al.}(2004){Watson}, {Hjorth}, {Levan}, {Jakobsson},
  {O'Brien}, {Osborne}, {Pedersen}, {Reeves}, {Tedds}, {Vaughan}, {Ward}, \&
  {Willingale}}]{wats04}
{Watson}, D. {et~al.} 2004, \apjl, 605, L101

\bibitem[{{Weidinger} {et~al.}(2003){Weidinger}, {U}, {Hjorth}, {Gorosabel},
  {Klose}, \& {Tanvir}}]{weid_gcn_2215}
{Weidinger}, M., {U}, J.~P., {Hjorth}, J., {Gorosabel}, J., {Klose}, S., \&
  {Tanvir}, N. 2003, {GCN Report} 2215

\bibitem[{{Wiersema} {et~al.}(2004){Wiersema}, {C}, {Rol}, {Vreeswijk}, \&
  {M}}]{wier_gcn_2800}
{Wiersema}, K., {C}, R.~L., {Rol}, E., {Vreeswijk}, P., \& {M}, R.~A. 2004,
  {GCN Report} 2800

\bibitem[{{Woosley}(1993)}]{woos93}
{Woosley}, S.~E. 1993, \apj, 405, 273

\bibitem[{{Xu} {et~al.}(2005){Xu}, {Dai}, \& {Liang}}]{xu05}
{Xu}, D., {Dai}, Z.~G., \& {Liang}, E.~W. 2005, {astro-ph/0501458}

\bibitem[{{Yost} {et~al.}(2002){Yost}, {Frail}, {Harrison}, {Sari}, {Reichart},
  {Bloom}, {Kulkarni}, {Moriarty-Schieven}, {Djorgovski}, {Price}, {Goodrich},
  {Larkin}, {Walter}, {Shnote}, {Fox}, {Taylor}, {Berger}, \&
  {Galama}}]{yost02}
{Yost}, S.~A. {et~al.} 2002, \apj, 577, 155

\bibitem[{{Yost} {et~al.}(2003){Yost}, {Harrison}, {Sari}, \& {Frail}}]{yost03}
{Yost}, S.~A., {Harrison}, F.~A., {Sari}, R., \& {Frail}, D.~A. 2003, \apj,
  597, 459

\bibitem[{{Zhang} {et~al.}(2004){Zhang}, {Dai}, {Lloyd-Ronning}, \& {M{\'
  e}sz{\' a}ros}}]{zhanb04a}
{Zhang}, B., {Dai}, X., {Lloyd-Ronning}, N.~M., \& {M{\' e}sz{\' a}ros}, P.
  2004, \apjl, 601, L119

\bibitem[{{Zhang} \& {M{\' e}sz{\' a}ros}(2002)}]{zhanb02a}
{Zhang}, B., \& {M{\' e}sz{\' a}ros}, P. 2002, \apj, 571, 876

\end{thebibliography}
\end{document}